\documentclass[pre,aps,onecolumn,superscriptaddress,showpacs,nofootinbib]{revtex4-1}%{revtex4}
\usepackage{amssymb}
\usepackage{epsfig}
\usepackage{amsmath}
\usepackage{subfigure}
\usepackage{graphicx}
\usepackage{textcomp}
\usepackage{url}
\usepackage{float}
\usepackage{color}
\usepackage{cases}
\usepackage[normalem]{ulem}
\usepackage{enumitem} %enumitem is for the enumerate environment without indentation

\usepackage[colorlinks=true, urlcolor=blue, anchorcolor=blue, citecolor=blue,filecolor=blue,linkcolor=blue,menucolor=blue%,pagecolor=blue
]{hyperref}

\usepackage{color}
\definecolor{Blue}{rgb}{0.00, 0.00, 0.80}
\definecolor{Red}{rgb}{0.80, 0.00, 0.00}
\definecolor{Green}{rgb}{0.00, 0.50, 0.00}
\definecolor{Magenta}{rgb}{0.80, 0.00, 0.80}

\usepackage{fixmath}
\usepackage{epsfig}
\usepackage{textcomp}
\usepackage{epstopdf}

\newcommand{\antiquad}{\!\!\!\!\!\!\!\!}
\newcommand{\nn}{\nonumber}
\newcommand{\be}{\begin{equation}}
\newcommand{\ee}{\end{equation}}
\newcommand{\bea}{\begin{eqnarray}}
\newcommand{\eea}{\end{eqnarray}}

\begin{document}

%\title{Limiting behavior of the Tracy-Widom distribution at large Dyson index}

\title{The Tracy-Widom distribution at large Dyson index}

\author{Alain Comtet}
\affiliation{LPTMS, Universit\'e Paris Saclay, CNRS, 91405 Orsay, France}

\author{Pierre Le Doussal}
\address{Laboratoire de Physique de l'Ecole Normale Sup\'erieure, CNRS, ENS and PSL Universit\'e, Sorbonne Universit\'e, Universit\'e Paris Cit\'e,
24 rue Lhomond, 75005 Paris, France}

\author{Naftali R. Smith}
\email{naftalismith@gmail.com}
\affiliation{Racah Institute of Physics, Hebrew University of Jerusalem, Jerusalem 91904, Israel}

%\pacs{05.40.-a, 05.70.Np, 68.35.Ct}

\begin{abstract}
% We study the Tracy-Widom (TW) distribution $f_\beta(a)$ in the limit of large Dyson index $\beta \to +\infty$. By applying the saddle-point approximation on an associated problem of energy levels $E=-a$ of a quantum Hamiltonian defined by the stochastic Airy operator,
% we show that, at large $\beta$, the probability density function for the TW distribution scales as $f_\beta(a) \sim e^{-\beta s(E=-a)}$. We compute the large-deviation function $s(E)$ exactly by numerically solving the associated saddle-point equation, and calculate $s(E)$ analytically in limiting cases. The saddle-point approximation we employ amounts to applying the optimal fluctuation method to find the most likely realization of the stochastic Airy operator (SAO), conditioned on its ground-state energy being $E$.
% We extend these results to the $i$th largest eigenvalue ($i=1,2,3,\dots$) $a_i$ and to joint distributions of multiple eigenvalues.
% We also present an alternative derivation for calculating $s(E)$, by applying the weak-noise theory to the representation of the TW distribution in terms of a Ricatti diffusion process that is related to the SAO.
% \\

We study the Tracy-Widom (TW) distribution $f_\beta(a)$ in the limit of large Dyson index $\beta \to +\infty$.
This distribution describes the fluctuations of the rescaled largest eigenvalue $a_1$ of
the Gaussian (alias Hermite) ensemble (G$\beta$E) of (infinitely) large random matrices. 
We show that, at large $\beta$, its probability density function takes the large deviation form $f_\beta(a) \sim e^{-\beta \Phi(a)}$.
While the typical deviation of $a_1$ around its mean is Gaussian of
variance $O(1/\beta)$, this large deviation form describes the probability of rare events
with deviation $O(1)$, and governs the behavior of the higher cumulants.
We obtain the rate function $\Phi(a)$ as a solution of a Painlev\'{e} II equation.
We derive explicit formula for its large argument behavior, and for the lowest 
cumulants, up to order 4. We compute $\Phi(a)$ numerically for all $a$
and compare with exact numerical computations of the TW distribution at finite $\beta$.
These results are obtained by applying saddle-point approximations 
to an associated problem of energy levels $E=-a$, for a 
random quantum Hamiltonian defined by the stochastic Airy operator (SAO).
We employ two complementary approaches: (i) we use the optimal fluctuation method to 
find the most likely realization of the noise in the SAO, conditioned 
on its ground-state energy being $E$ (ii) we apply the weak-noise theory 
to the representation of the TW distribution 
in terms of a Ricatti diffusion process associated to the SAO.
We extend our results to the full Airy point process $a_1>a_2>\dots$
which describes all edge eigenvalues of the G$\beta$E, and
correspond to (minus) the higher energy levels of the SAO,
obtaining large deviation forms for the marginal distribution of $a_i$, 
the joint distributions, and the gap distributions.
\end{abstract}

\maketitle

\begin{widetext}

{%
	%\singlespacing
	\hypersetup{linkcolor=black}
	\tableofcontents
}

% {\red TODO: Baruch's comments about some of his papers we cited:

% A quick comment on the literature: in [73] we didn't use the OFM.
% Rather, we
% calculated a large-deviation asymptotic of P(H,T) at long times, using a
% previously existing exact
% representation of the solution. In particular, we proved that the
% $H^{5/2}$ tail,
% previously found at short times, remains there at H>>T  at arbitrary
% long times T>>1.

% One more literature comment: papers [90] and [91] did NOT deal with
% time-dependent potentials.
% }

\section{Introduction}

\subsection{Background}

Random matrix theory (RMT) continues to attract interest nearly a century after Wishart's celebrated work  \cite{wishart1928generalised}. One of the reasons for this continued interest is the useful applications of RMT to many topics \cite{mehta2004random,forrester,akemann2011oxford,BouchaudPottersBook}, including nuclear physics \cite{wigner1958distribution}, stochastic surface growth 
\cite{Johansson2000,majumdar2007courseH,krug} and trapped fermions \cite{marino2014phase,castillo2014spectral,dean2019noninteracting}.
Of central importance are statistical properties of the eigenvalues of random matrices \cite{mehta2004random,forrester}. These include the density of eigenvalues and their correlations, and also the extreme value statistics, i.e., the distribution of the largest eigenvalue of a random matrix, all of which have been shown to converge to universal limiting forms as the size of the matrix goes to infinity.

In particular, for $N \times N$ random matrices sampled from the Gaussian ensembles with Dyson index $\beta$, the distribution of the maximal eigenvalue (properly shifted and rescaled) converges, in the limit $N\to\infty$, to the Tracy-Widom 
distributions \cite{TW94,TW96, TracyReview}, which too depend on $\beta$.
 These distributions have appeared in many other contexts as well,
 ranging from statistical physics and probability theory to surface growth models \cite{Sph,Takeuchi,PLDKPZReview} and biological sequence matching problems \cite{TracyReview,forrester,krug}.
 % (for reviews see~\cite{AD,TracyReview,Deiftuni,maj-review,forrester,krug} and references thereof).
Although originally studied for the particular cases of the Gaussian Orthogonal Ensemble (GOE), Gaussian Unitary Ensemble (GUE) and Gaussian Symplectic Ensemble (GSE), corresponding to $\beta=1,2,4$ respectively \cite{TW94, TW96}, the Gaussian ensembles were subsequently extended to arbitrary $\beta > 0$ \cite{Dumitriu2002}, as were the Tracy-Widom distributions. 
The general ensemble, called G$\beta$E, involves $N \times N$ real symmetric tridiagonal matrices, with  independent and identically distributed (i.i.d.) Gaussian random variables on the diagonal, and independent inhomogeneous chi random variables on the upper (respectively lower) diagonal.

The joint probability density function (PDF) of the eigenvalues of the G$\beta$E ensemble is a remarkably
simple generalization to any $\beta>0$ of the Wigner-Dyson formula for the classical ensembles $\beta=1,2,4$,
namely
\be
\label{GbetaEjPDF}
P(\lambda_{1},\dots,\lambda_{N})\propto\prod_{i<j}\left|\lambda_{i}-\lambda_{j}\right|^{\beta}\exp\left(-\frac{\beta N}{4}\sum_{i}\lambda_{i}^{2}\right) \, .
\ee
Rewriting this joint PDF as $\sim \exp(- \beta {\cal H}[\lambda])$ reveals the
equivalent standard description as a classical Coulomb gas 
in canonical equilibrium at inverse temperature $\beta$, of energy function ${\cal H}[\lambda]$. 
In the large $N$ limit the eigenvalue density converges to the Wigner semicircle
$\rho_N(\lambda) := \frac{1}{N} \sum_i \delta(\lambda-\lambda_i) \to \rho_{\rm sc}(\lambda) = \frac{1}{2 \pi} \sqrt{(4-\lambda^2)_+}$, where we denote $(x)_+=\max(0,x)$,
with support $[-2,2]$. It is customary to choose the order $\lambda_1 > \lambda_2 > \dots$.
Focusing on the upper edge, the largest eigenvalues accumulate around $2$
with $O(N^{-2/3})$ deviations, i.e for $N \to + \infty$ and any fixed set of $i \geq 1$ they jointly behave as
\be  \label{APPDefinition}
\lambda_i = 2 + \frac{a_i}{N^{2/3}} \quad , \quad a_i \equiv N^{2/3}(\lambda_i-2) \, .
\ee 
The set of $\{ a_i \}_{i \geq 1}$ defines (in the limit $N \to +\infty$) the so-called Airy soft edge point process (APP), with $a_1>a_2> \dots$,
which is an infinite point process on $\mathbb{R}$. By definition, the largest point, $a_1$ 
is distributed according to the Tracy Widom $\beta$ distribution, denoted TW$\beta$, 
of PDF denoted $f_{\beta}(a)=F_{\beta}'(a)$.
For $\beta=1,2,4$ its cumulative distribution function (CDF) 
takes the well known forms, see e.g. \cite{RamirezRiderVirag2011} p 928,
\bea \label{TWexplicit}
&& F_{2}(a)=F(a)^{2}~,~~F_{1}(a)=F(a)E(a)~,~~F_{4}\left(2^{-2/3}a\right)=\frac{1}{2}\left[E(a)+\frac{1}{E(a)}\right]F(a) \, ,
 \\[1mm]
&& F(a)=e^{-\int_a^{+\infty} dt (t-a) q(t)^2 }
~,~~ E(a)=e^{-\frac{1}{2}\int_a^{+\infty} dt q(t) } \quad , %\quad q''(t)=t q(t) + 2 q(t)^3 \quad , \quad q(t) \underset{t \to +\infty}{\simeq}  \text{Ai}(t) 
\eea  
in terms of the Hastings-McLeod solution to the Painlev\'{e} II equation,  
\be
q''(t)=t q(t) + 2 q(t)^3 \quad , \quad q(t) \underset{t \to +\infty}{\simeq}  \text{Ai}(t) \, ,
\ee
where $\text{Ai}(t)$ is the Airy function. For $\beta=2$ the APP is a determinantal
point process, i.e. all its multi-point correlations can be expressed as determinants of the so-called
Airy kernel, and for $\beta=1,4$ it is a Pfaffian point process. 
This leads for $\beta=1,2,4$ to alternative expressions of $F_\beta(a)$ in terms of Fredholm determinants 
and Fredholm Pfaffians. 
For general $\beta$ however there is no explicit formula, and the APP is quite complicated to describe, being a subject of intense research in physics and mathematics. There are
explicit results for the right \cite{DV13,BorotNadal2012} and 
left \cite{BENM11} tails of the TW$\beta$ distribution
\be \label{tailsTWbeta}
1- F_\beta(a) \underset{a \to +\infty}{\simeq} c_\beta \, 
\frac{ e^{- \frac{2}{3} \beta a^{3/2} } }{a^{3 \beta/4} }\\
\quad , \quad  F_\beta(a) \underset{a \to -\infty}{\simeq} \tau_\beta \,
\frac{ e^{ - \beta \frac{|a|^3}{24} + \frac{\sqrt{2}}{3} (\frac{\beta}{2} -1) |a|^{3/2} } }{|a|^{\frac{1}{8} (3- \frac{2}{\beta} - \frac{\beta}{2})}}
\ee  
where the latter agrees with the higher order expansions 
obtained in \cite{BaikLeftTail} for $\beta=1,2,4$. 
For a review, and expressions of constants $c_\beta$, $\tau_\beta$ see 
\cite[Chap 3]{NadalThesis}. 
In \cite{BorotNadal2012}, a non-rigorous result is obtained for $c_\beta$ with general $\beta$, which reads
\be
\label{cBetaNadal}
c_\beta = \frac{\Gamma\left(\beta/2\right)}{\left(4\beta\right)^{\beta/2}2\pi} \, .
\ee

The G$\beta$E has another remarkable
property, it naturally extends to eigenvalue dynamics. Indeed, adding a time direction $t$,
the JPDF \eqref{GbetaEjPDF}
is also the stationary measure of the so-called $\beta$-Dyson Brownian motion ($\beta$-DBM)
\cite{mehta2004random,forrester},
$\{ \lambda_i(t) \}_{1 \leq i \leq N}$, which is simply the
overdamped dynamics associated to ${\cal H}[\lambda]$.
The $\beta$-DBM also has an edge limit $\{ a_i(t) \}_{i \geq 1}$,
called the Airy$_\beta$ line ensemble, or the extended APP
\cite{PrahoferSpohnAiry, Johansson03, OsadaDBMEdge2014,CorwinHammondALE,Landon2020ALEBeta,HuangALEBeta}.

Since the original definitions of the TW$\beta$ distributions in terms of %limiting forms for the distribution of 
%maximal 
the largest eigenvalues of random matrices, 
alternative representations %for the TW$\beta$ distributions 
have been found. One such representation yields the TW$\beta$ in terms of the distribution of the smallest %minimal 
eigenvalue of the ``stochastic Airy operator'' (SAO). This continuum operator arises 
as the edge limit of the G$\beta$E random tridiagonal matrices \cite{EdelmanSutton2007}. 
The SAO takes the 
%which takes the 
form of a Schr\"{o}dinger operator on the half-line, in presence
of a linear confining potential, and of an additional white
noise random potential of variance $4/\beta$ \cite{Edelman03Presentation, RamirezRiderVirag2011}.
In its equivalent Ricatti equation formulation it provides 
another useful 
representation of TW$\beta$ in terms of the probability that a particle, diffusing in presence of a time-dependent potential does not escape to infinity up to a given time \cite{BloemendalVirag,ViragReview,ESW14}.
These representations of $f_\beta(a)$, i.e. of the PDF of $a_1$, %can be 
immediately extend %ed 
to study the 
distribution of all points $a_i$ of the Airy point process \eqref{APPDefinition}.  
%distribution of the $i$th largest eigenvalue of random matrices.

Since analytical results for the G$\beta$E with $\beta \neq 1,2,4$ are
hard to obtain, it is natural to study the limit of large $\beta$ where some simplification
occurs. It corresponds to the low-temperature limit of the classical Coulomb gas mentioned 
above, but may also be interpreted as describing the ground state quantum
probability of trapped fermions with strong repulsive interactions \cite{SutherlandBook, Calogero1975, SLMS21}. For $\beta \gg 1$ the eigenvalues $\lambda_i$ 
form a ``crystal'' (confined in a quadratic potential), and converge, upon rescaling, 
to the zeroes of the Hermite polynomial $H_N$ (whose density
in the bulk converges at large $N$ to the semi-circle).
As for phonons in a standard crystal,
the typical fluctuations away from 
their equilibrium positions are  
Gaussian of order $O\left(1/\sqrt{\beta}\right)$. 
The covariance matrix $\langle \lambda_i \lambda_j \rangle_c =O(1/\beta)$ 
has been computed at large $N$ in the bulk of the semi-circle in 
\cite{Dumitriu2005,JW16},  which is related to the power-spectrum description of random matrices \cite{Relano2002, Relano2008, Riser2017, Kanzieper2023PhysD, Forrester2025PhysD}.
More recently in 
\cite{GorinInfiniteBeta,Touzo},  the calculation of the covariance matrix was 
extended to the dynamical covariance of the $\beta$-DBM. 
The edge limit, i.e. the fluctuations of the APP
at large $\beta$ were first studied in \cite{EPS14}.
It was shown that the $a_i$'s converge to the zeroes of
the Airy function, with $O(1/\sqrt{\beta})$ Gaussian fluctuations,
and the one point variance was obtained, see also \cite{AHV21}. 
The full covariance matrix $\langle a_i a_j \rangle_c$
(and its dynamical extension) 
was obtained more recently
in \cite{GorinInfiniteBeta,Touzo}.
Although the characterization of the APP as a Gaussian process at large $\beta$ holds
for typical fluctuations, it does not say anything about the atypical
large fluctuations. This information is carried in the 
higher order cumulants, which to our knowledge
have not been studied for the TW$\beta$ distribution $f_\beta(a)$, nor 
for the APP. In fact the full question of its large deviations
has not been addressed. Given the universality of this distribution it is an important outstanding problem.

The aim of this paper is to study the limiting scaling behavior at $\beta \to +\infty$ of the Tracy-Widom distribution and to obtain its large deviation form.
% The main goal of this paper is to study the limiting scaling behavior at $\beta \to +\infty$ of the Tracy-Widom distribution, which corresponds to the low-temperature limit of a Coulomb gas \cite{Touzo, Dumitriu2005, EPS14, AHV21}, but may also be interpreted as describing trapped fermions with strong repulsive interactions \cite{SutherlandBook, Calogero1975, SLMS21}. 
We achieve this goal using two different methods: (i) by applying to the stochastic Airy operator
the saddle-point approach that was derived in Ref.~\cite{SmithGroundState24} for general weakly-disordered quantum potentials.
(ii) by applying the weak-noise theory (WNT) to the ``diffusion'' (Ricatti) representation of the Tracy-Widom distribution. 
We find that as $\beta \to +\infty$, the TW$\beta$ distribution $f_\beta(a)$ becomes narrower and narrower, and fluctuations of $a$ from its typical value become exponentially unlikely as a function of $\beta$,
i.e. $f_\beta(a) \sim e^{- \beta \Phi(a)}$. This exponential scaling is given by a large-deviation principle. We obtain an exact characterization of 
the associated large deviation function $\Phi(a)$. Remarkably, it can again be
expressed in terms the solution of a Painlev\'{e} II equation, although it
is quite different from the result \eqref{TWexplicit} for $\beta=1,2,4$. 
This allows for an analytical determination in limiting cases, 
and for a numerical evaluation elsewhere.
We extend our results to some other large deviation properties of the APP, associated to 
the PDF of any of the $a_i$, and to the distribution of the gaps (note that for $\beta=2$ these
were studied in e.g. \cite{GregGaps}).

One should note that our results are different to those obtained by the standard minimization of the Coulomb gas free energy, a technique that has often been used to study large deviations in random matrix theory for large but finite $N$ \cite{MV09, MNSV09, MNSV11}. In particular our work is quite different from studying the large deviations of $\lambda_1$ at fixed $\beta$ and large $N$, which instead
takes the forms $P(\lambda_1) \sim \exp\left(- \beta N \Phi_+(\lambda_1-2)\right)$ and
$P(\lambda_1) \sim \exp\left(- \beta N^2 \Phi_-(|\lambda_1-2|)\right)$ for 
the right and left tails respectively \cite{DM06,DM08,MV09,BenArousLargest,FyodorovPLDTrivialization}. Indeed, the scaling behavior we find is different to the one that is typically obtained from the Coulomb gas method (and in particular, it cannot depend on $N$ since the only control parameter in our problem is $\beta$).
Importantly, our results are also different from the tail asymptotics of $f_\beta(a)$
at large $|a|$ and fixed $\beta$ displayed in \eqref{tailsTWbeta}, although we show that
they are consistent with them, through matching of $\Phi(a)$ at large $|a|$. 

Before presenting our results in the next subsection, let us give a bit more 
background for completeness. Some properties of the APP, in particular its
large deviations at general fixed $\beta$, and their connections to the 
1D KPZ equation, were studied in a number of papers, see e.g. \cite{BorodinGorinMomentMatch,GorinMoments1,GorinMoments2,PLDAlex4routes,PLDCorwinPRL,PLDProlhacAiry}.
The $\beta$ ensembles have also been constructed for Laguerre-Wishart and Jacobi ensembles
\cite{Dumitriu2002}, and the stochastic operator approach has been extended to the 
hard edge \cite{EdelmanSutton2007,RamirezRider2009,SAOHardToSoft}, to random matrices with spikes 
\cite{BloemendalVirag,BloemendalVirag2,ViragReview},
and very recently to block tridiagonal matrices \cite{RiderBlock}.
On the dynamics side, a characterization of the Laplace transform for the full Airy$_\beta$ line ensemble
was obtained in \cite{GorinLaplaceAiryLineEnsemble24}. 
Finally note that another limit of the G$\beta$E which has been studied is $\beta \to 0$,
in particular $\beta=c/N$, although in that limit the density extends to
the full real axis \cite{BouchaudGuionnet2012}. Correspondingly, the APP for 
$\beta \to 0$ converges to a Poisson point process \cite{DumazLabbePoisson}.

\subsection{Summary of main results} %and structure of the paper}

Let us state our main results. %briefly summarize our main findings.
For reasons which will become clear below we define $E_i=-a_i$
minus the APP and use the notation $E=-a$. Likewise we denote $\Phi(a)=s(E=-a)$
the rate function. 
First we show that in the limit of large Dyson index $\beta \gg 1$, the TW$\beta$ distribution,
i.e. the PDF $f_\beta\left(a\right)$ of $a_1$,
satisfies the following large-deviation principle scaling form
% \be \label{LDPintro}
% f_\beta\left(a\right)\sim e^{-\beta \, s\left(E = -a\right)} \quad , \quad 
% s(E) = - \lim_{\beta \to +\infty} \frac{1}{\beta} \ln f_\beta(a=-E) 
% \ee 
 \be \label{LDPintro}
f_{\beta}\left(a\right)\sim e^{-\beta\,\Phi\left(a\right)}\quad,\quad\Phi\left(a\right)=s(-a)=-\lim_{\beta\to+\infty}\frac{1}{\beta}\ln f_{\beta}(a) \, , 
\ee
where the rate function $s(E)$ and its derivative%
% \cite{footnote1} 
\footnote{Varying \eqref{NonLinearPhiEqIntro} with respect to $E$, multiplying
by $\phi$, and integrating from $x=0$ to $x=+\infty$, 
yields the expression for $s'(E)$ from the one
of $s(E)$.}
 $s'(E)$ are given by 
\be
\label{Sofphintro}
s(E) =\frac{1}{8}\int_{0}^{+\infty}\phi_E\left(x\right)^4 \, dx\,,
\quad  s'(E)= \frac{1}{4} \sigma_E \int_{0}^{+\infty}\phi_E\left(x\right)^2  dx \,,
\quad  \sigma_E= \pm 1= \text{sgn}\left(E + \zeta_1 \right)  \,,
\ee
where $\phi(x)=\phi_E(x)$ is the solution for $x \in \mathbb{R}^+$ of the Painlev\'{e} II type equation
\be
\label{NonLinearPhiEqIntro}
-\phi''\left(x\right)+\left[x+ \sigma_E \phi(x)^{2}\right]\phi\left(x\right)=E \phi \left(x\right)\,,
\ee
with Dirichlet boundary condition at the origin, $\phi(0)=0$, which vanishes
at infinity $\phi(x \to + \infty) = 0$, i.e. $\phi(x) \simeq k {\rm Ai}(x-E)$
for some $k$, and such that it has no other zero for $x>0$.
Here $\zeta_1$ is the first zero of the Airy function $\text{Ai}(x)$,
$\zeta_1 =-2.33811\dots$, which coincides with the typical value%
\footnote{In the large-$\beta$ limit, the distribution of $a_1$ is very narrowly peaked around its maximum; Therefore all possible interpretations of its `typical value' (e.g., the mean of $a_1$, or the maximum of its PDF, or the median, etc) coincide in the leading order.}
of $a_1$, $a^{\rm typ}=\zeta_1$.
Note that $\sigma_E \phi(x)^{2} \propto (E+\zeta_1)$ for small $|E+\zeta_1|$ 
and
there is no non-analyticity of $s(E)$ around $E=-\zeta_1$. 
\\

The asymptotic behaviors of $s(E)$ are found to be 
\be 
\label{sOfEAsymptotics}
s(E)\simeq\begin{cases}
\frac{2}{3}(-E)^{3/2}\quad, & E\to-\infty\,,\\[2mm]
\frac{1}{2 C_{2}}(E+\zeta_{1})^{2}\quad, & \left|E+\zeta_{1}\right|\ll1\,,\\[2mm]
\frac{1}{24}E^{3}\quad, & E\to+\infty\,,
\end{cases}
\ee 
where $C_2$ is the reduced variance (see below). The asymptotic behaviors
%which
for $|E| \gg 1$ are consistent with the known leading behaviors displayed in
\eqref{tailsTWbeta} for the 
%of the upper and lower 
tails of the TW distribution valid for any $\beta$ (recalling that $E=-a$).
 We also obtain subleading corrections in each of the limits in Eq.~\eqref{sOfEAsymptotics}, 
see Eq.~\eqref{sSolRightTailSubleading}, Sec.~\ref{subsec:Typical} and Eq.~\eqref{SSolLeftTailGenerali} respectively,
and compare with known results and numerical computations of TW$\beta$ at finite $\beta$.
The tail $E \to -\infty$
is dominated by the soliton solution of the stationary non-linear Schr\"{o}dinger equation,
%\url{https://en.wikipedia.org/wiki/Nonlinear_Schr%C3%B6dinger_equation}
$\phi_E(x) \sim \sqrt{-2 E}/\cosh(\sqrt{-2 E}(x-x_0))$,
obtained from \eqref{NonLinearPhiEqIntro} by neglecting the linear potential term $x$.
The tail $E \to +\infty$ is dominated by the ``Thomas Fermi/semi-classical'' solution 
$\phi_E(x) \sim \sqrt{(E-x)_+}$,
obtained from \eqref{NonLinearPhiEqIntro} 
by neglecting the second derivative. Similarly, we find that 
the cumulant generating function admits the large $\beta$ limit
\be 
\label{muDef}
\left\langle e^{-\beta\lambda a_{1}}\right\rangle =\left\langle e^{\beta\lambda E_{1}}\right\rangle \sim e^{\beta\mu(\lambda)}\quad,\quad\mu(\lambda)=\lim_{\beta\to+\infty}\frac{1}{\beta}\ln\left\langle e^{\beta\lambda E_1}\right\rangle =\max_{E\in\mathbb{R}}\left(\lambda E-s(E)\right)
\ee 
so that the cumulants for $n \geq 2$ behave as 
\be \label{cuma1n} 
\left\langle a_{1}^{n}\right\rangle _{c}=\left\langle \left(-E_{1}\right)^{n}\right\rangle _{c}\simeq\frac{C_{n}}{\beta^{n-1}}
\ee 
The typical fluctuations are thus Gaussian,
and the reduced variance $C_2 \approx 1.6697$ is given in \eqref{tildes2sol} and \eqref{VaraSol},
in agreement with Refs. \cite{Dumitriu2005, EPS14, AHV21}. In addition we 
obtain the reduced third cumulant, 
$C_3 \approx 0.743785$ given analytically in \eqref{tildes3sol}, \eqref{VaraSol}, see also 
\eqref{1point3rdCumulant} and \eqref{1point3rdCumulant2},
as well as the reduced fourth cumulant, $C_4\approx 0.5576$, given 
analytically in \eqref{4cumapp} and \eqref{4cumapp2}.
The subleading correction in the $1/\beta$ expansion to the variance
is also obtained analytically in \eqref{varianceSubleading}  and in 
\eqref{varianceSubleading2} (as well as the
correction to the mean, in agreement with previous results).
Finally, we compute the rate function $s(E)$ at all $E$ by numerically solving Eq.~\eqref{NonLinearPhiEqIntro} by a shooting method, 
and find good numerical agreement with the TW$\beta$ distribution computed 
as the solution to a boundary-value PDE problem (using the representation obtained by Bloemendal in \cite{BloemendalThesis}), see Fig.~\ref{figsOfEAiry}.

These results extend to the marginal PDF $f_\beta^{(i)}(a)$ of any other
point of the APP, $a_i=-E_i$ for $i \geq 2$. We show that it obeys the same
large deviation principle \eqref{LDPintro} at large $\beta$, with a
rate function $s_i(E=-a)$. This rate function is given by the same formula
\eqref{Sofphintro} where $\phi_E(x)$ satisfies again \eqref{NonLinearPhiEqIntro}
with the same boundary condition but with the constraint of having exactly $i$ zeroes (including
the one at $x=0$). To leading order in the limit $\beta \to +\infty$, the typical value equals to the mean,
with $a_i^{\rm typ} = \langle a_i \rangle = \zeta_i$, where $\zeta_i$ is the $i$-th zero of the Airy function, and $\sigma_E=\text{sgn}\left(E + \zeta_i \right)$.
%+ o(1/\beta)$ 

Finally, we extend most of these results to other observables of interest related to the $\text{Airy}_\beta$ point process, such as the joint distribution of pairs of eigenvalues and the distribution of gaps between eigenvalues, with a similar exponential scaling (as a function of $\beta$) in all cases.

\subsection{Outline of the paper}

% is obtained by evaluating the ``action'' over the solution to a saddle-point equation which is closely related to the Painlev\'{e} II equation. This solution describes the most likely realization of the stochastic Airy operator (SAO) conditioned on its largest eigenvalue, as well as the most likely realization of a related Ricatti diffusion process conditioned on the process blowing up at a given time.
% We compute the $s(E)$ at all $E$ by numerically solving the saddle-point equations and analytically at limiting values of $E$, and find good agreement with TW$\beta$ distributions with large $\beta$, see Fig.~\ref{figsOfEAiry}.
% In particular, this enables us to calculate the third cumulant of $a$ (at large $\beta$).
% We extend most of these results to other observables of interest related to the $\text{Airy}_\beta$ point process, such as the distribution of the $i$th largest eigenvalue of the SAO, the joint distribution of pairs of eigenvalues and the distribution of gaps between eigenvalues, with a similar exponential scaling (as a function of $\beta$) in all cases.

The paper is organized as follows. In Sec.~\ref{sec:StochasticAiry} we recall the representation of the Tracy-Widom distribution in terms of the smallest eigenvalue of the Stochastic Airy operator. In Sec.~\ref{sec:SaddlePointGeneral} we derive the saddle-point approximation for statistics of energy levels of general weakly-disordered potentials. In Sec.~\ref{sec:SaddlePointTW} we derive the large-$\beta$ behavior of the TW$\beta$ distribution by applying the saddle-point approximation to the stochastic Airy operator, and analytically calculate the limiting behaviors of the large-deviation function that describe the far tails of the distribution $a \to \pm \infty$ and typical fluctuations, $a \simeq \left\langle a\right\rangle $.
In Sec.~\ref{sec:pairs} we study the large $\beta$ behavior of the joint distribution of pairs of eigenvalues and of the gap between two eigenvalues.
In Sec.~\ref{sec:WNT} %\sout{and \ref{sec:Ablowitz}} 
we present an alternative derivation of some these results by applying the WNT to the ``diffusion'' representation of TW$\beta$. In Sec.~\ref{sec:discussion} we summarize and discuss our main findings.
Some technical details related to the derivations are given in the Appendices.

\bigskip

\section{Representation of the Tracy-Widom distribution in terms of the stochastic Airy operator}
\label{sec:StochasticAiry}

It was proved in \cite{Edelman03Presentation, RamirezRiderVirag2011} that the eigenvalues $E_i$ of the following stochastic Airy operator, defined for $x>0$ as
\be 
\label{HSAOdef}
H_{\rm SAO} = - \partial_x^2 + x + \frac{2}{\sqrt{\beta}} \, \eta(x) 
\ee 
with vanishing (Dirichlet) boundary condition at $x=0$ and
where $\eta(x)$ is a centered Gaussian white noise, with
\be
\left\langle \eta \left(x\right)\right\rangle =0,\quad
\left\langle \eta\left(x\right)\eta\left(x'\right)\right\rangle =\delta\left(x-x'\right)
\ee 
are given by
\be 
E_i = - a_i 
\ee 
where the $a_i$ form the Airy point process whose definition was recalled in the Introduction,
see \eqref{APPDefinition}, as the (right) edge limit of the G$\beta$E eigenvalues. 
The $a_i$'s form an infinite set and are ordered
as $a_1>a_2>\dots$, so that the eigenvalues form an increasing sequence, $E_1<E_2<\dots$.

% the Airy soft edge point process, which for $\beta=2$ is a DPP,} obtained as the edge limit $a_i \equiv N^{2/3}(\lambda_i-2)$ 
% of the G$\beta$E with eigenvalues joint PDF 
% \be
% \label{GbetaEjPDF}
% P(\lambda_{1},\dots,\lambda_{N})\propto\prod_{i<j}\left|\lambda_{i}-\lambda_{j}\right|^{\beta}\exp\left(-\frac{\beta N}{4}\sum_{i}\lambda_{i}^{2}\right)
% \ee
% which has limit support $[-2,2]$. It is customary to choose $\lambda_1 > \lambda_2 > \dots$
% and $a_1>a_2>\dots$. The lowest eigenvalue of $H_{\rm SAO}$ is thus $E_1=- a_1$. 
% The distribution of $a_1$ is TW$_\beta$, the $\beta$-TW distribution. Its CDF
% is denoted $F_\beta(s)$. For $\beta=1,2,4$ it has
% the usual form in terms of solutions to the Painlev\'{e} equation, see \cite{RamirezRiderVirag2011} p 928. 

Naturally, the eigenvalue problem 
\be
\label{SchrodingerHSAO}
(H_{{\rm SAO}}\psi)(x) =-\psi''\left(x\right)+\left[x+{\frac{2}{\sqrt\beta}}\,\eta(x)\right]\psi\left(x\right)=E\psi\left(x\right)
\ee
with Dirichlet boundary condition $\psi(0)=0$ may be interpreted as the time-independent Schr\"{o}dinger equation for a single quantum particle in $d=1$ dimension (with units chosen such that $\hbar^2 / 2m = 1$), in a potential $V(x) = V_0(x) + V_1(x)$ with a deterministic part
\be
\label{V0def}
V_{0}\left(x\right)=\begin{cases}
x, & x>0,\\[1mm]
+ \infty, & x<0.
\end{cases}
\ee
and additional disorder $V_1(x) = \left(2/\sqrt{\beta} \right)  \eta(x)$. Within this interpretation, the eigenvalues $E_i$ are the energy levels.
 It is worth noting that spectral properties of Schr\"{o}dinger operators with white-noise potentials have been studied for many decades, see e.g. \cite{Halperin65, VV95}.

The Tracy-Widom distribution TW$\beta$ thus describes the distribution (up to a sign) of the ground-state energy $E_1$ of the disordered potential $V(x)$. In the limit $\beta \gg 1$ which interests us here the disorder intensity is small, so typically the energy levels $E_i$ of $V(x)$ will be very close to those of $V_0(x)$. However, atypical values of $E_i$, driven by unlikely realizations of the disorder, are still possible, and it is our goal to calculate their probability distribution.

\bigskip

\section{Saddle-point approximation for general background potential $V_0(x)$}
\label{sec:SaddlePointGeneral}

In this section, we derive the saddle-point equation that gives the large-$\beta$ behavior of the distribution of the energy levels for a general $V_0(x)$ on the real line.
We subsequently specialize on the particular case \eqref{V0def} (corresponding to the Tracy-Widom distribution) in the following section% \cite{footnote2} 
\footnote{For sake of generality we first treat here the problem on the full line, specializing later to \eqref{V0def}, but one can perform the same steps directly on the half-line, with Dirichlet boundary conditions. All integrals then are on $x \in [0,+\infty[$.}.

Our starting point is the probability (density) for a given realization of the white noise,
\be
\mathcal{P}\left[\eta\right]\sim e^{-\frac{1}{2}\int_{-\infty}^{+\infty}\eta^{2}dx} \, .
\ee
This may be expressed in terms of the quantum potential $V(x)$, by inverting the relation
$V(x) = V_0(x) + \left(2/\sqrt{\beta} \right) \eta(x)$,
leading to
\bea
\label{PVdef}
&&\mathcal{P}\left[V\right]\sim e^{-\beta \mathcal{S}\left[V\right]} \, ,\\[1mm]
\label{sVdef}
&&\mathcal{S}\left[V\right]=\frac{1}{8}\int_{-\infty}^{+\infty}\left[V(x)-V_{0}(x)\right]^{2}dx\,.
\eea
For brevity we will denote the PDF $P_i(E)$ of the $i$th energy level $E_i$,
and its associated large deviation
function $s_i(E)$, respectively as $P(E_i)$ and $s(E_i)$.
The PDF 
%The distribution of the $i$th energy level 
$P(E_i)$ may be written as a path integral. This would correspond to summing the probabilities \eqref{PVdef} of all realizations of the quantum potential $V(x)$ for which the $i$th energy level equals $E_i$.

In the limit $\beta\to\infty$, this path integral may be evaluated, in the leading order, by using a saddle-point approximation. The procedure was performed in Ref.~\cite{SmithGroundState24} in a similar context, but we give the derivation here for completeness.
The saddle-point approximation immediately results in the scaling behavior
\be
\label{PEScaling}
P\left(E_{i}\right)\sim e^{-\beta s\left(E_{i}\right)}, \qquad \beta \gg 1
\ee
%(we do not explicitly denote the $i$ dependence of $P$ and $s$  for the sake of brevity).
Here the large-deviation function $s(E_i)$ does not depend on $\beta$, and it is given by the minimum of the action functional \eqref{sVdef} over all possible potentials $V(x)$, constrained on the value of the $i$th energy level, $\mathcal{E}_i[V]=E_i$.
We incorporate the latter constraint into the minimization procedure by introducing a Lagrange multiplier $\lambda$, and minimizing the modified action functional
\be
\label{Slambdadef}
\mathcal{S}_{\lambda}\left[V\right]=\mathcal{S}\left[V\right]-\lambda\left(\mathcal{E}_{i}\left[V\right]-E_{i}\right) \, .
\ee

It is difficult to write the functional $\mathcal{E}_{i}\left[V\right]$ explicitly. However, one can easily express the leading order variation of the energy levels $\delta \mathcal{E}_{i}$ in terms of the variation of the potential $\delta V(x)$, and the wave function $\psi_i(x)$ that corresponds to the $i$th energy level for the potential $V(x)$, normalized in such a way that% \cite{footnote:real} 
\footnote{We choose the wave functions to be real, which can be done since they are wave functions of bound energy states.}
\be
\label{psiNormalization}
\int_{-\infty}^\infty \psi_i(x)^2  dx = 1.
\ee
Indeed, by using first-order perturbation theory, one finds that
\be
\delta \mathcal{E}_{i}=\int_{-\infty}^{+\infty}\delta V\left(x\right)\psi_{i}(x)^{2} dx +O\left(\delta V^{2}\right) \, .
\ee
 which, in terms of functional
derivatives can be written as $\delta \mathcal{E}_{i}\left[V\right]/\delta V(x)=\psi_{i}(x)^{2} $.

The variation of the action functional \eqref{sVdef} is simply given by
\be
\delta\mathcal{S}=\frac{1}{4}\int_{-\infty}^{+\infty}\left[V(x)-V_{0}(x)\right]\delta V\left(x\right)dx+O\left(\delta V^{2}\right) \, .
\ee
Putting the last two equations together, we find that the variation in the modified action functional \eqref{Slambdadef} is
\be
\delta\mathcal{S}_{\lambda}=\frac{1}{4}\int_{-\infty}^{+\infty}\left[V(x)-V_{0}(x)-4\lambda\psi_{i}(x)^{2}\right]\delta V\left(x\right)dx+O\left(\delta V^{2}\right)\,.
\ee
For the optimal $V(x)$, this variation must vanish in the leading order for any $\delta V(x)$, implying that
\be
\label{V1ProptoPsisq}
V(x)-V_{0}(x) = 4\lambda\psi_{i}(x)^{2} \, .
\ee
Plugging this into the Schr\"{o}dinger equation, we obtain
\be
\label{NonLinearPsiEq}
-\psi_{i}''\left(x\right)+\left[V_{0}\left(x\right)+4\lambda\psi_{i}(x)^{2}\right]\psi_{i}\left(x\right)=E_{i}\psi_{i}\left(x\right) \, .
\ee
Eq.~\eqref{NonLinearPsiEq} must be solved for $\psi_i(x)$ subject to the normalization condition \eqref{psiNormalization}.

It is convenient to rescale the wave function % 
\be \label{phipsi} 
\phi_{i}\left(x\right)=2\sqrt{\left|\lambda\right|} \, \psi_{i}\left(x\right) \, , 
\ee
so that Eq.~\eqref{NonLinearPsiEq} becomes
\be
\label{NonLinearPhiEq}
-\phi_{i}''\left(x\right)+\left[V_{0}\left(x\right)+\text{sgn}\left(\lambda\right)\phi_{i}(x)^{2}\right]\phi_{i}\left(x\right)=E_{i}\phi_{i}\left(x\right) \, ,
\ee
where %
\be
\text{sgn}(x)=\begin{cases}
1, & x>0,\\[1mm]
-1, & x<0
\end{cases}
\ee
is the sign function.
This equation has to be solved subject to the boundary conditions $\phi_i(x \to \pm \infty) = 0$ which follow from normalizibility.
Moreover, since $\phi_i(x)$ is the wavefunction of the $i$th energy level, one must require that it has exactly $i$ zeros. For the particular case \eqref{V0def} that corresponds to the stochastic Airy operator (hence to the Airy point process and to the TW$\beta$ distribution), this includes the zero at $x=0$.
Once Eq.~\eqref{NonLinearPhiEq} has been solved, one can then obtain the optimal realization of the disorder $V(x)$ from Eq.~\eqref{V1ProptoPsisq}.  Plugging this expression into the action functional \eqref{sVdef}, or equivalently in \eqref{Slambdadef},
we obtain the optimal action in terms of the wave function, which identifies with the large-deviation function as 
\be
\label{Sofphi}
s(E_i)= \mathcal{S}[V]=\frac{1}{8}\int_{-\infty}^{+\infty}\phi_{i}^{4}\left(x\right)dx\,.
\ee
%we obtain the large-deviation function $s(E_i) = \mathcal{S}$.
It is useful to note that the large-deviation function and the Lagrange multiplier are related to each other through
\be
\label{dsdELambda}
s'(E_i) = \frac{ds}{dE_i}  = \lambda \, .
\ee
This relation follows from the fact that $E_i$ and $\lambda$ are conjugate variables, see e.g. Ref.~\cite{Vivoetal}. One can also see it directly by 
writing that $s(E_i)={\cal S}_{\lambda}[V]$ at the optimum $V(x)$ and $\lambda$ (which
are functions of $E_i$) and taking the derivative of 
\eqref{Slambdadef} w.r.t. $E_i$, using the conditions that
$\mathcal{S}_{\lambda}$ must be stationary w.r.t. $V(x)$ and $\lambda$.
As is usual at a saddle point, only the explicit dependence in $E$ remains.

The minimum of $s(E_i)$, at which $s$ is expected to vanish, should be attained at $E_i = E_i^{(0)}$, where $E_i^{(0)}$ is the $i$th energy level of the potential $V_0(x)$ in the absence of disorder.
Using this together with the relation \eqref{dsdELambda}, one finds that
\be
\label{sgnLambda}
\text{sgn}(\lambda) = \text{sgn}\left(E_i-E_i^{(0)}\right) \, .
\ee
This relation is useful because  it can be used in Eq.~\eqref{NonLinearPhiEq} to eliminate $\lambda$ from the equation altogether.

\bigskip

\section{Large-$\beta$ behavior of the Tracy-Widom distribution}
\label{sec:SaddlePointTW}

Let us now return to the particular case of interest here, corresponding to the quantum potential \eqref{V0def}.
First of all, we find that at large $\beta$, the Tracy-Widom distribution  (for $a_1$,
and its generalization to any $a_i$ of the APP)
follows the scaling behavior
\be
\label{PaScaling}
f_\beta\left(a_{i}\right)\sim e^{-\beta s\left(E_i = -a_{i}\right)}, \qquad \beta \gg 1,
\ee
which is Eq.~\eqref{PEScaling} after plugging in $E_i = -a_i$.
In order to calculate the large-deviation function $s(E_i)$, we must solve Eq.~\eqref{NonLinearPhiEq} for $x>0$, where it reads [using \eqref{sgnLambda}]
\be
\label{NonLinearPhiEq2}
-\phi_{i}''\left(x\right)+\left[x+\text{sgn}\left(E_{i}-E_{i}^{\left(0\right)}\right)\phi_{i}(x)^{2}\right]\phi_{i}\left(x\right)=E_{i}\phi_{i}\left(x\right)\,,
\ee
with a Dirichlet boundary condition at the origin, $\phi_i(0)=0$.
It is an elementary exercise in quantum mechanics to show that the energy levels of the potential \eqref{V0def} with zero disorder are given by
\be
E_{i}^{\left(0\right)} = -\zeta_i \, , \quad i=1,2,\dots,
\ee
where $\zeta_i$ are the zeroes of the Airy function $\text{Ai}(x)$,
$\zeta_1 =-2.33811$,
$\zeta_2 = - 4.08795$, etc.
In what follows, it will also be useful to give the corresponding (normalized) eigenfunctions. These are defined on the positive half line $x \ge 0$, and given by
\be
\label{psii0}
\psi_{i}^{\left(0\right)}\left(x\right)=\frac{\text{Ai}\left(x+\zeta_{i}\right)}{\text{Ai}'\left(\zeta_{i}\right)} \, .
\ee
Finally, $s(E_i)$ is obtained by evaluating the integral \eqref{Sofphi} (on the integration domain $0<x<\infty$) 
\be \label{sEi}
s(E_i)= \frac{1}{8}\int_{0}^{+\infty}\phi_{i}^{4}\left(x\right)dx\,,
\ee
Using \eqref{NonLinearPhiEq2} one can also show that (see Appendix \ref{app:der})
\be \label{sEider}
s'(E_i)= \frac{1}{4} \text{sgn}\left(E_{i}-E_{i}^{\left(0\right)}\right) 
\int_{0}^{+\infty}\phi_{i}^{2}\left(x\right)dx\,.
\ee

Eq.~\eqref{NonLinearPhiEq2} can in general be solved numerically \cite{SmithGroundState24}, by using the shooting method, see e.g. Ref.~\cite{SB80}  (for a more detailed discussion of numerical solutions to the Painlev\'{e} II equation, see \cite{FW14, FW15}). The shooting method reduces the boundary-value problem to an initial value problem. The shooting parameter we employ is $\phi_i'(0)$.
One then computes the integral \eqref{Sofphi} numerically on the solution $\phi_i(x)$, to obtain $s(E_i)$ at all values of $E_i$. Indeed, in Fig.~\ref{figsOfEAiry} we show that this numerical method for calculating $s(E_i)$ yields results which are in excellent agreement with a direct numerical calculation of the Tracy-Widom distributions with $\beta=10,20$, which we obtained using the numerical method from Ref.~\cite{BloemendalThesis}  (note that this numerical method was further developed and used in \cite{Computation}, however, to our knowledge, the large-deviation behaviors have not been studied numerically before).

Importantly, however, there are three limits in which $s(E_i)$ can be calculated analytically \cite{SmithGroundState24}: $E_i \to \pm \infty$ 
which, as we show below, match smoothly with the tails \eqref{tailsTWbeta} of the Tracy-Widom distribution, 
and $E_i \simeq E_i^{(0)}$ which describes the typical fluctuations. In the latter limit, one can systematically calculate $s(E_i)$ as a power series around $E_i = E_i^{(0)}$ using perturbation theory, and obtain the leading-order behavior of the cumulants of the distribution at $\beta \to +\infty$.
We now perform the analytic calculation of $s(E_i)$ in each of these limits, whose results are also plotted in Fig.~\ref{figsOfEAiry}, and in particular we obtain the second, third and fourth cumulants.

Let us also mention a few identities relating derivatives and/or integrals of the wavefunction $\phi_i$ and of the large-deviation function $s(E_i)$, which will be useful in the following. We show in Appendix \ref{app:der} that these identities follow from Eq.~\eqref{NonLinearPhiEq2}.
The first one gives  
\be \label{phider0}
\phi_i'(0)^2= \int_0^{+\infty} dx \, \phi_i(x)^2  = 4 |s'(E_i)| 
\quad , \quad \psi'_i(0)^2 = 1 
\ee 
(note also that $\phi_i''(0)=0$). Next one has
\bea 
% \label{2id}
\label{sEphiIdentity1}
&& \frac{2}{3} E_i s'(E_i) - s(E_i) = \frac{\sigma_{E_i}}{4} \int_0^{+\infty} dx x \phi_i(x)^2 \, ,\\[1mm]
\label{sEphiIdentity2}
&& E_i s'(E_i) - 3 s(E_i) = \frac{3 \sigma_{E_i}}{4} \int_0^{+\infty} dx  \phi_i'(x)^2 \, ,
\eea 
where we recall the definition \eqref{Sofphintro} of the notation $\sigma_E$.

\begin{figure}
\includegraphics[width=0.47\textwidth,clip=]{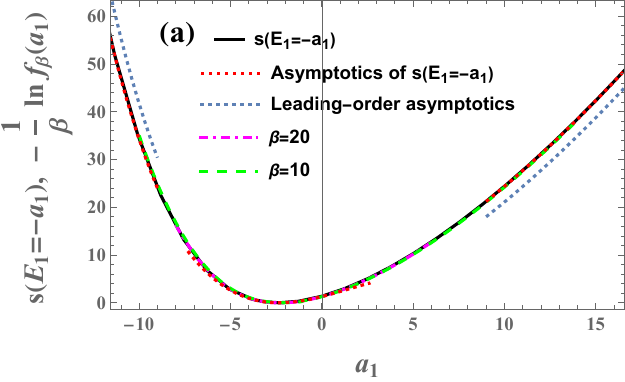}
 \hspace{2mm}
\includegraphics[width=0.47\textwidth,clip=]{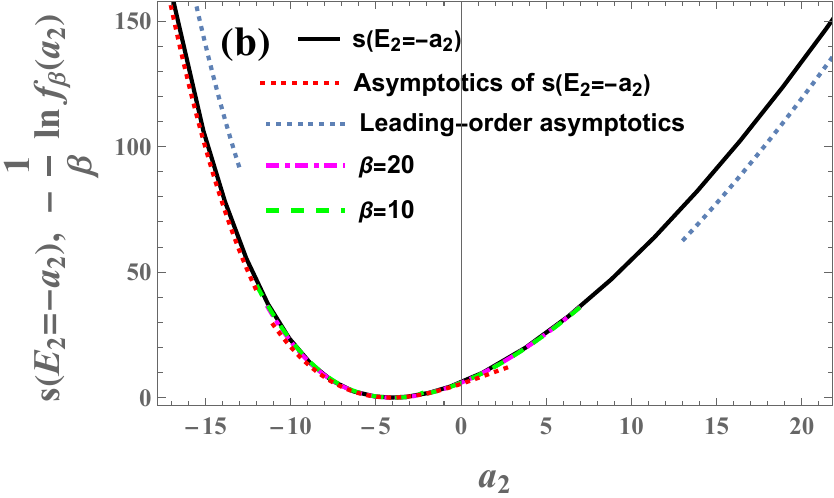}
\caption{ (a) Solid line: The large deviation function $\Phi(a_1)= s(E_1=-a_1)$ as a function of the largest eigenvalue $a_1 = -E_1$, where $E_1$ is the ground state energy of the potential \eqref{V0def}. $s(E_1=-a_1)$ is computed on numerical solutions to Eq.~\eqref{NonLinearPhiEq}.
Red dotted lines: the asymptotic behaviors (which include subleading corrections) \eqref{sSolRightTailSubleading}, \eqref{SSolLeftTail} and \eqref{sofECubicOrder} (with $i=1$) of $s(a_1)$ at $a_1 \to \pm\infty$ and $a_1 \simeq \zeta_1$, respectively.
Blue dotted lines: the leading-order asymptotic behaviors \eqref{sSolRightTail} and \eqref{SSolLeftTailLeadingOrder} at $a_i \to \pm\infty$, respectively.
Also plotted are numerical computations of $-\frac{1}{\beta} \ln f_{\beta}(a_1)$ where $f_{\beta}(a_1)$ is the Tracy-Widom distribution, for relatively large Dyson indices, $\beta=10,20$ (dashed and dot-dashed lines, respectively). $f_\beta(a_1)$ is computed using the numerical method from Ref.~\cite{BloemendalThesis}. One can see that our theory describes the distributions very well.
(b) similar to (a) but for the second largest eigenvalue $a_2 = -E_2$. In (b), the red dotted lines at $a_2 \to \pm\infty$ correspond to Eqs.~\eqref{sSolRightTaili}  and \eqref{SSolLeftTailGenerali} (with $i=2$).}
% {\red TODO: can we calculate the subleading order theory for $a_2 \to \infty$ (excited state)?}}
\label{figsOfEAiry}
\end{figure}

\subsection{$a_i \to +\infty$ tail: Localized solution}
\label{sec:Soliton}

For a general smooth potential $V_0(x)$, the $E_i \to -\infty$ (i.e. $a_i \to  +\infty$) tail is dominated by realizations of the disorder which are strongly localized around the minimum of $V_0(x)$ \cite{SmithGroundState24}. $V_0(x)$ may then be approximated by a constant, $V_0(x) \simeq V_0(x_{\min})$, for $-\infty < x < \infty$.
In our case, $V_0(x)$  is given by \eqref{V0def} and is not smooth at $x=0$. 
Nevertheless, the solution is expected to be localized around a point a little to the right of $x=0$, and in the leading order, the localization will be strong enough to justify approximating $V_0(x) \simeq 0$ everywhere, i.e., approximating the system as infinite in both directions (the Dirichlet boundary condition becomes unimportant at the leading order).

Under this approximation, Eq.~\eqref{NonLinearPhiEq} may be written as
\be
\label{phiNewton}
\phi_{i}''\left(x\right)=-E_{i}\phi_{i}\left(x\right)-\phi_{i}(x)^{3}\,,
\ee
where we used \eqref{sgnLambda} to plug in $\text{sgn}(\lambda)=-1$.
Using a mechanical analogy, we interpret Eq.~\eqref{phiNewton} as Newton's second law for a particle of unit mass moving in an effective  double well potential
\be
\label{Ueffdef}
U_{\text{eff}}\left(\phi_{i}\right)=\frac{1}{2}E_{i}\phi_{i}^{2}+\frac{1}{4}\phi_{i}^{4} \, .
\ee
Here $\phi_i$ and $x$ play the roles of position and time, respectively.
Using the conservation of ``energy'', we find that
$U_{\text{eff}}\left(\phi_{i}\right) + \frac{1}{2} \phi_{i}'\left(x\right)^2$
is constant.
From the boundary conditions $\phi_i(x \to \pm \infty) = 0$, this constant must vanish, i.e.,
\be
\frac{1}{2}E_{i}\phi_{i}^{2}+\frac{1}{4}\phi_{i}^{4}+\frac{1}{2}\phi_{i}'\left(x\right)^{2}=0 \, .
\ee
We rewrite this equation in the form
\be
\label{dphidx}
\frac{d\phi_{i}}{\sqrt{-E_{i}\phi_{i}^{2}-\frac{1}{2}\phi_{i}^{4}}}=\pm dx \, .
\ee
We now solve for the ground state $i=1$, and treat the excited states $i=2,3,\dots$ below.
We integrate Eq.~\eqref{dphidx} it to obtain
\be
% \frac{1}{\sqrt{-E_{i}}}\coth^{-1}\left(\sqrt{\frac{2E_{i}}{2E_{i}+\phi_{i}^{2}}}\right)=\pm\left(x-x_{0}\right)\,,
\frac{1}{\sqrt{-E_{1}}}\coth^{-1}\left(\sqrt{\frac{2E_{1}}{2E_{1}+\phi_{1}^{2}}}\right)=\pm\left(x-x_{0}\right)\,,
\ee
which we solve to obtain the wave function
\be
\label{phiSolLocalized}
% \phi_{i}(x)=\frac{\sqrt{-2E_{i}}}{\cosh\left(\sqrt{-E_{i}}\left(x-x_{0}\right)\right)}
\phi_{1}(x)=\frac{\sqrt{-2E_{1}}}{\cosh\left(\sqrt{-E_{1}}\left(x-x_{0}\right)\right)} \, .
\ee
Eq.~\eqref{phiSolLocalized} describes a wave function that is strongly localized, with a characteristic width $\sim 1/\sqrt{-E_i}$, around a point $x_0$ which is so far undetermined. If $V_0(x)$ were smooth, then $x_0$ would be its minimum \cite{SmithGroundState24}.
For our case, $x_0$ must be taken to be in the range $1/\sqrt{-E_i} \ll x_0 \ll 1$: This is far enough from the origin for the Dirichlet boundary condition to be approximately satisfied, and yet close enough to the origin for the approximation $V(x_0) \simeq 0$ to hold.
In Fig.~\ref{figphi1ofxSoliton}(a), we compare the prediction \eqref{phiSolLocalized} with an exact, numerical solution of Eq.~\eqref{NonLinearPhiEq}, for $E_1 = -126$. The agreement is excellent, although $x_0$ is found by fitting to the numerical data. It would thus be interesting to develop a subleading-order theory for $E_1 \to -\infty$, in which one would (in particular) obtain a theoretical prediction for $x_0$,  and we shall do so shortly. But first,
let us evaluate the action on the leading-order solution. Plugging \eqref{phiSolLocalized} into \eqref{Sofphi} we obtain
\be
\label{sSolRightTail}
% s(E_i) \simeq \frac{1}{8}\int_{-\infty}^{+\infty}\frac{4E_{i}^{2}}{\cosh^{4}\left(\sqrt{-E_{i}}\left(x-x_{0}\right)\right)}dx=\frac{\left(-E_{i}\right)^{3/2}}{2}\int_{-\infty}^{+\infty}\frac{dy}{\cosh^{4}y}=\frac{2\left(-E_{i}\right)^{3/2}}{3}\,,
s(E_{1})\simeq\frac{1}{8}\int_{-\infty}^{+\infty}\frac{4E_{1}^{2}}{\cosh^{4}\left(\sqrt{-E_{1}}\left(x-x_{0}\right)\right)}dx=\frac{\left(-E_{1}\right)^{3/2}}{2}\int_{-\infty}^{+\infty}\frac{dy}{\cosh^{4}y}=\frac{2\left(-E_{1}\right)^{3/2}}{3}\,.
\ee
\\

\begin{figure}
\includegraphics[width=0.47\textwidth,clip=]{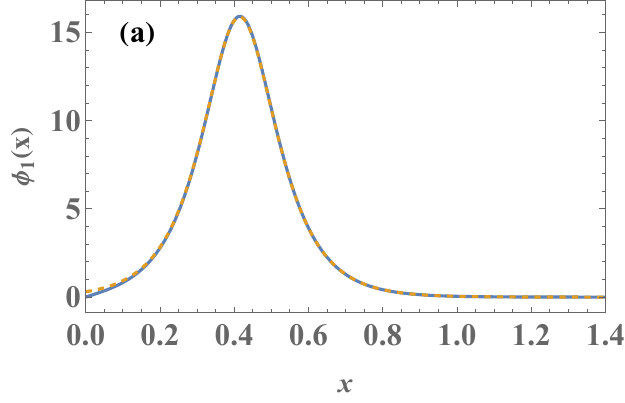}
\hspace{2mm}
\includegraphics[width=0.47\textwidth,clip=]{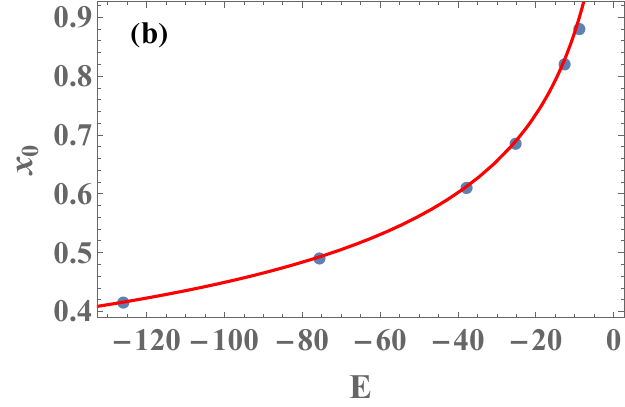}
\caption{
(a) Solid line: Exact (numerically-obtained) solution $\phi_1(x)$ to Eq.~\eqref{NonLinearPhiEq} for ground-state energy $E_1 = -126$.
Dashed line: the (approximate) theoretical prediction \eqref{phiSolLocalized}, where $x_0 = 0.415$ was initially obtained as a fitting parameter but then found to agree with our formula \eqref{x0ofEAsymptotic}.
(b) $x_0$ (which we determined numerically as the local maximum of $|\phi_1(x)|$) as a function of $E_1$ (markers), which show excellent agreement with the $E_1 \to -\infty$ analytic prediction \eqref{x0ofEAsymptotic} (solid line).}
\label{figphi1ofxSoliton}
\end{figure}

% \begin{figure}
% \includegraphics[width=0.47\textwidth,clip=]{figx0vsE.pdf}
% \caption{$x_0$ (which we determined numerically as the local maximum of $|\phi_1(x)|$) as a function of $E_1$ (markers), which shows excellent agreement with the $E_1 \to -\infty$ analytic prediction \eqref{x0ofEAsymptotic} (solid line).}
% \label{figx0ofE}
% \end{figure}

To obtain the subleading correction we first determine $x_0$. 
For this we need to study the behavior of $\phi_1(x)$ near $x=0$.
First, using \eqref{phider0} we see that to leading order $\phi_1'(0) \simeq 2 |E_1|^{1/4}$.
%\sout{Next examining \eqref{NonLinearPhiEq2} near $x=0$, we see that there is a regime where the non-linear term $\phi_1^3$ can be neglected, and}
 Next, let us examine \eqref{NonLinearPhiEq2} near $x=0$. Due to the Dirichlet boundary condition at $x=0$ we see that there is a regime where the non-linear term $\phi_1^3$ can be neglected. Moreover,
one can approximate the term $x + |E_1| \approx |E_1|$ (this is confirmed by examining the series expansion near $x=0$),
so that we need to solve $\phi_1''=|E_1| \phi_1$ 
with $\phi_1(0)=0$ and $\phi_1'(0) \simeq 2 |E_1|^{1/4}$ which leads to
% {\red N: I changed to the notations $x \ll x_0$ and $(x_0-x) \sqrt{|E|_i} \gg 1 $ instead of $x \lesssim x_0$ and $(x_0-x) \sqrt{|E|_i} \gtrsim 1 $ (repsecitvely) in the following two equations.}
\be 
\label{phi1sinh}
\phi_1(x) \simeq \frac{2}{|E_1|^{1/4}} \sinh\left( \sqrt{|E_1|} x \right) \quad , \quad x \ll x_0 \, .
\ee
The value of $x_0$ can be determined by matching the two expressions \eqref{phiSolLocalized} and \eqref{phi1sinh} in their joint regime of validity $1/\sqrt{|E_1}\ll x \ll x_0$. Indeed,
% On the other hand 
one has, from \eqref{phiSolLocalized} 
\be 
\phi_1(x) \simeq 2 \sqrt{2} \sqrt{|E_1|} e^{- (x_0-x) \sqrt{|E_1|}}
\quad , \quad (x_0-x) \sqrt{|E_1|} \gg 1 
\ee 
Both are proportional to $e^{\sqrt{|E_1|} x}$ and the matching
of the prefactor gives 
$\frac{1}{2 \sqrt{2} |E_1|^{3/4}} \simeq e^{- x_0 \sqrt{|E_1|}}$
which leads to 
\be 
\label{x0ofEAsymptotic}
x_0 \simeq \frac{3}{4} \frac{\ln (4 |E_1|)}{\sqrt{|E_1|}}.
\ee 
The prediction \eqref{x0ofEAsymptotic} displays excellent agreement with our exact results which are based on a numerical solution to Eq.~\eqref{NonLinearPhiEq2}, see Fig.~\ref{figphi1ofxSoliton}. In particular, it agrees with the value that we used in Fig.~\ref{figphi1ofxSoliton}(a) for the particular case $E_1 = -126$.

% {\red Remark: P: one can probably do the matching even better since we are just solving again
% the double well classical motion with an energy which is $1/2 \phi_i'(0)^2= 2 |s'(E_i)|$
% hence we can use the exact formula. My guess is that it will give the
% same value of $x_0$ but maybe we should check it}

 Importantly, determining $x_0$ enables us to obtain corrections to the leading-order result \eqref{sSolRightTail} for the tail of the rate function. Indeed, now we can use the  identity \eqref{sEphiIdentity1}, and, replacing  Eq.~\eqref{phiSolLocalized} in the integrand, we %first identity in \eqref{2id},
obtain 
% {\red N: I flipped the sign of the expression $\frac{3}{4} \ln \left(4 |E_1|\right)$}
\be 
\frac{2}{3} E_1 s'(E_1) - s(E_1) = - \frac{1}{4} \int dx x \phi(x)^2 
\simeq - \frac{1}{2} x_0 \sqrt{|E_1|} \int_{-\infty}^{+\infty}  \frac{dy}{\cosh(y)^2} 
= - x_0 \sqrt{|E_1|} \simeq  - \frac{3}{4} \ln \left(4 |E_1|\right) \, ,
\ee
where we used the value of $x_0$ from \eqref{x0ofEAsymptotic}.
Integrating, and taking into account the known leading term \eqref{sSolRightTail}, we find 
% {\red N: I changed $o(|E_1|) \to o(1)$ in the following equation, and flipped some signs}
\be
\label{sSolRightTailSubleading}
% s(E_1) = \frac{2}{3} (- E_1)^{3/2} - \frac{3}{4} \ln(-E_1) + c + o(|E_1|) \quad , \quad c=\frac{3}{4} \ln 4 - \frac{1}{2} 
s(E_1) = \frac{2}{3} \left(- E_1\right)^{3/2} + \frac{3}{4} \ln\left(- E_1\right) + c + o(1) \quad , \quad c=\frac{3}{4} \ln 4 + \frac{1}{2} 
\ee 
which perfectly matches the leading and subleading terms in \eqref{tailsTWbeta}. 
  Indeed, taking the large-$\beta$ limit in the expression~\eqref{cBetaNadal} for $c_{\beta}$, one obtains
$c_{\beta}\sim e^{-\beta\left(\frac{3}{4}\ln4+\frac{1}{2}\right)}$,
in perfect agreement with our constant $c$ in \eqref{sSolRightTailSubleading}.
\\

It is important to note that the solution \eqref{phiSolLocalized} does not vanish at any $x>0$; As a result, it may only describe the ground state $i=1$. Wave functions of the excited states ($i=2,3,\dots$) are (approximately) given by sums of solutions of the type \eqref{phiSolLocalized} with multiple $x_0$'s and alternating signs, corresponding to the motion of a classical particle in the potential \eqref{Ueffdef} with ``energy'' that is slightly larger than zero. For example, the first excited state is given by
\be
\label{phiSolExcited}
\phi_{2}(x) \simeq \frac{\sqrt{-2E_{2}}}{\cosh\left(\sqrt{-E_{2}}\left(x-x_{0,1}\right)\right)}-\frac{\sqrt{-2E_{2}}}{\cosh\left(\sqrt{-E_{2}}\left(x-x_{0,2}\right)\right)} \, ,
\ee
where $1/\sqrt{-E_i} \ll x_{0,1} , x_{0,2} , x_{0,2} - x_{0,1} \ll 1$
The classical motion in the potential \eqref{Ueffdef} dies down at $x \to +\infty$ due to subleading corrections. 
% \sout{to the approximation made here (i.e., due to the small effect of the potential $V_0(x)$ that was altogether neglected above). Furthermore, the value of $x_0$ for the ground state, or of $x_{0,j}$ for the excited states, may also be calculated from subleading corrections.  We do not attempt to perform the subleading-order calculation here.}
The values of $x_{0,j}$ for the excited states can, in principle, be calculated in a similar manner to the calculation of $x_0$ (for the ground state) which we performed above. The calculation, which we do not attempt to perform here, would involve matching between solutions of the type \eqref{phi1sinh} from linear theory in the vicinity of each of the $x_{0,j}$'s, and solutions of the type \eqref{phiSolLocalized} sufficiently far from the $x_{0,j}$'s.

% Finally, we evaluate the action on the solution. For the ground state, plugging \eqref{phiSolLocalized} into \eqref{Sofphi} we obtain
% \be
% \label{sSolRightTail}
% % s(E_i) \simeq \frac{1}{8}\int_{-\infty}^{+\infty}\frac{4E_{i}^{2}}{\cosh^{4}\left(\sqrt{-E_{i}}\left(x-x_{0}\right)\right)}dx=\frac{\left(-E_{i}\right)^{3/2}}{2}\int_{-\infty}^{+\infty}\frac{dy}{\cosh^{4}y}=\frac{2\left(-E_{i}\right)^{3/2}}{3}\,,
% s(E_{1})\simeq\frac{1}{8}\int_{-\infty}^{+\infty}\frac{4E_{1}^{2}}{\cosh^{4}\left(\sqrt{-E_{1}}\left(x-x_{0}\right)\right)}dx=\frac{\left(-E_{1}\right)^{3/2}}{2}\int_{-\infty}^{+\infty}\frac{dy}{\cosh^{4}y}=\frac{2\left(-E_{1}\right)^{3/2}}{3}\,.
% \ee
For the excited states, in the leading order, the contributions of each of the localized solutions around the $x_{0,j}$'s to the action add up, and since there are $i$ such terms, one obtains in general
\be
\label{sSolRightTaili}
s(E_{i})\simeq \frac{2 i \left(-E_{1}\right)^{3/2}}{3}\,.
\ee
leading to the asymptotic behavior
\be
\label{PSolRightTail}
f_\beta\left(a_{i}\right)\sim e^{-2\beta i a_{i}^{3/2} \! /3} \, , \quad a_i \to +\infty.
\ee

It is worth noting that, as has been observed in several other contexts %as often is the case when applying the OFM 
\cite{SMP,CorwinGhosal,LDKPZandKP2019, Zilber2019, Agranov2020, Smith22OU, MukerjeeSmith23CH, Smith24Absx}, the OFM is expected to be valid sufficiently far in the tails of the distribution even if the noise is not weak, since the action there is very large. We thus expect the asymptotic behavior \eqref{PSolRightTail} to be valid at $a_i \to +\infty$ even if $\beta$ is not large, and indeed, for $i=1$ Eq.~\eqref{PSolRightTail} is in perfect agreement with the well-known results for the right tail of the Tracy-Widom distribution, see e.g., Refs.~\cite{RamirezRiderVirag2011, DV13}. For $i \geq 2$ we are not aware of any rigorous result.
Note that in the regime $\lambda_i - 2=O(1)$, a simple Coulomb gas argument (see e.g.~\cite{MS14}) leads to a large deviation
probability 
$\exp\left(-\beta Ni\Phi_{+}\left(\lambda_{i}-2\right)\right)$ (assuming that the optimum
configuration is $\lambda_1 \approx \dots \approx \lambda_i$). 
It matches the above result if one takes $\lambda_i - 2=a_i N^{2/3}$ 
and uses that $\Phi_+(x) \simeq \frac{2}{3} x^{3/2}$ for $x \ll 1$.

\subsection{$a_i \to -\infty$ tail: Large-scale solution}
\label{subsec:ThomasFermi}

In the limit $E_i \to + \infty$ ($a_i \to -\infty$), the behavior is very different: The spatial scale of the solution to Eq.~\eqref{NonLinearPhiEq2} is  much larger than one. In the leading order, $\phi_i(x)$ is nonnegligible, and slowly varying, in the region $0<x<E_i$, and negligible at $x>E_i$ \cite{SmithGroundState24}. As a result, at $0<x<E_i$ the term in \eqref{NonLinearPhiEq2} with the second derivative may be neglected, and (using that $\text{sgn}\left(E_{i}-E_{i}^{\left(0\right)}\right)=+1$)
%$\text{sgn}(\lambda)=1$) 
the equation becomes trivial to solve, and its solution is given by
\be
\label{phiSolLargeE}
\phi_{i}(x)\simeq\begin{cases}
\sqrt{E_{i}-x}\,, & 0<x<E_{i}\,,\\[1mm]
0 \, , & x>E_{i} \, .
\end{cases}
\ee
The solution \eqref{phiSolLargeE} is not valid very close to $x=0$. Near the origin, as we observed in numerical solutions,  there is a narrow boundary layer in which the solution adapts to the boundary condition $\phi(0)=0$, see Fig.~\ref{figphi1ofxTF}(a). Moreover, for the excited states ($i=2,3,\dots$), the solution crosses the origin $i-1$ times within this boundary layer, see e.g. Fig.~\ref{figphi1ofxTF}(b) for the first excited state.
% \sout{See Sec.~\ref{sec:Ablowitz} below for an estimate of the width of this boundary layer, and associated correction term to the action.}
Physically, it is perhaps more intuitive to understand this solution in terms of the corresponding optimal realization of the potential $V(x)$, which, using \eqref{V1ProptoPsisq}, is given by
\be
\label{VxPositiveTail}
V(x)\simeq\begin{cases}
E_i\,, & 0<x<E_{i}\,,\\[1mm]
x\,, & x>E_{i}\,.
\end{cases}
\ee
The disorder thus increases the minimal value of the potential to $E_i$.
One can now immediately calculate the action \eqref{Sofphi} to obtain
\be
\label{SSolLeftTailLeadingOrder}
s\left(E_{i}\right)\simeq\frac{1}{8}\int_{0}^{E_{i}}\left(E_{i}-x\right)^{2}dx=\frac{E_{i}^{3}}{24} \, .
\ee
Note that this result does not (explicitly) depend on $i$. The dependence on $i$ will be apparent in the subleading corrections to Eq.~\eqref{SSolLeftTailLeadingOrder},  see below.
%\sout{which we do not attempt to calculate analytically here.}
Our leading order result \eqref{SSolLeftTailLeadingOrder}, 
corresponds to the following asymptotic behavior for the distribution of $a_i$ in the APP 
\be
\label{PSolLeftTail}
f_{\beta}\left(a_{i}\right)\sim e^{-\beta\left(-a_{i}\right)^{3}\!/24}\,, \quad a_i \to -\infty \, .
\ee
obtained here in the limit where $\beta$ is taken to be large before $|a_i|$.
For $i=1$ Eq.~\eqref{PSolLeftTail} is in perfect agreement with 
the leading, fixed $\beta$, left tail asymptotics \eqref{tailsTWbeta}
of the Tracy-Widom distribution. 

An interesting question is whether we can obtain the subleading term in 
the large $E_i$ asymptotics of $s(E_i)$. Indeed we
see that there is a subleading term $\sim |a|^{3/2}$ which is proportional to $\beta$ 
in the fixed $\beta$ left tail result 
\eqref{tailsTWbeta}, hence if matching holds we should expect
\begin {equation}
\label{SSolLeftTail}
s(E_1) \simeq \frac{E_1^3}{24}-\frac{\sqrt 2}{6} E_1^{3/2} \, .
\end {equation}

\begin{figure}
% \includegraphics[width=0.47\textwidth,clip=]{figPhi1ofxEMinus100.pdf}
%  \hspace{2mm}
\includegraphics[width=0.47\textwidth,clip=]{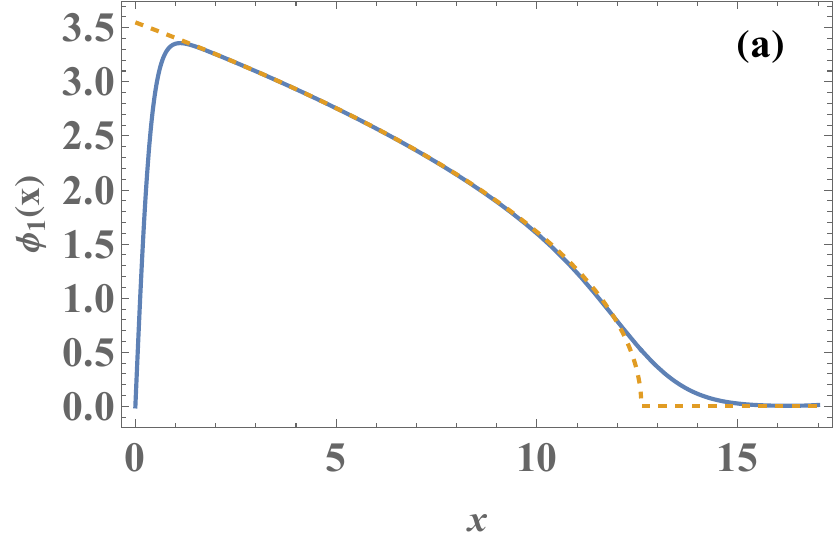}
\hspace{2mm}
\includegraphics[width=0.47\textwidth,clip=]{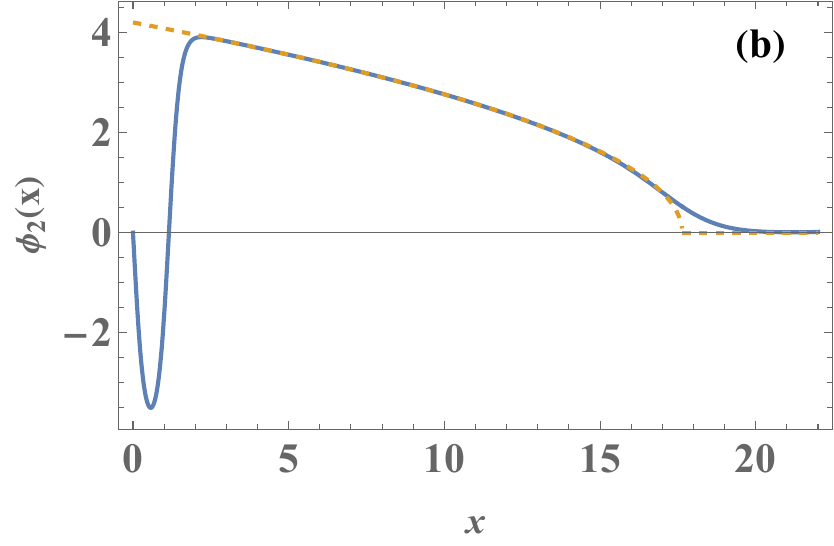}
\caption{(a) Solid line: Exact (numerically-obtained) solution $\phi_1(x)$ to Eq.~\eqref{NonLinearPhiEq} for ground-state energy $E_1 = 12.6$.
Dashed line: the (approximate) theoretical prediction \eqref{phiSolLargeE}.
(b) Similarly for the first excited state, with $E_2 = 17.6$.}
\label{figphi1ofxTF}
\end{figure}

We now show how to derive the subleading term in \eqref{SSolLeftTail} by studying
% To obtain the subleading behavior we study 
the boundary layer for $x$ near zero. We start with $i=1$ and 
subsequently consider $i >1 $. 
We employ a matched asymptotic expansion. In the region $x \ll E_1$,
% In that region 
one can neglect the $x$ term in \eqref{NonLinearPhiEq2}, 
% For $i=1$ 
which yields the equation
$- \phi_1'' - E_1 \phi_1 + \phi_1^3 = 0$. Inserting
the boundary layer form 
\be
\phi_1(x)=\sqrt{E_1} \, f\left(x \sqrt{E_1}\right)\,,
\ee
we obtain
\be
- f''(y) - f(y) + f(y)^3 = 0 \, ,
\ee
which is to be solved subject to the boundary conditions $f(0)=0$ and 
% with boundary condition
$f(+\infty)=1$ to match the outer solution \eqref{phiSolLargeE} in the joint region of validity $1/\sqrt{E_1} \ll x \ll E_1$.
Using a similar mechanical analogy to the one from the previous subsection, this becomes equivalent to solving
\be
\label{fyEnergyConservation}
\frac{1}{2} f'(y)^2 + \frac{1}{2} f(y)^2 -  \frac{1}{4}  f(y)^4= \frac{1}{4} \, ,
\ee
which is the conservation of ``mechanical energy'' equation, where the constant on the right-hand side is determined from the boundary condition at $y \to +\infty$. Note that, due to the boundary condition at $y=0$, we obtain $f'(0) = 1/\sqrt{2}$, i.e.,
$\phi_1'(0)=2 \sqrt{s'(E_1)} \simeq E_1/\sqrt{2}$, 
in agreement with \eqref{sEider}.
% This is equivalent to solving
% \be 
%  \frac{1}{2} \phi_1'(x)^2 + \frac{1}{2} E_1 \phi_1(x)^2 -  \frac{1}{4}  \phi_1(x)^4= \frac{1}{4} E_1^2 
% \ee 
% where at $\phi(0)=0$ which gives $\phi_1'(0)=2 \sqrt{s'(E_1)} \simeq E_1/\sqrt{2}$ 
% in agreement with \eqref{sEider} 
% and 
% at $x=+\infty$ it gives $\phi_1'(x) \to 0$ and $\phi_1(x) \to \sqrt{E_1}$ {\red P: one should
% rewrite this equation for the BL function $f$ to be mathematically correct, I am being sloppy here.}
We now solve Eq.~\eqref{fyEnergyConservation} with the Dirichlet boundary condition at the origin to obtain
\be 
\int_0^{f} \frac{d\tilde{f}}{1-\tilde{f}^2} = y/\sqrt{2} \quad , \quad 
 f(y)= \tanh\left( y /\sqrt{2}\right)  \, .
\ee 
% One finds 
% \be 
% \int_0^{\phi_1} \frac{d\phi}{E_1-\phi^2} = x/\sqrt{2} \quad , \quad 
% \phi_1(x) = \sqrt{E_1}  f\left(x \sqrt{E_1}\right) 
% \quad , \quad f(y)= \tanh\left( y /\sqrt{2}\right) 
% \ee 
We now use the relation \eqref{sEphiIdentity2}, and plugging in the boundary layer solution%
\footnote{The contribution of the outer solution \eqref{phiSolLargeE} to the integral in \eqref{sEphiIdentity2}, is of order $O(1)$ and we therefore neglect it.},
we obtain
%second relation in \eqref{2id}
\be 
\label{identity2Soliton}
E_1 s'(E_1) - 3 s(E_1) = \frac{3}{4} \int_0^{+\infty} dx \phi_1'(x)^2
\simeq  \frac{3}{4} E_1^{3/2} \int_0^{+\infty} dy f'(y)^2 = \frac{3}{4} \frac{\sqrt{2}}{3} E_1^{3/2} \, .
\ee 
%\sout{since one can check that the leading behavior of the r.h.s is entirely given by the boundary layer.} 
 Taking into account the known leading-order behavior \eqref{SSolLeftTailLeadingOrder},
this demonstrates \eqref{SSolLeftTail}.
% \be 
% \label{SSolLeftTailnew}
% s(E_1) \simeq \frac{E_1^3}{24}-\frac{\sqrt 2}{6} E_1^{3/2} \, .
% \end {equation}

% {\red P: To get the higher $i$ we would need to find boundary layer solutions 
% with more oscillations. One sees how to do it if one solves simply
% $- \phi_i'' - E_i \phi_i = 0$ i.e. one has 
% $\phi_k(x)= (\sqrt{E_k}/k) \sin(k \sqrt{E_k} x)$, which is
% the correct solution for $\phi_k(x)$ small but
% when one adds the $\phi^3$ term I do not see yet 
% how the above calculation (energy conservation) works, do you?.}\\
% {\magenta N: I think that, to get additional oscillations, one must go to higher order in the theory. However, it is not necessary to carry out the calculation explicitly. Basically, the function $f(y)$ will start at $f(0)=0$, and then go back and forth between the values $+1$ and $-1$, lingering for a long time at each one of these values, and taking the form of $\tanh$ when going between them. As a result, the $i$-dependence of the integral will be $\int \phi_i'(x)^2 dx \propto 2i-1$, so I believe that in general
% $s(E_i) \simeq \frac{E_i^3}{24}-(2i-1)\frac{\sqrt 2}{6} E_i^{3/2}$.
% This seems to agree reasonably well with numerics for $i=2$.\\
% On second thought, I think this is related to what Alain wrote to us in one of his last emails, where he asked if I could send him the extension of the solution to negative $x$.}

 Let us now briefly describe the structure of the boundary layer for the excited states, $i=2,3,\dots$, and thus obtain the subleading corrections for them.
 For the excited states, the boundary layer consists of oscillations of the wave function $\phi_i(x)$ between the two values $\pm\sqrt{E_i}$, see Fig.~\ref{figphi1ofxTF} for the first excited state.
 Each of the $i-1$ half-oscillations, i.e., each transition between $\phi_i(x) \simeq \pm \sqrt{E_i}$ to $\phi_i(x) \simeq \mp \sqrt{E_i}$, is described by the same hyperbolic tangent scaling function $f(y)$ given above, extended to the entire real line $-\infty < y < +\infty$, such that $f(\pm \infty) = \pm 1$. In addition, there is the initial quarter-oscillation beginning at $x=0$ (as in the ground state).
 The half-oscillations contribute twice as much as the quarter-oscillation to the integral in \eqref{sEphiIdentity2}, and thus, the generalization of \eqref{identity2Soliton} to all $i=1,2,\dots$ is
 $E_i s'(E_i) - 3 s(E_i) \simeq (2i-1)\frac{3}{4} \frac{\sqrt{2}}{3} E_i^{3/2}$,
 from which we obtain the subleading correction to the asymptotic behavior of the rate function
\be
\label{SSolLeftTailGenerali}
s(E_i) \simeq \frac{E_i^3}{24}-(2i-1)\frac{\sqrt 2}{6} E_i^{3/2} \, .
\ee
in a similar manner to the calculation above for the particular case $i=1$.
For the first excited state, $i=2$, Eq.~\eqref{SSolLeftTailGenerali} shows good agreement with numerical results, see Fig.~\ref{figsOfEAiry}(b).
%  So we obtain, for the first excited state,
%  \begin {equation}
% \label{SSolLeftTaili2}
% s(E_2) \simeq \frac{E_2^3}{24}- \frac{1}{\sqrt{2}} E_2^{3/2} \, .
% \end {equation}

\subsection{Typical fluctuations $a_i \simeq \left\langle a_{i}\right\rangle $}
\label{subsec:Typical}

% {\red P: I find some of the formula below a bit heavy notationally. It would be
% better to introduce your coefficients $A,B$ (maybe give them more generic names like
% them $\tilde s_2, \tilde s_3$ ?) and make some of these
% formula less heavy. 
% Since $\mu(\lambda)$ has been introduced 
% you may want to use 
% \be 
% s'(E)= \lambda   \quad , \quad \mu'(\lambda) = E  \quad \Rightarrow \mu''(\lambda)= \frac{1}{\lambda} s'(\lambda)
% \ee 
% which implies that 
% $s(\lambda)= \sum_{n \geq 2} \frac{\tilde s_n}{n!} \lambda^n = \sum_{n \geq 2} \frac{\kappa_n}{n (n-2)!} 
% \lambda^n$, hence the cumulants are $\kappa_n = \frac{\tilde s_n}{n-1}$ and maybe they can 
% be introduced earlier, that would allow to get rid of this Appendix which I find
% also a bit heavy.

% {\bf Important note:} I am defining $s(E)=\sum_{n} \frac{s_n}{n!} (E-E_{\rm typ})^n$
% in Appendix that seems a good notation, hence $\tilde s_n$ would be more appropriate
% for $s(\lambda)$ (although it is less fundamental because beyond order 3 
% $s$ is not a function of $\lambda$ alone, see Appendix)
% }

% {\magenta N: I deleted some redundant equations in this subsection, but I thought it best to leave the final results in a relatively explicit form (i.e., without using some shorter notation for the inner product), I think that the way it currently is makes it easier for readers who will wish to compare their results with ours, without reading our paper in detail... But if you prefer, I will change the notation.}

The behavior of $s(E_i)$ around the typical value $E_i \simeq E_i^{(0)}$ describes the distribution of the typical fluctuations of $E_i$. 
Since $s(E_i)$ is minimal at $E_i = E_i^{(0)}$ and since $ds/dE_i=\lambda$, the regime $E_i \simeq E_i^{(0)}$ corresponds to $\lambda \to 0$. To study the typical regime it is thus natural to compute $s$ as a power series in $\lambda$, as we do below. Before doing so, it is useful to relate the coefficients of this series to the 
cumulants of $E_i$. 
To this aim let us recall the definition Eq.~\eqref{muDef} of the cumulant generating function (CGF), extended to describe any level $E_i$  (we suppress its dependence on $i$ for brevity).
From its series expansion $\mu(\lambda)=\sum_{n \geq 1} \frac{\kappa_n}{n!} \lambda^n$,
one obtains the cumulants of $E_i$, i.e. 
$\langle (E_i-E_i^{(0)})^n \rangle_c \simeq \frac{\kappa_n}{\beta^{n-1}}$
(as in \eqref{cuma1n}, extended to any $i$, with $C_n=(-1)^n \kappa_n$). 
Now, using the relations
\be 
\frac{ds}{dE_i}= \lambda   \quad , \quad \frac{d\mu}{d\lambda} = E_i
\ee
between the rate function and the CGF one finds that
\be
\frac{d^2\mu}{d\lambda^2}= \frac{1}{\lambda} \frac{ds}{d\lambda} \, .
\ee
Hence the perturbative expansion of $s$ as a power series in $\lambda$ reads
\be
s(\lambda)= \sum_{n \geq 2} \frac{\tilde s_n}{n!} \lambda^n = \sum_{n \geq 2} \frac{\kappa_n}{n (n-2)!} \lambda^n \, ,
\ee
so that the reduced cumulants of $E_i$ can be retrieved from
this expansion as $\kappa_n = \frac{\tilde s_n}{n-1}$.

At $\lambda \to 0$, the potential $V_0(x)$ is dominant, while the disorder $V_1(x)$ is a small perturbation. %, as can be seen, e.g., from Eq.~\eqref{V1ProptoPsisq}.
It is convenient to work with the normalized wave function $\psi_i(x)$.
In the leading order, it is simply given by the zero-noise one $\psi_i^{(0)}(x)$, and we obtain the subleading correction $\psi_{i}\left(x\right)\simeq\psi_{i}^{\left(0\right)}\left(x\right)+\psi_{i}^{\left(1\right)}\left(x\right)$ by applying first-order perturbation theory
% \bea
% \label{psiPerturbation}
% \psi_{i}^{\left(1\right)}\left(x\right)&=&\sum_{j\ne i}\frac{\int_{-\infty}^{+\infty}V_{1}\left(y\right)\psi_{i}^{\left(0\right)}(y)\psi_{j}^{\left(0\right)}(y)dy}{E_{i}^{\left(0\right)}-E_{j}^{\left(0\right)}}\psi_{j}^{\left(0\right)}\left(x\right)+O\left(\lambda^{2}\right)=\nn\\
% &=&\sum_{j\ne i}\frac{\int_{-\infty}^{+\infty}4\lambda\psi_{i}^{\left(0\right)}(y)^{3}\psi_{j}^{\left(0\right)}(y)dy}{E_{i}^{\left(0\right)}-E_{j}^{\left(0\right)}}\psi_{j}^{\left(0\right)}\left(x\right)+O\left(\lambda^{2}\right) \, ,
% \eea
\be
\label{psiPerturbation}
\psi_{i}^{\left(1\right)}\left(x\right)=\sum_{j\ne i}\frac{\left\langle \psi_{j}^{\left(0\right)}|V_{1}|\psi_{i}^{\left(0\right)}\right\rangle }{E_{i}^{\left(0\right)}-E_{j}^{\left(0\right)}}\psi_{j}^{\left(0\right)}\left(x\right)+O\left(\lambda^{2}\right)=\sum_{j\ne i}\frac{\int_{-\infty}^{+\infty}4\lambda\psi_{i}^{\left(0\right)}(y)^{3}\psi_{j}^{\left(0\right)}(y)dy}{E_{i}^{\left(0\right)}-E_{j}^{\left(0\right)}}\psi_{j}^{\left(0\right)}\left(x\right)+O\left(\lambda^{2}\right)\,,
\ee
where we used the leading-order approximation $V_{1}\left(x\right)\simeq4\lambda\psi_{i}^{\left(0\right)}\left(x\right)^{2}$ of Eq.~\eqref{V1ProptoPsisq}.
%\sout{Eq.~\eqref{V1ProptoPsisq} to move from the first line of the equation to the second.}
We now use the relation $\phi_i = 2\sqrt{|\lambda|} \,  \psi_i$ to rewrite Eq.~\eqref{Sofphi} in terms of the normalized wave function, and using \eqref{psiPerturbation}, we obtain $s$ as a function of $\lambda$, up to cubic order:
% {\red N: I checked the signs in the following equations and they are indeed correct, thank you for noticing the mistakes.}
% \bea
% \label{sOfLambdaPerturbation}
% s&=&2\lambda^{2}\int_{-\infty}^{+\infty}\psi_{i}^{4}\left(x\right)dx\simeq2\lambda^{2}\int_{-\infty}^{+\infty}\psi_{i}^{\left(0\right)}\left(x\right)^{4}dx+8\lambda^{2}\int_{-\infty}^{+\infty}\psi_{i}^{\left(0\right)}\left(x\right)^{3}\psi_{i}^{\left(1\right)}\left(x\right)dx\nn\\
% &\simeq&2\lambda^{2}\int_{-\infty}^{+\infty}\psi_{i}^{\left(0\right)}\left(x\right)^{4}dx+32\lambda^{3}\sum_{j\ne i}\frac{\left[\int_{-\infty}^{+\infty}\psi_{i}^{\left(0\right)}(y)^{3}\psi_{j}^{\left(0\right)}(y)dy\right]^{2}}{E_{i}^{\left(0\right)}-E_{j}^{\left(0\right)}} \, .
% \eea
\bea
\label{sOfLambdaPerturbation}
s&=&2\lambda^{2}\int_{-\infty}^{+\infty}\psi_{i}^{4}\left(x\right)dx = \frac{\tilde{s}_{2}\lambda^{2}}{2!}+\frac{\tilde{s}_{3}\lambda^{3}}{3!}+O\left(\lambda^{4}\right) \, ,\\[1mm]
\label{tildes2sol}
\frac{\tilde{s}_{2}}{2!}&=&2\int_{-\infty}^{+\infty}\psi_{i}^{\left(0\right)}\left(x\right)^{4}dx =  \frac{2\int_{0}^{+\infty}\text{Ai}\left(x+\zeta_{i}\right)^{4}dx}{\text{Ai}'\left(\zeta_{i}\right)^{4}} \, ,\\[1mm]
\label{tildes3sol}
\frac{\tilde{s}_{3}}{3!}&=&32\sum_{j\ne i}\frac{\left[\int_{-\infty}^{+\infty}\psi_{i}^{\left(0\right)}(y)^{3}\psi_{j}^{\left(0\right)}(y)dy\right]^{2}}{E_{i}^{\left(0\right)}-E_{j}^{\left(0\right)}} = 32\sum_{j\ne i}\frac{\left[\int_{-\infty}^{+\infty}\text{Ai}^{3}(y+\zeta_{i})\text{Ai}(y+\zeta_{j})dy\right]^{2}}{ \left(\zeta_{j}-\zeta_{i}\right) \text{Ai}'\left(\zeta_{i}\right)^{6}\text{Ai}'\left(\zeta_{j}\right)^{2}} \,.
\eea
from which we immediately obtain the three lowest cumulants of the Tracy-Widom distribution at $\beta \gg 1$, describing the typical-fluctuations regime: In this regime, $f_\beta(a_i)$ is Gaussian with mean
$\left\langle a_{i}\right\rangle  \simeq -E_i^{(0)} = \zeta_i$ and second and third cumulants given by
\be
\label{VaraSol}
% \text{Var}\left(a_{i}\right)\simeq\frac{4\int_{0}^{+\infty}\text{Ai}\left(x+\zeta_{i}\right)^{4}dx}{\beta\text{Ai}'\left(\zeta_{i}\right)^{4}} \, ,
\text{Var}\left(a_{i}\right)\simeq\frac{\tilde{s}_{2}}{\beta}\,,\qquad\left\langle a_{i}^{3}\right\rangle _{c}\simeq-\frac{\tilde{s}_{3}}{2\beta^{2}}\,,
\ee
respectively, the variance being in perfect agreement with Refs.~\cite{Dumitriu2005, EPS14, AHV21,Touzo}.
% (note that in Ref.~\cite{EPS14}, the leading-order correction to the mean, of order $\sim 1/\beta$, is also given).
% The third cumulant is given by
% \be
% \label{ThirdCumulant}
% \left\langle a_{i}^{3}\right\rangle _{c}\simeq\frac{96}{\beta^{2}}\sum_{j\ne i}\frac{\left[\int_{-\infty}^{+\infty}\text{Ai}^{3}(y+\zeta_{i})\text{Ai}(y+\zeta_{j})dy\right]^{2}}{\left(\zeta_{i}-\zeta_{j}\right)\text{Ai}'\left(\zeta_{i}\right)^{6}\text{Ai}'\left(\zeta_{j}\right)^{2}} \, .
% \ee
%
From here we can calculate the skewness
\be
\gamma_{1}=\frac{\left\langle a_{i}^{3}\right\rangle _{c}}{\left[\text{Var}\left(a_{i}\right)\right]^{3/2}}\simeq
- \frac{\tilde{s}_{3}}{2\tilde{s}_{2}^{3/2}\sqrt{\beta}} \, .
\ee
For the maximal eigenvalue, $i=1$, a numerical computation of these expressions%
\footnote{ We found that the sum in \eqref{tildes3sol} converges quite rapidly, and that truncating it after the first 100 terms was sufficient to obtain the level of accuracy that we report.}
gives
\be
\text{Var}\left(a_{1}\right)\simeq\frac{1.6697}{\beta}\,,\quad\left\langle a_{1}^{3}\right\rangle _{c}\simeq\frac{0.744736}{\beta^{2}},\quad\gamma_{1}\simeq\frac{0.345169}{\sqrt{\beta}} \, .
\ee
In fact we find that even for $\beta=1,2,4$ these predictions are in reasonable agreement with exact (numerical) results \cite{Bornemann10}, see Table \ref{table:MeanVarianceSkewness}.

More generally, it follows from the scaling \eqref{PaScaling} (together with the assumption that the rate function is analytic at its minimum) that the $n$th cumulant of $a_i$ scales as
\be
\left\langle a_{i}^{n}\right\rangle _{c}\simeq C_{n}\beta^{1-n} 
\ee
at large $\beta$, where $C_n = (-1)^n \kappa_n$ is a constant. In particular, the fourth cumulant scales as $\left\langle a_{i}^{4}\right\rangle _{c}\simeq C_{4}\beta^{-3}$
where $C_4$ can be obtained by pushing the perturbative procedure performed above to the next order. This, in turn, implies that the excess kurtosis scales as
\be
\gamma_{2}=\frac{\left\langle a_{i}^{4}\right\rangle _{c}}{\left[\text{Var}\left(a_{i}\right)\right]^{2}}\sim\beta^{-1}
\ee
as $\beta \to +\infty$.

In Appendix \ref{app:thirdcum}, we use a different method to obtain the fourth cumulant and the excess kurtosis,
\be
\label{FourthCumKurtosis}
\left\langle a_{1}^{4}\right\rangle _{c}\simeq \frac{0.55761182}{\beta^3} \, ,\qquad \gamma_{2}\simeq \frac{0.20000341}{\beta}
% \left\langle a_{1}^{4}\right\rangle _{c}\simeq \frac{0.55761403}{\beta^3} \, ,\qquad \gamma_{2}\simeq \frac{0.2000042}{\beta}
\ee
(the analytic value of $C_4$, in terms of Airy integrals, is given in Appendix \ref{app:thirdcum}, see Eq.~\eqref{4cumapp}).
The value $0.2$ of the coefficient  that gives the kurtosis is also in fair agreement with the known results for
$\beta=1,2,4$ \cite{Bornemann10}, see Table \ref{table:MeanVarianceSkewness}.
We also obtain the subleading corrections to the mean and variance of the $a_i$'s, see Eqs.~\eqref{meanCorrection} and \eqref{varianceSubleading} respectively (the correction to 
the mean was also obtained in \cite{EPS14, Touzo}).
In Appendix \ref{app:perturbation} we rederive some of these results directly from the cubic equation \eqref{NonLinearPhiEq2}.

\begin{table}
 \begin{tabular}{ | l | l | l | l | l | l | l | l | l |}
    \hline
Dyson index $\beta$ & Mean & $\zeta_1 + 1.12/\beta$ & Variance & $1.6697/\beta$ & Skewness & $0.344728/ \! \sqrt{\beta}$ & Kurtosis & $0.2/\beta$\\ \hline
1 &  $-1.20653$ &  $-1.22$ & $1.60778$ & $1.6697$ & $0.29346$  & $0.344728$ & $0.16524$  & $0.2$ \\ \hline
2 & $-1.77109$ &  $-1.78$ & $ 0.81319$ & $0.8348$ & $0.22408$  & $0.244071$ & $0.093448$ & $0.1$ \\ \hline
4 & $-2.05520$ & $-2.06$ & $0.41091$ & $0.41742$ & $0.16550$  & $0.172584$ & $0.049195$ & $0.05$ \\ \hline
% $\beta \to +\infty$ & $-2.33811$ &  $1.6697/\beta$ & & $0.344728/\sqrt{\beta}$  & \\ \hline
    \hline
    \end{tabular}
    \caption{Mean, variance skewness and kurtosis of Tracy-Widom distributions with Dyson indices $\beta=1,2,4$, compared to the predicted the large-$\beta$ asymptotic behaviors. The exact numerical values are taken from Ref.~\cite{Bornemann10}. Results for the mean and variance were already compared to their large-$\beta$ asymptotic behaviors in \cite{EPS14}, and their comparisons are given here too for completeness.
    % {\red TODO: add subleading correction to the variance?}
    }
\label{table:MeanVarianceSkewness}
\end{table}

Finally, by applying the Legendre transform to the power-series expansion of the CGF $\mu(\lambda)$, 
one obtains the asymptotic behavior of the rate function $s(E_i)$ around the typical value,
\be
\label{sofEPowerSeries}
s\left(E_{i}\right)=\sum_{n=2}^{\infty}\frac{s_{n}}{n!}\left(E_{i}-E_{i}^{\left(0\right)}\right)^{n}\,,
\ee
where the first few coefficients are related to those from the expansion of $\mu(\lambda)$ via (see Appendix \ref{appendix:Legendre})
\be 
\label{sOfKappa}
s_2= \frac{1}{\kappa_2} \quad , \quad s_3 = - \frac{\kappa_3}{\kappa_2^3} 
\quad , \quad s_4 = \frac{3 \kappa_3^2 - \kappa_2 \kappa_4}{\kappa_2^5} \, .
\ee
To cubic order, we obtain 
\be
\label{sofECubicOrder}
s\left(E_{i}\right)\simeq\frac{1}{2\tilde{s}_{2}}\left(E_{i}-E_{i}^{\left(0\right)}\right)^{2} 
- \frac{\tilde{s}_{3}}{12\tilde{s}_{2}^{3}}\left(E_{i}-E_{i}^{\left(0\right)}\right)^{3}
\ee
A comparison of the exact $s(E_i)$ and its expansion \eqref{sofECubicOrder} up to cubic order is given in Fig.~\ref{figsOfEAiry}, for $i=1,2$.

Incidentally, the formulas \eqref{sEphiIdentity1} and \eqref{sEphiIdentity2} (which we did not use so far in the derivation of the cumulants) can be applied in the typical-fluctuations regime to yield nontrivial identities that involve the Airy functions, see Appendix \ref{app:identities}.

\bigskip

\section{Joint distributions of pairs of eigenvalues and gap distributions}
\label{sec:pairs}

\subsection{Pairs of eigenvalues}

Using a similar formalism to the one presented above, one can study the joint distribution of pairs of eigenvalues $a_i ,a_j$, which we will denote $f_\beta(a_i, a_j)$ (and similarly one can study the joint distributions of any number $k$ of eigenvalues). The action functional \eqref{sVdef} will have to be minimized under two constraints, which are the values $E_i = -a_i$ and $E_j = -a_j$ of the $i$th and $j$th energy levels, respectively, of the potential $V(x)$. We incorporate these two constraints by  extending the definition of the modified action functional 
\eqref{Slambdadef} to include now two 
Lagrange multipliers, which we denote by $\lambda_i$ and $\lambda_j$ respectively. Following a very similar derivation to the one given above for the distribution of a single eigenvalue, the saddle-point equations are the Schr\"{o}dinger equations for the (normalized) wave functions
\bea
\label{psiieq}
-\psi_{i}''\left(x\right)+\left[V_{0}\left(x\right)+V_{1}\left(x\right)\right]\psi_{i}\left(x\right)&=&E_{i}\psi_{i}\left(x\right)\,,\\[1mm]
\label{psijeq}
-
\psi_{j}''\left(x\right)+\left[V_{0}\left(x\right)+V_{1}\left(x\right)\right]\psi_{j}\left(x\right)&=&E_{j}\psi_{j}\left(x\right)\,,
\eea
where the optimal realization of the disorder is related to the wavefunctions through 
\be
\label{V1psiipsij}
V_1(x) = V(x)-V_{0}(x) = 4\lambda_{i}\psi_{i}(x)^{2}+4\lambda_{j}\psi_{j}(x)^{2} \, .
\ee
Eqs.~\eqref{psiieq}-\eqref{V1psiipsij} must be solved under the boundary conditions $\psi_i(0) = \psi_j(0) = \psi_i(\infty) = \psi_j(\infty) = 0$, and in addition $\psi_i(x)$ and $\psi_j(x)$ must vanish at exactly $i$ and $j$ points respectively (including at $x=0$).
The joint distribution is then given by
\be
\label{PaScalingij}
f_\beta\left(a_{i}, a_{j}\right)\sim e^{-\beta s\left(E_i = -a_{i}, E_j = -a_{j}\right)}, \qquad \beta \gg 1,
\ee
where the large-deviation function $s\left(E_i, E_j\right)$ is obtained by evaluating the action \eqref{sVdef}. If desired, the $\lambda_i$'s may be eliminated from the equations by rescaling $\phi_i = 2\sqrt{|\lambda_i|} \, \psi_i$ as we did for the case of a single energy level.

Eqs.~\eqref{psiieq}-\eqref{V1psiipsij} are, again, difficult to solve analytically except in certain limiting cases (see below), so in general one must solve them numerically.
The shooting method, which we used in the case of the distribution of a single energy level, is less convenient to use for solving Eqs.~\eqref{psiieq}-\eqref{V1psiipsij} since it would require the use of more than one shooting parameter.
Instead, we solve these equations using an iterations algorithm. The algorithm takes, as an input, the values of $\lambda_i$ and $\lambda_j$, and an initial guess $V_{1,0}(x)$ for the optimal realization of the disorder $V_1(x)$ (one can choose for instance $V_{1,0}(x)=0$). Given a candidate $V_{1,n}(x)$ for the optimal realization of the disorder at th $n$th iteration of the algorithm, we perform the $(n+1)$th iteration as follows. First, compute the wave functions $\psi_{i,n}(x)$ and $\psi_{j,n}(x)$ corresponding to the $i$th and $j$th energy levels (respectively) of the Hamiltonian $-d^2 / dx^2 + V_0(x) + V_{1,n}(x)$, through a numerical diagonalization. The realization of the disorder at the next iteration is then computed through
\be\label{Vnplus1}
V_{1,n+1}(x) = 4\lambda_{i}\psi_{i,n}(x)^{2}+4\lambda_{j}\psi_{j,n}(x)^{2}.    
\ee
The initial condition for this algorithm may be chosen to be zero disorder, $V_{1,0}(x) = 0$.
We find that the algorithm usually converges after $\sim 10$ iterations.
Once the algorithm has converged, one numerically computes $E_i$, $E_j$ and the action $s$ by using the numerically-found $V_1(x)$.

We found that the algorithm performs very well for small values of $\lambda_i, \lambda_j$. However, for large values of $\lambda_i, \lambda_j$, the algorithm tends to develop an instability. In these cases, in the second step of each iteration, we replace the right-hand side of Eq.~\eqref{Vnplus1} by a convex combination of itself with the realization of the disorder at the previous iteration, i.e.,
\be\label{Vnplus1Convex}
V_{1,n+1}(x) = \alpha V_{1,n+1}(x) + (1-\alpha) [4\lambda_{i}\psi_{i,n}(x)^{2}+4\lambda_{j}\psi_{j,n}(x)^{2}] \, ,
\ee
with $0 < \alpha < 1$  (for $\alpha \to 0$, this reduces to the original version of the algorithm). This modification restores the stability of the algorithm (if $\alpha$ is sufficiently large), but it increases the number of iterations until convergence. To ensure stabilitiy, it is sometimes necessary to choose $\alpha$ very close to 1, i.e., $1-\alpha \ll 1$. In this case, a reasonable way to test for convergence is to check if there is any significant change in the potential or the wavefunctions after $O(1/(1-\alpha))$ consecutive steps.
Our algorithm is inspired by the Chernykh-Stepanov algorithm \cite{CS01} which has been extensively used for solving saddle-point equations in several contexts, including the macroscopic fluctuation theory of lattice gases \cite{YK20, BSM22a, BSM22b} and in  studies of the Kardar-Parisi-Zhang equation \cite{MKV, SMV2019, TSM23}.

Analytically, we are able to solve the saddle-point equations in the typical fluctuations regime, using a similar method to the way we did it for a single energy level.
In the leading order, we approximate the (normalized) eigenfunctions by the zero-noise ones
$\psi_i(x) \simeq \psi_i^{(0)}(x)$.
Using Eq.~\eqref{V1psiipsij}, we obtain the optimal realization of the noise (in the leading order),
\be
V_{1}\left(x\right)\simeq4\left[\lambda_{i}\psi_{i}^{\left(0\right)}\left(x\right)^{2}+\lambda_{j}\psi_{j}^{\left(0\right)}\left(x\right)^{2}\right]\,.
\ee
Now applying first-order perturbation theory, we calculate the leading-order effect of the disorder on the energy levels:
\be
E_{i}-E_{i}^{\left(0\right)}\simeq\int V_{1}\left(x\right)\psi_{i}^{\left(0\right)}\left(x\right)^{2}dx\simeq4\int\left[\lambda_{i}\psi_{i}^{\left(0\right)}\left(x\right)^{2}+\lambda_{j}\psi_{j}^{\left(0\right)}\left(x\right)^{2}\right]\psi_{i}^{\left(0\right)}\left(x\right)^{2}dx
\ee
(and similarly for $E_j$).
This result can be rewritten in the form
\be
\label{EiEjlambdas}
\left(\begin{array}{c}
E_{i}-E_{i}^{\left(0\right)}\\[1mm]
E_{j}-E_{j}^{\left(0\right)}
\end{array}\right)\simeq \mathcal{C}\left(\begin{array}{c}
\lambda_{i}\\[1mm]
\lambda_{j}
\end{array}\right)\,,\quad \mathcal{C}=\left(\begin{array}{cc}
C_{ii} & C_{ij}\\[1mm]
C_{ji} & C_{jj}
\end{array}\right)\,,
\ee
with 
\be
C_{kl}=4\int\psi_{k}^{\left(0\right)}\left(x\right)^{2}\psi_{l}^{\left(0\right)}\left(x\right)^{2}dx=\frac{4\int_0^\infty \text{Ai}\left(x+\zeta_{k}\right)^{2}\text{Ai}\left(x+\zeta_{l}\right)^{2}dx}{\text{Ai}'\left(\zeta_{k}\right)^{2}\text{Ai}'\left(\zeta_{l}\right)^{2}} \, .
\ee

From Eq.~\eqref{EiEjlambdas}, one can immediately extract the quadratic behavior of $s(E_i, E_j)$ around the point $\left(E_{i}^{\left(0\right)}, E_{j}^{\left(0\right)} \right)$, by using the relations
$\partial s / \partial E_i = \lambda_i$, $\partial s / \partial E_j = \lambda_j$
[these relations are simple extensions of \eqref{dsdELambda}].
One finds
\be
\label{sEiEjTypical}
s \simeq\frac{1}{2}\left(\begin{array}{cc}
E_{i}-E_{i}^{\left(0\right)} & E_{j}-E_{j}^{\left(0\right)}\end{array}\right)\mathcal{C}^{-1}\left(\begin{array}{c}
E_{i}-E_{i}^{\left(0\right)}\\[1mm]
E_{j}-E_{j}^{\left(0\right)}
\end{array}\right)
\ee
where $\mathcal{C}^{-1}$ is the inverse of the matrix $\mathcal{C}$.
Putting this together with Eq.~\eqref{PaScalingij}, we find that typical joint fluctuations of pairs of energy levels follow a multivariate Gaussian distribution, with covariance matrix $\mathcal{C} / \beta$, i.e., the variance of each energy level is given by Eq.~\eqref{VaraSol} and the covariance of each pair of energy levels is given by 
\be
\label{CovaraicneSol}
\text{Cov}\left(a_{i},a_{j}\right)\simeq\frac{C_{ij}}{\beta}=\frac{4\int_{0}^{+\infty}\text{Ai}\left(x+\zeta_{i}\right)^{2}\text{Ai}\left(x+\zeta_{j}\right)^{2}dx}{\beta\text{Ai}'\left(\zeta_{i}\right)^{2}\text{Ai}'\left(\zeta_{j}\right)^{2}}\,.
\ee
Eq.~\eqref{CovaraicneSol} exactly coincides with Eq.~(F6) in Appendix F of 
\cite{Touzo}, %It is Eq. (I6) in the arXiv version
which was obtained by a direct application of standard quantum mechanics first-order perturbation theory on the SAO operator. We have extended that calculation to obtain the general third and fourth cumulant 
$\langle a_i a_j a_k \rangle_c$ and $\langle a_i a_j a_k a_\ell \rangle_c$ in Appendix \ref{app:thirdcum}  and in Appendix \ref{app:Greenfunction}.

\subsection{Gap distributions}

From the joint distribution $f_\beta(a_i, a_j)$ one can extract the distribution of their difference 
$a_{ij} = a_i - a_j$, which we denote by $f_\beta(a_{ij})$. At $\beta \gg 1$, our OFM formalism yields again the scaling 
\be
\label{PaScalingDeltaij}
f_\beta\left(a_{ij}\right)\sim e^{-\beta s\left(E_{ij} = -a_{ij}\right)}, \qquad \beta \gg 1,
\ee
where $E_{ij} = E_i - E_j$ is the gap between energy levels of the potential $V(x)$.
The LDF $s\left(E_{ij}\right)$ is obtained by minimizing the action \eqref{sVdef} constrained on the value of $E_{ij}$. Introducing a Lagrange multiplier $\lambda$ to enforce this constraint (see Appendix \ref{app:gap} for a derivation), one obtains the saddle-point equations \eqref{psiieq}-\eqref{V1psiipsij} with 
%{\red N: I flipped the signs here}
$\lambda_i = -\lambda_j = \lambda$,
i.e., Eq.~\eqref{V1psiipsij} is replaced by
% {\red P: I do not understand the set up. It does not seem possible to
% write some closed Schrodinger like equation directly for the marginal of $E_{ij}$ is that correct?
% You still have two eigenvalue equations one for $E_i$ and one for $E_j$? Can they be decoupled
% in sum and difference variables? How do you integrate over $E_i+E_j$ at fixed $E_{ij}$?
% Should I write $s(E_{ij})= \min_{E_i,E_j,E_i-E_j=E_{ij}} s(E_i,E_j)$ i.e. I can only
% get it from the two level PDF - or could there be a shortcut? At the very least
% you should explain this a bit better}
\be
\label{V1psiipsijlambda}
V_{1}(x)=4\lambda\left[\psi_{i}(x)^{2}-\psi_{j}(x)^{2}\right]\,.
\ee
% {\red what is the role of this potential? You solve with it and
% then compute what ? which integral? or you use $s'=\lambda$ ? please explain better}.
The values of $E_i$ and $E_j$ in Eqs.~\eqref{psiieq} and \eqref{psijeq} are a priori unknown, but determined by $E_{ij}$ and $\lambda$.
Besides solving the saddle-point equations numerically using the algorithm described above, one can obtain $s(E_{ij})$ analytically in limiting cases, as we now elaborate. For simplicity, we will consider the case $i=2, j=1$, i.e., we study the distribution of the gap $E_{21}$ between the two lowest energy levels.

% \subsubsection{$E_{21} \to +\infty$}

\begin{enumerate}
    \item 
$E_{21} \to +\infty$:
As we saw from the asymmetry of the distribution of a single energy level, fluctuations with unusually high energy levels are far less likely than those with unusually low energy levels. It is thus natural to expect that the most likely way for the system to create a fluctuation with $E_{21} \gg 1$ is by $E_1$ being unusually low, while $E_2$ remains relatively close to its typical value.
In the leading order, we therefore expect the behavior to coincide with the $E_1 \to -\infty$ tail \eqref{sSolRightTail}  of the single-energy-level result above, i.e., we expect that
\be
\label{sDeltaERightTail}
s(E_{21})\simeq \frac{2 E_{21}^{3/2}}{3}\,.
\ee

% \subsubsection{$E_{21} \to 0^+$}

\item
$E_{21} \to 0^+$:
This limit is actually easier to analyze using the definition of the Tracy-Widom distribution as describing the edge behavior of a G$\beta$E random matrix. 
The eigenvalues of the matrix behave statistically as a classical gas with logarithmic interactions, in thermal equilibrium at inverse temperature $\beta$, see Eq.~\eqref{GbetaEjPDF}. In the limit $E_{21} \ll 1$, the dominant term is the logarithmic interaction between $E_2$ and $E_1$, leading to
\be
\label{sDeltaELeftTail}
s(E_{21}) \simeq -\ln (E_{21}) \, .
\ee
It would be interesting to reproduce Eq.~\eqref{sDeltaELeftTail} from the solution to our saddle-point equations in the limit $\lambda \to -\infty$, and/or to obtain subleading corrections to Eq.~\eqref{sDeltaELeftTail}.

% {\red P: Naftali you can comment that unless you have some new idea}

% {\red N: The numerics seem to indicate that the subleading corrections to Eq.~\eqref{sDeltaELeftTail} are surprisingly small, they appear to be $o(1)$. Pierre do you have any intuition for why this should be the case? P: No and I do not even know how to derive the Wigner-Dyson probability $E_{21}^\beta$
% from the method of this paper, do you ? How can that be the probability
% that 
% some atypical random potential (how does it look like?) 
% has two almost degenerate energy levels. The usual folklore is
% that in a SR correlated random potential the level spacing distribution is
% Poissonian. Basically because the overlap of two localized
% states is zero. So how do you create this $E_{21}^\beta$? 
% You need to create some overlap. How ? Maybe the linear potential plays an
% important role?
% It looks like a very interesting question, if you have any idea it would be worth mentioning it.
% It very very vaguely reminds me some
% papers on 2 level systems (a double well 2 by 2 matrix with random matrix elements),
% see e.g. arXiv:2101.02787 or the Chalker Gurarie type of papers, this one https://arxiv.org/pdf/cond-mat/0204630
% or the following one (hard to read)
% Bosonic excitations in random media. But OK I cannot make any precise connection.
% Maybe Alain will have some idea. Else I can ask Virag when the paper is out,
% maybe he will find some argument. 
% }

\begin{figure}
\includegraphics[width=0.47\textwidth,clip=]{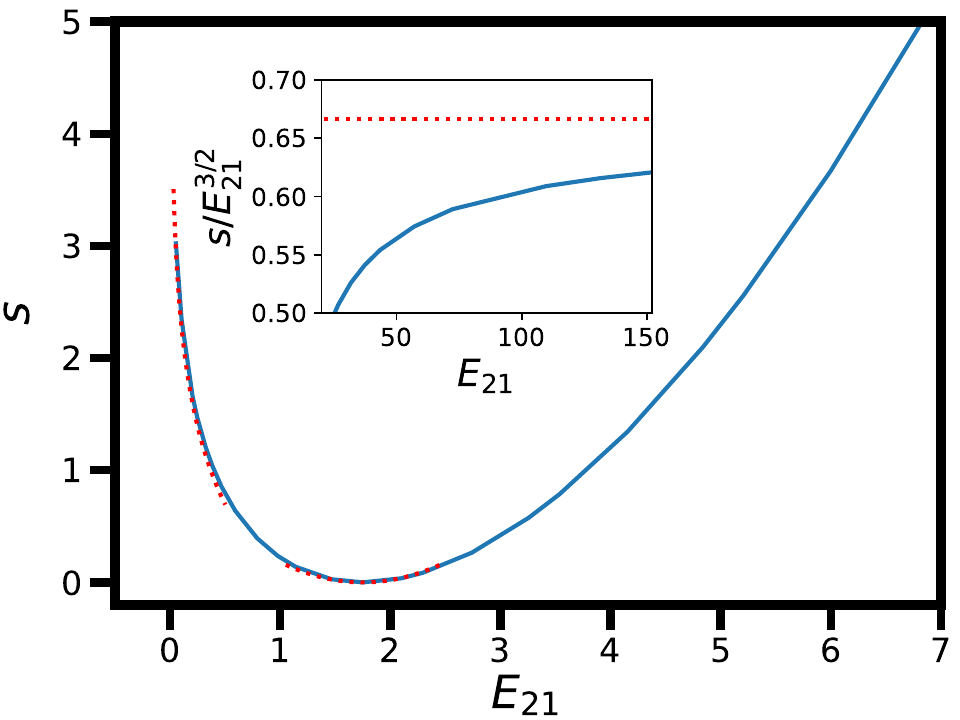}
\caption{Solid line: The large-deviation function $s(E_{21})$ that describes the distribution of the gap $E_{21} =  -a_{21}$ 
between the two largest eigenvalues in the $\text{Airy}_\beta$ point process at $\beta \gg 1$, see Eq.~\eqref{PaScalingDeltaij}. The inset shows the regime $E_{21} \gg 1$.
Dotted lines are the asymptotic behaviors \eqref{sDeltaERightTail}, \eqref{sDeltaELeftTail} and \eqref{sDeltaETypical}, describing the right tail, left tail and typical fluctuations of the distribution, respectively.}
\label{figSofaGap}
\end{figure}

% One way to create a very small energy gap is for the realization of the potential to be similar to the one Eq.~\eqref{VxPositiveTail} from the $E_i \to +\infty$ tail of the single-level problem. In this case, the action is $s \simeq E_1 ^3 / 24$. The energy gap $E_{21}$ is very small and can be approximated by the gap between the lowest two energy levels of a particle-in-a-box potential

% \subsubsection{Typical fluctuations of $E_{21}$}
\item 
Typical fluctuations of $E_{21}$:
We have already shown that, at $\beta \gg 1$, pairs of eigenvalues follow a multivariate Gaussian distribution. It follows that the gap distribution is Gaussian, and all that remains is to calculate its mean and variance.
The mean is given (in the leading order at $\beta \gg 1$) by the gap at zero-noise,
$\left\langle E_{21}\right\rangle  \simeq E_{21}^{\left(0\right)} = E_{2}^{\left(0\right)} - E_{1}^{\left(0\right)}$, while the variance is given by
\bea
 \text{Var}\left(a_2-a_1\right) &=&
\text{Var}\left(E_{21}\right)=\text{Var}\left(E_{1}\right)+\text{Var}\left(E_{2}\right)-2\text{Cov}\left(E_{1},E_{2}\right) \nn\\
&\simeq&\frac{C_{11}+C_{22}-2C_{12}}{\beta}= {\frac{4\int_{0}^{+\infty}\left[\text{Ai}\left(x+\zeta_{1}\right)^{2}-\text{Ai}\left(x+\zeta_{2}\right)^{2}\right]^{2}dx}{\beta\text{Ai}'\left(\zeta_{2}\right)^{2}\text{Ai}'\left(\zeta_{2}\right)^{2}} }=\frac{1.57193\dots}{\beta}\,,
\eea
where, to remind the reader, the coefficients $C_{ij}$ are given in Eq.~\eqref{CovaraicneSol}.
The corresponding asymptotic behavior of $s(E_{21})$ is
\be
\label{sDeltaETypical}
s(E_{21}) \simeq \frac{\left(E_{21}-E_{21}^{\left(0\right)}\right)^{2}}{2\left(C_{11}+C_{22}-2C_{12}\right)} \, .
\ee
In Fig.~\ref{figSofaGap} we plot the exact large-deviation function, computed through numerical solutions to the saddle-point equations, together with its asymptotic behaviors \eqref{sDeltaERightTail}, \eqref{sDeltaELeftTail} and \eqref{sDeltaETypical}, with good agreement.
In Eq.~\eqref{ThirdCumulantGap} of Appendix \ref{app:thirdcum},
 and in Appendix \ref{app:Greenfunction},
we obtain the third cumulant of the gap distribution, from which one can obtain the subleading correction to Eq.~\eqref{sDeltaETypical}, which is cubic in $E_{21}-E_{21}^{\left(0\right)}$.
% \be 
% \overline{ (a_i - a_j )^2 }^c = \frac{4}{\beta^2} 
% \int_0^{+\infty} dx \, \left( \frac{{\rm Ai}(x+ \zeta_i)^2}{{\rm Ai}'(\zeta_i)^2}
% - \frac{{\rm Ai}(x+ \zeta_j)^2}{{\rm Ai}'(\zeta_j)^2} \right)^2 
% \ee 

\end{enumerate}

\bigskip

% \section{Physical picture}
\section{Alternative derivation of the large-$\beta$ behavior of the Tracy-Widom distribution using the diffusion representation}
\label{sec:WNT}
%Alain

\subsection{Ricatti diffusion and optimal path}

In this Section we give an alternative derivation of our main result using the diffusion representation.
The stochastic Airy operator \eqref{HSAOdef} can be analyzed in terms of the Riccati variable $z=\frac{\psi'}{\psi}$, where $\psi = \psi_i(x)$ is the wave function. Plugging this in Eq.~\eqref{SchrodingerHSAO}, one finds that this variable satisfies
\begin {equation}
\label{Langevinz}
z'(x)= -z^2+x-E+\frac{2}{\sqrt \beta}\eta(x) 
\end {equation}
(with $x \ge 0$). It is helpful to think of the spatial variable $x$ as time, so we will change the notation $x \to t$ in most of the following derivation. With  this interpretation the Riccati equation describes the motion of a fictitious overdamped Brownian particle, moving in the time dependent potential 
\be
\label{Uofzt}
U(z,t)= \frac{z^3}{3}-(t-E)z 
\ee
which for $t-E>0$ has a stable point $z_0=\sqrt{t-E}$ and an unstable one 
$z_1=-\sqrt{t-E}$ through which the particle can escape, see Fig.~\ref{figUofzt}. From the Dirichlet boundary condition of the 
stochastic Airy operator the initial position of 
the particle is at $z(0)=+\infty$. Then the particle immediately rolls downhill and $z(t)$ becomes finite for any small finite time.

The nodes of the wave function correspond to times at which the Riccati variable blows up. The event ``$z(t)$ does not explode'' thus corresponds to a solution $\psi(x)$ which has no nodes except the one at $x=0$, as is the case for the ground-state wavefunction of the SAO. In accordance with Sturm oscillation theorem, one can relate the ground state probability distribution to the probability of non-explosion.
\begin {equation}
P\left(E\le E_{1}\right)=P\left[z(t)\text{ never explodes }\forall t>0\right].
\end {equation}
The eigenvalue problem \eqref{SchrodingerHSAO} is thus cast into a problem of escape over a time-dependent barrier $U(z,t)$. 
% Such problems appear in particular in the context of transport processes and have been extensively studied in the physics literature \cite{Getfert2010, KM25, Meerson25}.
Qualitatively similar problems, involving escape over (possibly time-dependendent) barriers and/or finite-time blowup of Brownian particles under the effect of external forces have been extensively studied in the physics literature, see e.g. \cite{Getfert2010, KM25, Meerson25}.
In this context a central object  is the notion of optimal path. To characterize it we first  study the time evolution of the barrier in the two regimes $E>0$ and $E<0$:
\begin{enumerate}
    \item For $E>0$ , at time $ t=0$, $U(z,0)$ is a smooth monotonous function. One therefore expects that a particle which is released from $+\infty$ at time $t=0$ will rapidly fall in the potential well but, as time goes on, a barrier is formed which prevents the particle to escape from the unstable point $z_1=-\sqrt{t-E}$.
\item  For $E<0$, $U(z,t)$ exhibits a potential barrier $\forall t>0$, therefore the escape mechanism is similar to a standard activation process with an Arrhenius behaviour.
This observation qualitatively accounts for the sharp asymmetry of the tails of the large deviation function discussed in section \ref{sec:SaddlePointTW}.
\end{enumerate}
% 1) For $E>0$ , at time $ t=0$, $U(z,0)$ is a smooth monotonous function. One therefore expects that a particle which is released from $+\infty$ at time $t=0$ will rapidly fall in the potential well but, as time goes on, a barrier is formed which prevents the particle to escape from the unstable point $z_1=-\sqrt{t-E}$.\\
% 2) For $E<0$, $U(z,t)$ exhibits a potential barrier $\forall t>0$, therefore the escape mechanism is similar to a standard activation process with an Arrhenius behaviour.
% This observation qualitatively accounts for the sharp asymmetry of the tails of the large deviation function discussed in section \ref{sec:SaddlePointTW}.

\begin{figure}
\includegraphics[width=0.4\textwidth,clip=]{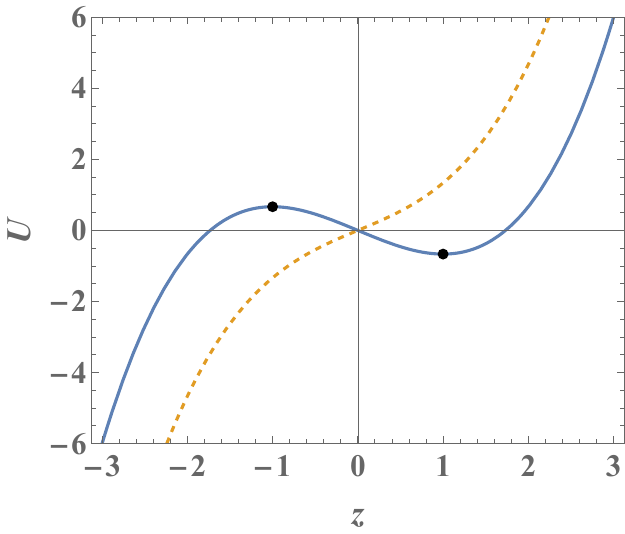}
\caption{The time-dependent potential $U(z,t)$ as a function of $t$ for $t-E > 0$ (solid line) and $t-E < 0$ (dashed line), see Eq.~\eqref{Uofzt}. The fat dots correspond to the points $z=z_0$ and $z=z_1$ described in the text.}
\label{figUofzt}
\end{figure}

The transition probability $P(z_1, t_1\vert z_0,t_0)$ of the process is given by the path integral
\begin {equation}
\label{pathIntegral}
P(z_1, t_1\vert z_0, 0)=\int_{q(0)=z_0}^{q(t_1)=z_1}  Dq (t) e^{-\beta  S[q]} \, ,
\end {equation}
where the weight of the path is the Onsager-Machlup functional corresponding to Eq.~\eqref{Langevinz},
% {\red P: is the last term present for Ito?}
\begin {equation}
\label{BetaSdef}
\beta S[q] =\int_{0}^{t_{1}}\left[\frac{\beta}{8}(\dot{q}+E+q^{2}-t)^{2}-q\right]dt \, .
\end {equation}
For $\beta$ large, we evaluate the path integral \eqref{pathIntegral} via the saddle-point approximation, and find that it is dominated by the optimal path $q(t)$: The path  which minimizes the first term in the action \eqref{BetaSdef}.
It satisfies the Euler-Lagrange equation
\begin {equation}
\label{EulerLagrangeq}
\ddot q= 2q^3-2(t-E)q+1 \, .
\end {equation}
The optimal path $q(t)$ is defined for $0 < t < \infty$ with the following boundary conditions:
\begin{itemize}
\item At time $ t=0 $ the Dirichlet boundary condition $\psi(0)=0$ requires that $q(0)=+\infty$.
\item The non-explosion condition implies that it takes an infinite time  to reach the unstable point, i.e. 
\begin {equation}
\lim_{t\to\infty} q(t) \simeq -\sqrt{t-E} \, .
\end {equation}
\end{itemize}

After finding the optimal path $q(t)$, the probability distribution of the ground state is such that
\begin {equation}
P(E<E_{1})\sim\exp\left[-\frac{\beta}{8}\int_{0}^{+\infty}\left(\dot{q}+E+q^{2}-t\right)^{2}dt\right] \, ,
\end {equation}
with exponential accuracy.
It is convenient to pass to the Hamiltonian formalism and rewrite the equation of motion \eqref{EulerLagrangeq} as a first order system  
\begin{eqnarray}
\dot p & = & -\frac{\partial H}{\partial q} \, , \\[1mm]
\dot q & = & \frac{\partial H}{\partial p} \, ,
\end{eqnarray}
 where 
\begin {equation}
H= 2p^2+p(t-E-q^2) \, .
\end {equation}
This gives
\begin{eqnarray}
\dot p & = & 2pq \, , \\[1mm]
\dot q & = & 4p+t-E-q^2 \, .
\end{eqnarray}
 We now set $p=\sigma \frac{v^2}{2}$, where $\sigma=\pm 1$ is for
the moment an arbitrary sign which will be fixed below. One obtains 
$q=\dot v/v$ as well as
\begin {equation}
\label{NonLinearvEq}
\ddot v= 2 \sigma v^3+(t-E)v \, \quad , \quad  \sigma=\pm 1 .
\end {equation}
One thus recovers the Painlevé 2 equation%
\footnote{More generally, the Painlevé 2 equation may contain an additional additive constant term which leads to additional types of solutions, see e.g. \cite{Troy18}, but we do not encounter this term here.}.
Expressing the action \eqref{BetaSdef} in terms of $v$ gives
\begin {equation}
S= \frac{1}{8}\int_0^{+\infty} (\dot q+E-q^2-t)^2 dt =\frac{1}{2}\int_0^{+\infty} v^4(t)dt \, .
\end {equation}
 Comparison with Eq.~\eqref{NonLinearPhiEq} shows that the equation \eqref{NonLinearvEq} satisfied by $v$ coincides with the nonlinear Schrödinger equation derived in section \ref{sec:SaddlePointGeneral}
provided we identify $ v=\sigma \frac{1}{\sqrt 2}\phi$ and $\sigma = \sigma_E=  {\rm sgn}(E + \zeta_1)$.
Furthermore the optimal action $S(E)$ identifies with the rate function $s(E)$ computed there  [see Eq.~\eqref{BetaSdef}].
\\

Before going any further we need to make some comments on this approach. Although it is in several respects similar to that of Virag et al \cite{BloemendalVirag,ViragReview,ESW14},
it differs on the following points:
\begin{itemize}
    \item  The use of the forward process instead of the backward one.
    \item  The semiclassical approach adapted to the study of  the weak noise limit $\beta\rightarrow\infty$.
\end{itemize}
% ${\cdot}$ The use of the forward process instead of the backward one.\\
% ${\cdot}$ The semiclassical approach adapted to the study of  the weak noise limit $\beta\rightarrow\infty$.\\
With these observations we may now proceed to analyze the solution $ v=\frac{1}{\sqrt 2}\phi$ . The trajectorial approach will  provide new insight in the boundary conditions and the nature of the solution, in particular near the boundary layer at $t=0$.
We begin by analyzing the boundary conditions satisfied by $v(t)$.

1) For small $t$, we solve Eq.~\eqref{EulerLagrangeq} through a Laurent expansion, and the result is
\begin {equation}
\label{qLaurent}
q(t)=\frac{1}{t}-\frac{E t}{3}+ \frac{t^2}{4}+ct^3  + \frac{E}{36} t^4  + O(t^5)
\end {equation}
where $c$ is an arbitrary constant. 
Therefore 
\begin {equation}
v^2(t)=2  \sigma  p(t)=\frac{ \sigma}{2}(\dot q+q^2-t+E)= \frac{ \sigma}{2} \left(5c+\frac{E^2}{9}\right)t^2+  O(t^4)
\end {equation}
This implies that $v(t)$ grows linearly for small $t$, i.e. $v(t)=a(E)t$.
In Appendix \ref{app:vt0} it is shown that $a^2(E)$ can be expressed in terms of the derivative of the classical action
\begin {equation}
a^2(E)=2  \sigma S'(E) =  2 |S'(E)|
\end {equation}
 Note that since $a^2(E)$ must be positive, it determines $\sigma={\rm sgn} (S'(E))$,
in agreement with section \ref{sec:SaddlePointGeneral}.
Note that for $E=-\zeta_1$ one must have $c=-\zeta_1^2/45$.

2) For $t\rightarrow + \infty$ the linearization of Eq.~\eqref{NonLinearvEq} gives the Airy equation, whose solution is $v(t)\simeq k\text{Ai}(t-E)$ for some $k$  which depends on $E$. 
 The tail of the Airy function $v(t) = \frac{\phi_1(t)}{\sqrt{2}} \propto e^{- \frac{2}{3} (t-E)^{3/2}}$ can be seen in Fig.~\ref{figphi1ofxTF} for $t> E_1$ ($t$ is called $x$ there).
This solution implies that $q(t)$ satisfies the required boundary condition at $t \to + \infty$, since
$q(t)=\frac{\dot v(t)}{v(t)}\simeq -(t-E)^{1/2}$.
This result confirms the physical picture discussed previously:
the particle approaches the unstable point without leaving the well.

\subsection{Relation with the Ablowitz-Segur solution }
% \label{sec:Ablowitz} 

In the case $\sigma=+1$ (i.e. for $E > - \zeta_1$)
our problem \eqref{NonLinearvEq} reduces to the study of the differential equation 
\begin {equation}
\ddot v= 2v^3+(t-E)v
\end {equation}
on the positive axis $t>0$ with boundary conditions
\begin{eqnarray}
v(0) & = &0\, , \\[1mm]
\displaystyle{\lim_{t\to\infty}}v(t) & = & 0\, .
\end{eqnarray}
The time translation $\tau=t-E $, $w(\tau)=v(t)$ leads to the Painlevé 2 equation 
\begin {equation}
\label{wPainleve}
\ddot w= 2w^3+\tau w
\end {equation}
on the half line $[-E, + \infty[$ with boundary conditions%
\footnote{The Painlev\'{e} II equation with this type of boundary condition was also studied in \cite{Bender}.}
% \cite{foonoteBender}
\begin{eqnarray}
w(-E) & = &0 \, ,
 \\[1mm]
\displaystyle{\lim_{\tau\to+\infty}}w(\tau) & = & 0 \, .
\end{eqnarray}
Hastings and McLeod \cite{HM80} proved that, at large time, any solution of \eqref{wPainleve}, extended to the full line, is asymptotic to $k \text{Ai}(\tau)$ for some $k$ (see NIST).
Among this one parameter family we have to find $k$ such that $w(-E)=0$. If we assume that $w(\tau)$ keeps a constant sign on $[-E,\infty[$, this means that $-E$ is the rightmost zero of the extended solution.
This selects the so-called Ablowitz Segur solutions $w_{k} (\tau)$ with $\vert k\vert <1 $ 
\cite{AblowitzSegur} characterized by
\bea
w_{k}(\tau) &\sim& k \text{Ai}(\tau) \text{ as } \tau\rightarrow\infty \,, \\[1mm]
w_{k}(\tau) &=& \frac{d}{(-\tau)^{1/4}} \sin (\varphi(\tau)-\theta_0)\text{ as } \tau\rightarrow -\infty
\eea
% \begin {equation}
% w_{k}(\tau)\sim k \text{Ai}(\tau) \text{ as } \tau\rightarrow\infty
% \end {equation}
% \begin {equation}
% w_{k}(\tau)= \frac{d}{(-\tau)^{1/4}} \sin (\phi(\tau)-\theta_0)\text{ as } \tau\rightarrow -\infty
% \end {equation}
where $\varphi(\tau)=\frac{2}{3} \vert \tau\vert ^{3/2}-\frac{3}{4}d^2\ln \vert \tau\vert $.
The constants $d$, $\theta_0$ are given in terms of $k$ by connection formulas (see \cite{Clarkson16}, NIST). 

%{\red N: Everything from here onwards is only approximately correct, in the limit where $E \gg 1$, no?}
Numerical solutions of Rosales \cite{Rosales78} suggest that the rightmost zero $\tau=-E$ corresponds to values of $k$ very close to the critical value $k=1$. This regime was studied in detail in \cite{Bothner}.
As shown in Rosales, the solution remains close to the parabola $ 2 w^2+\tau=0$ (see Fig. 
\ref{figBoundaryLayer}). Phrased differently, this means that the physical solution $v(t)= w(t-E)$ is very close to the Thomas-Fermi solution $v_{\text{TF}}(t) $ discussed in section \ref{subsec:ThomasFermi}
% \begin{align*} 
%  v_{\text{TF}}(t) &= \sqrt {\frac{E-t}{2}}& 0<t<E \\ 
%  v_{\text{TF}}(t) &= 0 & E<t
% \end{align*}
\be \label{TF} 
v_{\text{TF}}(t)=\begin{cases}
\sqrt{\frac{E-t}{2}}\,, & 0<t<E,\\
0\,, & E<t\,.
\end{cases}
\ee
Fig.~\ref{figBoundaryLayer}(a) is a sketch of the solution $w(\tau)$. Note that it remains close to the parabola $ 2 w^2+\tau=0$. The physical solution defined on the interval $[-E,\infty[$ is shown in the solid line.

\begin{figure}
% \includegraphics[width=0.47\textwidth,clip=]{Fig 1 Alain.pdf}
% % \includegraphics[width=0.47\textwidth,clip=]{sOfELinearPotentialHardWall.pdf}
%  \hspace{2mm}
% % \includegraphics[width=0.49\textwidth,clip=]{sOfEvsTracyWidoms.pdf}
% \includegraphics[width=0.47\textwidth,clip=]{Fig 2 Alain.pdf}
\includegraphics[width=0.32\textwidth,clip=]{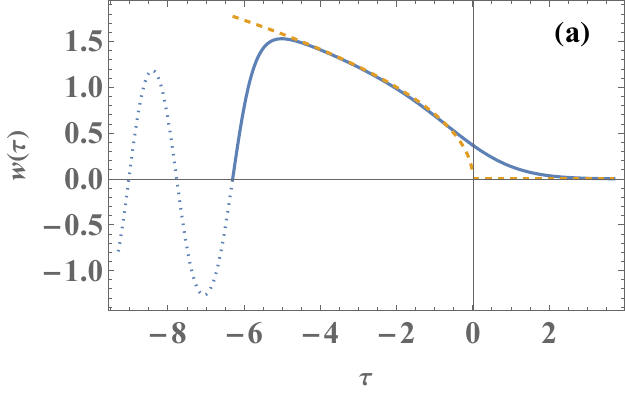}
\hspace{1mm}
\includegraphics[width=0.32\textwidth,clip=]{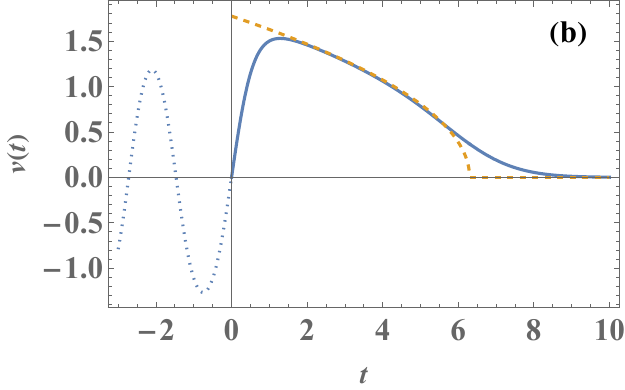}
\hspace{1mm}
\includegraphics[width=0.32\textwidth,clip=]{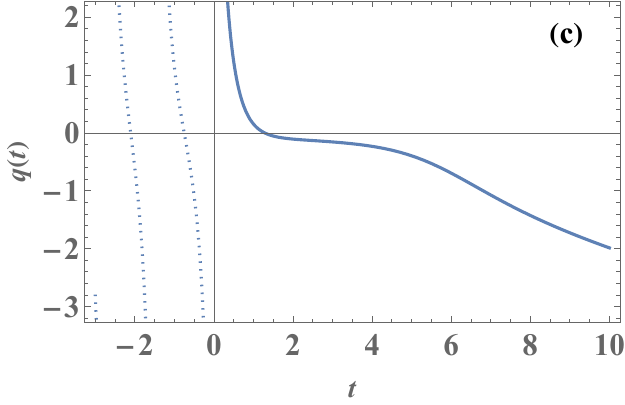}
\caption{ 
(a) The Ablowitz-Segur solution $w(\tau)$ for a value of $k$ slightly less than $1$, i.e.  $k\simeq1-5.88\times 10^{-6}$, corresponding to $E=6.3$, see also Ref.~\cite{Rosales78}.
The solid line is the physical solution which we use here. The value of $-E$ can be read from the first negative zero.
 The dotted line is the (non-physical) continuation of the solution to $\tau < -E$. The dashed line corresponds to the Thomas-Fermi approximation \eqref{TF}.
(b) A translation of the Ablowitz-Segur solution 
in terms of the function $v(t)$ 
(c) a sketch of the corresponding trajectory $q(t)$.}
\label{figBoundaryLayer}
\end{figure}

 The physical interpretation of the extended solution $v(t)=w(t-E)$ 
  [plotted in Fig.~\ref{figBoundaryLayer}(b)] is best understood from a plot of $q(t)=\frac{v'(t)}{v(t)}$ shown in Fig.~\ref{figBoundaryLayer}(c). The interpretation is the following one.
  For $t<0$ and $E>0$, the potential $U(z,t)$ in \eqref{Uofzt} in which the particle moves is an increasing function of $z$, see Fig.~\ref{figUofzt}.
  The system has no equilibrium point and therefore the particle rolls down to $-\infty$ and reappears immediately at $+\infty$. The first attempt for which the particle does not escape to infinity is realized  at $t=0$. This corresponds to the physical solution for $t>0$.

In this context we now return to the discussion of the boundary layer of \ref{subsec:ThomasFermi}.
We first compute the action using the Thomas-Fermi approximate solution $v_{\text{TF}}(t)$ given in \eqref{TF}.
It reads 
\begin {equation}
\label{SofETF}
S(E)=\frac{1}{2}\int_0^{+\infty} v^4(t) dt \simeq \frac{1}{2}\int_0^{E} v_{\text{TF}}^4(t) dt =\frac{E^3}{24} \, .
\end {equation}
The slope at the origin is
\begin {equation}
v'(0)=a(E)=\sqrt{2s'(E)}=\frac{E}{2} \, .
\end {equation}

As mentioned in Section \ref{subsec:ThomasFermi} the Thomas Fermi solution $ v_{\text{TF}}(t)$  does not satisfy the boundary condition $v(0)=0$ unlike the true solution $v(t)=w(t-E)$. 
The trajectorial approach allows to estimate the width of the boundary layer.
Let us give a heuristic argument for the form of the subleading term.
The  (approximate) boundary layer solution $v(t) \simeq \frac{Et}{2}$ is thus linear in $t$
at small $t$. The true boundary layer solution derived in Section \ref{subsec:ThomasFermi} 
is
$v(t)=\sqrt{\frac{E}{2}}\tanh\left(t\sqrt{\frac{E}{2}}\right)$.
The time scale 
for which it matches with the Thomas Fermi solution is $t_0 \simeq \frac{3 \sqrt{2}}{4} 
\frac{\ln E}{\sqrt{E}}$. This is of the same order as the time scale $t_1$ over which the particle crosses the origin for the first time $q(t_1)=0$. In that regime $q(t)$ crosses over from 
$q(t)=\frac{\sqrt{2E}}{\sinh\left(t\sqrt{2E}\right)}$ for $t \ll t_0$ 
to $q(t) \simeq \frac{-1}{\sqrt{E-t}}$ for $t > t_1$. Finally, for $t > E$,
there is a large time regime  where $q(t) \simeq - \sqrt{t-E}$.

% It matches with the Thomas-Fermi solution on a scale $t_0$ such that
% $\frac{Et_0}{2}=\sqrt{\frac{E-t_0}{2}} $ which for large $E$ gives $t_0\sim \sqrt{ \frac{2}{ E}}$. This is precisely the time scale {\blue $t_1$} over which the particle crosses the origin for the first time. Indeed by using  the Laurent expansion $q(t)$, Eq.~\eqref{qLaurent}, one finds 
% $t_1=\sqrt {\frac{3}{E}}$.  Using this result one can compute the correction to the classical action by taking into account the boundary layer. {\blue To obtain an order-of-magnitude estimate for the subleading correction to the Thomas-Fermi action \eqref{SofETF}, let us assume} that the solution is linear for $0<t<t_0$ and Thomas-Fermi like for $t_0<t<E$. This gives the following expansion for large $E$
% \begin {equation}
% S(E) \simeq \frac{E^3}{24}-\frac{\sqrt 2}{10} E^{3/2} \, .
% \end {equation}
% This is to be compared with the exact result  for the tail of the Tracy Widom distribution at large $E$
% and at fixed $\beta$ obtained in \cite{BENM11}) 
% \begin {equation}
% S(E) \simeq \frac{E^3}{24}-\frac{\sqrt 2}{6} E^{3/2} \, ,
% \end {equation}
% and indeed we find that our calculation gives the correct order of magnitude for the subleading term.
%  A more accurate study of the boundary layer is performed in Section \ref{subsec:ThomasFermi}
% and the result, displayed in \eqref{SSolLeftTail}, and matches exactly the formula of \cite{BENM11}. 

\section{Discussion}
\label{sec:discussion}

To summarize, we have studied the TW$\beta$ distribution, 
and additional properties of the $\text{Airy}_\beta$ point process (APP)
 $\{ a_i \}_{i \geq 1}$,
such as joint distributions of pairs of eigenvalues and spectral gap distributions, in the limit of large Dyson index $\beta \to +\infty$.
We  showed that, in general, these distributions follow large-deviation principles of the type \eqref{PaScaling}, where the large-deviation functions $s$ are obtained by evaluating certain action integrals over the solutions to saddle-point equations that are closely related to the Painlev\'{e} II equation. We obtained two equivalent representations for the latter equations, one through the stochastic Airy operator (SAO), and a second through  its associated stochastic Ricatti equation. We solve these  saddle-point equations numerically in general, and analytically in certain limiting cases
 (upper and lower tails, typical region).
These solutions describe the most likely realization of the SAO conditioned on a given value of $a_i$.

 In the case of the TW$\beta$ distribution (i.e. the distribution of $a_1$)
we have shown that the large argument behavior of our large deviation function $s$ matches the known
fixed $\beta$ asymptotics for the tails, including some subdominant terms.
We find that the $n$-th cumulant of the TW$\beta$ distribution behaves, for $n \ge 2$, as  $\simeq C_n/\beta^{n-1}$, and calculated the reduced cumulants $C_n$ exactly 
for $n=2,3,4$ (as well as those of $a_i$ for all $i$, and some multipoint correlations). 
For $n=2$ it reproduces a known result, obtained in 
the pioneering work \cite{EPS14} which explored the edge of the G$\beta$E at large $\beta$.
The present paper extends significantly the results 
of that work by obtaining the full large deviation rate functions.
The remarkable occurence of the Painlev\'{e} II equation in this problem (which
also occurs for e.g. $\beta=2$) hints at some special integrability properties 
associated to the point $\beta=+\infty$, alluded to in \cite{GorinInfiniteBeta}.

It is reasonable to expect that our results might also be reached by analyzing the large-$\beta$ behavior of the G$\beta$E  at fixed $N$, maybe using techniques as in \cite{BorotNadal2012}, and then taking the large-$N$ limit towards the edge, i.e., that the limits $N \to +\infty$ and $\beta \to +\infty$ commute.  This program was in fact carried in \cite{Touzo,GorinInfiniteBeta}
but only at the level of typical, i.e. Gaussian fluctuations. To carry it through 
at the level of large deviations, i.e. keeping track of rare atypical events,
seems challenging.
It would also be interesting to study the joint distribution of $M$ eigenvalues for general $M$ (here we studied $M=1$ and $M=2$). In particular, the sum of the $M$ lowest eigenvalues of the SAO  gives the many-body ground state energy of $M$ noninteracting, spinless fermions in the potential \eqref{V0def}  in presence of additional white-noise disorder  (see \cite{PLDAlex4routes} for some results on that problem, and \cite{FermionsRandom} in the absence of linear potential).
The case $M \gg 1$, describing a Fermi gas with many particles, is of special interest \cite{dean2019noninteracting}.

It would be interesting to extend our results to point processes that describe the large-$N$ limiting behaviors of other standard  RMT ensembles such as the Wishart-Laguerre (where the limiting process is described by the stochastic Bessel operator \cite{EdelmanSutton2007, RamirezRider2009, ESW14}), as well as
circular and Ginibre ensembles. Finally it would be interesting to study the large $\beta$ large deviations
for the dynamics, i.e. for the full Airy line ensemble.

\bigskip
\begin{center}
\textbf{Declaration of competing interest}
\end{center}

The authors declare that they have no known competing financial interests or personal relationships that could have appeared
to influence the work reported in this paper.

\smallskip

\begin{center}
\textbf{Data availability}
\end{center}

Data will be made available on reasonable request.

\section*{Acknowledgments}

NRS is grateful to Satya N. Majumdar for bringing this problem to his attention.
PLD thanks V. Gorin, J. Huang, G. Schehr and L. Touzo for discussions related to the large $\beta$ limit
of random matrix ensembles.
We are grateful to Brian Sutton, Alex Bloemendal, Per-Olof Persson and Alan Edelman for useful discussions and advice, and for sharing their code for numerical computation of Tracy-Widom distributions.
NRS acknowledges support from the Israel Science Foundation (ISF) through Grant No. 2651/23, from the Binational Science Foundation (BSF) through Grant No. 2024162, and from the Golda Meir fellowship.
 PLD acknowledges support from ANR Grant No. ANR- 23-CE30-0020-01 EDIPS.

\bigskip

\appendix

\section{A few identities relating the rate function $s(E_i)$ and wavefunction $\phi_i(x)$}
\label{app:der}

We establish a few identities which are consequences of the saddle-point equation \eqref{NonLinearPhiEq2}. We denote
$\sigma_{E_i}=\text{sgn}\left(E_{i}-E_{i}^{\left(0\right)}\right)$.

Let us first show Eq. \eqref{sEider}.
Consider an infinitesimal variation $E_i \to E_i + \delta E_i$. From \eqref{sEi}
one has 
\be 
\label{deltasEi}
\delta s(E_i)  = \frac{1}{2}\int_{0}^{+\infty} dx \, \phi_{i}^{3}\left(x\right) \delta \phi_{i}(x) \,,
\ee
where $\delta \phi_{i}(x)$ is the variation to linear order of the solution $\phi_i(x)$
of \eqref{NonLinearPhiEq2}, i.e. it satisfies 
\be 
-\delta \phi_{i}''\left(x\right) + (x - E_i) \delta \phi_{i}\left(x\right) 
+ 3 \, \sigma_{E_i}  \phi_{i}(x)^{2} 
\delta \phi_i(x) = \delta E_i \phi_i(x) 
\ee
Let us multiply by $\phi_i(x)$ and integrate over $x \in [0,+\infty[$.
The first term can be integrated by parts twice and using $\phi_i(0)=0$, $\delta \phi_i(0)=0$ and the vanishing boundary conditions at $x\to+\infty$, the boundary terms vanish so we obtain
\be 
\label{integralOfphideltaphi}
 \int_{0}^{+\infty}dx\left[-\phi_{i}''\left(x\right)+\left(x-E_{i}\right)\phi_{i}\left(x\right)+3\,\sigma_{E_{i}}\phi_{i}(x)^{3}\right]\delta\phi_{i}(x)=\delta E_{i}\int_{0}^{+\infty}dx\,\phi_{i}(x)^{2}
\ee
Using that $\phi_i(x)$ satisfies \eqref{NonLinearPhiEq2} we see
that the left-hand side of Eq.~\eqref{integralOfphideltaphi} simplifies and one obtains 
\be
2 \, \sigma_{E_i} \int_0^{+\infty} dx  
\phi_{i}(x)^{3}   
\delta \phi_i(x) = \delta E_i \int_0^{+\infty} dx \, \phi_i(x)^2
\ee 
which, together with \eqref{deltasEi}, leads to \eqref{sEider}.

Let us now show Eq. \eqref{phider0}. The equation \eqref{NonLinearPhiEq2}
can be rewritten as
\be 
\frac{d}{dx}\left[\frac{1}{2}\phi_{i}'(x)^{2}+\frac{E_{i}}{2}\phi_{i}(x)^{2}-\frac{\sigma_{E_{i}}}{4}\phi_{i}(x)^{4}\right]=x\phi_{i}(x)\phi_{i}'(x) \, .\label{id2}
\ee 
Note that the left and right and sides of Eq.~\eqref{id2} may be interpreted as the rate of change of ``mechanical energy'', and rate of work due to the ``external force'' (see Section \ref{sec:Soliton} for more details on this mechanical analogy, but in the case where the external force is absent).
Integrating Eq.~\eqref{id2} for $x \in [0,+\infty[$, and using that $\phi_i,\phi_i'$ vanish at infinity 
and $\phi_i(0)=0$ we obtain
\be 
 \phi_i'(0)^2 = -  \int_0^{+\infty} dx \, x \frac{d}{dx} \phi_i(x)^2   = \int_0^{+\infty} dx \, \phi_i(x)^2 
\ee 
with no additional boundary terms since $\phi(x)$ decays faster than $1/\sqrt{x}$ at infinity.
Thanks to \eqref{sEider} we obtain another identity $\phi_i'(0)^2=4 |s'(E_i)|$.

We now establish two more useful identities, Eqs.~\eqref{sEphiIdentity1} and \eqref{sEphiIdentity2} of the main text. First  
multiplying \eqref{NonLinearPhiEq2} by $\phi(x)$, integrating and using integration by parts, one has
\be 
\int_0^{+\infty}  dx \phi_i'(x)^2 + \int_0^{+\infty} dx x \phi_i(x)^2 - E_i \int_0^{+\infty} dx  \phi_i(x)^2 + \sigma_{E_i} \int_0^{+\infty} dx  \phi_i(x)^4 = 0
\ee 
which, using the identities in the text \eqref{sEi}, \eqref{sEider}, leads to
\be \label{id01} 
\int_{0}^{+\infty}dx\phi_{i}'(x)^{2}+\int_{0}^{+\infty}dxx\phi_{i}(x)^{2}+4\sigma_{E_{i}}\left[2s\left(E_{i}\right)-E_{i}s'\left(E_{i}\right)\right]=0 \, .
\ee 
This implies in particular that $2 s(E_i) - E_i s'(E_i) >0$ for $E_i<E_i^{\rm typ}$
and $2 s(E_i) - E_i s'(E_i) <0$ for $E_i>E_i^{\rm typ}$.

Next, consider \eqref{id2}. Integrating it implies 
\be
\frac{1}{2} \phi_i'(x)^2 + \frac{E_i}{2} \phi_i(x)^2 - \frac{\sigma_{E_i}}{4}  \phi_i(x)^4 
= - \int_{x}^{+\infty} dy \, y \phi_i(y) \phi_i'(y) \, .
\ee 
Integrating over $x \in [0,+\infty[$ one obtains 
\bea 
&& \frac{1}{2} \int_0^{+\infty}  dx \phi_i'(x)^2 + \frac{E_i}{2} \int_0^{+\infty}  dx \phi_i(x)^2 - \frac{\sigma_{E_i}}{4}  \int_0^{+\infty}  dx \phi_i(x)^4 
= - \int_0^{+\infty}  dx \int_{x}^{+\infty} dy \, y \phi_i(y) \phi_i'(y) \nn\\
&& 
\qquad\qquad\qquad\qquad\qquad\qquad = 
- \int_0^{+\infty}  dx x^2 \phi_i(x) \phi_i'(x)
= \int_0^{+\infty}  dx x \phi_i(x)^2 
\eea  
which, using the identities in the text \eqref{sEi}, \eqref{sEider},
leads to 
\be \label{id02} 
\frac{1}{2} \int_0^{+\infty}  dx \phi_i'(x)^2 + 
2 \sigma_{E_i} \left[ E_i s'\left(E_i\right)  - s\left(E_i\right) \right]  = \int_0^{+\infty}  dx x \phi_i(x)^2  \, .
\ee   
We can now combine the two identities \eqref{id01} and \eqref{id02}. We obtain
the identities \eqref{sEphiIdentity1} and \eqref{sEphiIdentity2} of the main text. %\eqref{2id}.

\section{Third and fourth cumulants by direct perturbation theory
} \label{app:thirdcum}

\subsection{Third cumulant}

In this Appendix we obtain the third and fourth cumulants by direct perturbation theory of the
eigenvalues of the SAO. The second cumulant was obtained by this method in
Appendix F of \cite{Touzo}. Let us use compact notations and denote $\psi_i^{(0)}(x)=A_i(x):= \frac{{\rm Ai}(x+\zeta_i)}{{\rm Ai}'(\zeta_i)}$.  We use here standard quantum mechanics perturbation
theory, with explicit sums overs energy levels.
In \cite{EPS14} a slightly different approach was used, by first obtaining an
explicit expression for the Green's function. In Appendix \ref{app:Greenfunction}
we derive alternative expressions using that second method.

The second order perturbation theory for the $a_i$'s, Eq.~(F4) of \cite{Touzo} reads
\be  \label{aiperturbation} 
a_{i}=\zeta_{i}-\frac{2}{\sqrt{\beta}}\left\langle A_{i}^{2},w\right\rangle +\frac{4}{\beta}\sum_{i'\neq i}\frac{\left\langle A_{i}A_{i'},w\right\rangle ^{2}}{\zeta_{i}-\zeta_{i'}}+O\left(w^{3}\right)
\ee 
where $w(x)$ is the unit white noise (denoted $V_1(x)$ in the text) 
and $\langle f,g \rangle = \int_0^{+\infty} dx f(x) g(x)$.
%We denote $\langle f, 1 \rangle = \langle f \rangle = \int_0^{+\infty} dx f(x)$.
Here we denote noise averages by overline. We will use Wick's theorem to compute moments, which reads 
\bea 
\overline{\langle A , w \rangle \langle B , w \rangle } &=& \langle A, B \rangle \, , \label{wick2} \\[1mm]
\label{Wick2}
\overline{\langle A , w \rangle \langle B , w \rangle \langle C , w \rangle \langle D , w \rangle  }
&=& \langle A, B \rangle \langle C, D  \rangle + 
\langle A, C \rangle \langle B, D  \rangle + \langle A, D \rangle \langle B, C  \rangle \, , \label{wick4}
\eea  
and similarly for the sixth moment with 15 terms (15 pairings). 
From it one recovers the second cumulant, in the present notations,
as well as the second cumulant of the gap
\be 
\overline{a_{i}a_{j}}^{c}=\frac{4}{\beta^{2}}\left\langle A_{i}^{2},A_{j}^{2}\right\rangle \quad,\quad\overline{(a_{i}-a_{j})^{2}}^{c}=\frac{4}{\beta^{2}}\left\langle \left(A_{i}^{2}-A_{j}^{2}\right)^{2},1\right\rangle 
\ee 
which agrees with \eqref{CovaraicneSol}. Note that averaging
Eq. \eqref{aiperturbation} gives the $O(1/\beta)$ correction to
the mean value, given in Eq.~(F9) of \cite{Touzo}, and 
which we recall here
\be 
\label{meanCorrection}
\overline{a_{i}}=\zeta_{i}+\frac{4}{\beta}\sum_{i'\neq i}\frac{\left\langle A_{i}^{2},A_{i'}^{2}\right\rangle }{\zeta_{i}-\zeta_{i'}}\,.
\ee
For $i=1$, a numerical computation of the sum (which we obtained by truncating it after the first $5\times10^5$ terms) gives
\be
\label{meanCorrectioni1}
\overline{a_1} = \zeta_1 + \frac{1.12\dots}{\beta}
\ee
 Eq.~\eqref{meanCorrectioni1} appears to be in agreement with the result of Ref.~\cite{EPS14}, (where the coefficient obtained was $1.124813904\dots$). In Appendix \ref{app:Greenfunction} we show that our analytic expression \eqref{meanCorrection} coincides with the one given in \cite{EPS14}. Eq.~\eqref{meanCorrectioni1} is compared to known results for $\beta=1,2,4$ in Table \ref{table:MeanVarianceSkewness}, with good agreement (despite the fact that the values of $\beta$ are not large).

Next, it is easy to see,  using \eqref{Wick2} with $A=D$, that 
\be
\overline{\langle A,w\rangle^{2}\langle B,w\rangle\langle C,w\rangle}^{c}=\overline{\left(\langle A,w\rangle^{2}-\overline{\langle A,w\rangle^{2}}\right)\langle B,w\rangle\langle C,w\rangle}=2\langle A,B\rangle\langle A,C\rangle\,.
\label{diag1} 
\ee
Hence we obtain the third cumulant as
\bea 
\label{3point3rdCumulant}
\overline{ a_i a_j a_k }^c &=& C_{i;j,k} + C_{j;i,k} + C_{k;i,j} \, , \\[2mm]
C_{i;j,k} &=&\frac{16}{\beta^{2}}\overline{\sum_{i'\neq i}\frac{\left\langle A_{i}A_{i'},w\right\rangle ^{2}}{\zeta_{i}-\zeta_{i'}}\left\langle A_{j}^{2},w\right\rangle \left\langle A_{k}^{2},w\right\rangle }=\frac{32}{\beta^{2}}\sum_{i'\neq i}\frac{\left\langle A_{i}A_{i'},A_{j}^{2}\right\rangle \left\langle A_{i}A_{i'},A_{k}^{2}\right\rangle }{\zeta_{i}-\zeta_{i'}}\,.
\eea 
The equal point third cumulant is thus
\be
\label{1point3rdCumulant}
\overline{ a_i^3 }^c = 
\frac{96}{\beta^2}  \sum_{i' \neq i} 
\frac{\left\langle A_i^3 , A_{i'} \right\rangle^2 }{\zeta_i-\zeta_{i'}} 
\ee 
in agreement with \eqref{tildes3sol} of the main text (where we recall that $\kappa_n = \frac{\tilde s_n}{n-1}$), which was obtained by a different method.
We also obtain the two point third cumulant
\be \overline{a_{i}^{2}a_{j}}^{c}=2C_{i;i,j}+C_{j;i,i}=\frac{32}{\beta^{2}}\left(\sum_{i'\neq i}2\frac{\left\langle A_{i}^{3},A_{i'}\right\rangle \left\langle A_{i}A_{j}^{2},A_{i'}\right\rangle }{\zeta_{i}-\zeta_{i'}}+\sum_{i'\neq j}\frac{\left\langle A_{j}A_{i}^{2},A_{i'}\right\rangle ^{2}}{\zeta_{j}-\zeta_{i'}}\right)
\ee
as well as the third cumulant of the gap
% One finds 
% \bea 
% && \overline{ (a_i-a_j)^3 }^c 
% = \frac{32}{\beta^2}  \bigg(  3 \sum_{i' \neq i} 
% \frac{\langle A_i^3 , A_{i'} \rangle^2 }{\zeta_i-\zeta_{i'}}  - 
% 3 \sum_{i' \neq j} 
% \frac{\langle A_j^3 , A_{i'} \rangle^2 }{\zeta_j-\zeta_{i'}} 
% \\
% && - \sum_{i' \neq i} 
% 2 \frac{\langle A_i A_{i'}, A_i^2  \rangle \langle A_i A_{i'}, A_j^2  \rangle }{\zeta_i-\zeta_{i'}}
% - 
% \sum_{i' \neq j} 
% \frac{\langle A_j A_{i'}, A_i^2  \rangle^2  }{\zeta_j-\zeta_{i'}}
% + \sum_{i' \neq j} 
% 2 \frac{\langle A_j A_{i'}, A_j^2  \rangle \langle A_j A_{i'}, A_i^2  \rangle }{\zeta_j-\zeta_{i'}}
% + 
% \sum_{i' \neq i} 
% \frac{\langle A_i A_{i'}, A_j^2  \rangle^2  }{\zeta_i-\zeta_{i'}} \bigg) 
% \eea 
\be
\label{ThirdCumulantGap}
\overline{\left(a_{i}-a_{j}\right)^{3}}^{c}=\frac{32}{\beta^{2}}\left(f_{i,j}-f_{j,i}\right)\quad,\quad f_{i,j}=\sum_{i'\neq i}\frac{3\left\langle A_{i}^{3},A_{i'}\right\rangle ^{2}-2\left\langle A_{i}^{3},A_{i'}\right\rangle \left\langle A_{i}A_{j}^{2},A_{i'}\right\rangle +\left\langle A_{i}A_{j}^{2},A_{i'}\right\rangle ^{2}}{\zeta_{i}-\zeta_{i'}} \, .
% \left( 
% 3 \langle A_i^3 , A_{i'} \rangle^2 - 2 \langle A_i^3 , A_{i'} \rangle
% \langle A_i A_j^2 , A_{i'} \rangle + \langle A_i A_j^2 , A_{i'} \rangle^2 \right) 
\ee

\subsection{Fourth cumulant}

To compute the leading contribution to the fourth cumulant we need 
the next order in perturbation theory.
The term $\delta^{(3)} a_i = O(w^3)$ in \eqref{aiperturbation} can be written explicitly as
(see \cite{LandauLifschitzQuantum, WikiPerturbation})
\be \label{order3} 
\delta^{(3)}a_{i}=-\frac{8}{\beta^{3/2}}\sum_{i'\neq i}\sum_{i''\neq i}\frac{\langle A_{i}A_{i'},w\rangle\langle A_{i'}A_{i''},w\rangle\langle A_{i''}A_{i},w\rangle}{(\zeta_{i}-\zeta_{i'})(\zeta_{i}-\zeta_{i''})}+\frac{8}{\beta^{3/2}}\left\langle A_{i}^{2},w\right\rangle \sum_{i'\neq i}\frac{\left\langle A_{i}A_{i'},w\right\rangle ^{2}}{\left(\zeta_{i}-\zeta_{i'}\right)^{2}} \, .
\ee 
The fourth cumulant will arise from two types of terms, schematically,  
$\overline{w^2 w^2 w w}^c$ and $\overline{w^3 w w w}^c$, where $w^n$ corresponds to the
$n$-th order in perturbation theory. Let us start with the first contribution.
We need the identity
\be 
\overline{\left\langle A,w\right\rangle ^{2}\left\langle B,w\right\rangle ^{2}\left\langle C,w\right\rangle \left\langle D,w\right\rangle }^{c}=4\langle A,B\rangle\langle A,D\rangle\langle B,C\rangle+4\langle A,B\rangle\langle A,C\rangle\langle B,D\rangle \, .\label{diag2} 
% \\
% && \overline{\langle A, w \rangle^3 \langle B, w \rangle  \langle C, w \rangle
% \langle D, w \rangle }^c = 6 \langle A, B \rangle  \langle A, C \rangle \langle A, D \rangle 
% \label{diag3} 
\ee 
There is a simple diagrammatic representation using Wick's theorem
to obtain \eqref{diag1} and \eqref{diag2}. %, \eqref{diag3} 
In \eqref{diag1} one draws all connected graphs with 3 vertices: $A$ a degree 2 vertex (two lines attached to it) and $B,C$ each a degree one vertex. In \eqref{diag2} there are 4 vertices: $A$ and $B$ have degree 2 and $C$, $D$ degree one. 
%In \eqref{diag3} $A$ is a degree 2 vertex and $B$, $C$, $D$ are degree one.
The number of ways to connect gives the combinatoric factor for each graph.
If one prefers an algebraic method one recall the definition
\be  
\overline{X_{1}X_{2}X_{3}X_{4}}^{c}=\partial_{\xi_{1}}\partial_{\xi_{2}}\partial_{\xi_{3}}\partial_{\xi_{4}}|_{\xi_{i}=0}\ln Z[\xi]\quad,\quad Z[\xi]=\overline{\exp\left(\sum_{i=1}^{4}\xi_{i}X_{i}\right)} \, ,\label{log}
\ee 
where in this formula $Z[\xi]$ can equivalently be replaced by $\tilde Z[\xi]=
\overline{ \prod_{i=1}^4 (1 + \xi_i X_i)}$. 
To obtain e.g. \eqref{diag2}, one replaces $X_1=\langle A, w \rangle^2$, 
$X_2=\langle B, w \rangle^2$, $X_3=\langle C, w \rangle $ and 
$X_4=\langle D, w \rangle $, one expands the product in $\tilde Z[\xi]$ and 
computes its noise average 
using Wick's theorem (up to sixth moment), and finally one
expands the logarithm in \eqref{log}. 
This leads to the first contribution 
\bea 
\left.\overline{a_{i}a_{j}a_{k}a_{\ell}}^{c}\right|_{1} &=& \frac{256}{\beta^{3}}\left(C_{i,j;k,\ell}+C_{i,k;j,\ell}+C_{i,\ell;j,k}+C_{j,k;i,\ell}+C_{j,\ell;i,k}+C_{k,\ell;i,j}\right) \, ,\\
\label{C} C_{i ,j;k ,\ell} 
&=& \sum_{i' \neq i } \sum_{j' \neq j } \frac{ \langle A_i A_{i'}, A_j A_{j'} \rangle 
}{(\zeta_i - \zeta_{i'}) (\zeta_j - \zeta_{j'}) } 
\left( \langle A_i A_{i'}, A_k^2 \rangle \langle A_j A_{j'}, A_\ell^2 \rangle 
+ \langle A_i A_{i'}, A_\ell^2 \rangle \langle A_j A_{j'}, A_k^2 \rangle 
\right) \, .
\eea 

Let us consider now the terms of the form $\overline{w^3 w w w}^c$. 
We first compute the contributions from the first term in \eqref{order3}.
It will have the form 
\be
\left.\overline{a_{i}a_{j}a_{k}a_{\ell}}^{c}\right|_{2}=\frac{64}{\beta^{3}}\left(D_{i;j,k,\ell}+D_{j;i,k,\ell}+D_{k;i,j,\ell}+D_{\ell;i,j,k}\right) \, ,
\ee
where $D_{i; j, k ,\ell}=\overline{X_1 X_j X_k X_\ell}^c$ with the choice
\be 
X_1= \sum_{i' \neq i} \sum_{i'' \neq i}
\frac{ \langle A_i A_{i'} ,w \rangle \langle A_{i'} A_{i''} ,w \rangle \langle A_{i''} A_i ,w \rangle
}{(\zeta_i-\zeta_{i'}) (\zeta_i-\zeta_{i''})} \quad , \quad X_{j,k,\ell} = \langle A_{j,k,\ell}^2 ,w \rangle \, .
\ee 
We obtain the result by first noting that the above procedure gives the simpler cumulant
\be
% && \overline{\langle A, w \rangle^2 \langle B, w \rangle^2  \langle C, w \rangle
% \langle D, w \rangle }^c = 4 \langle A, B \rangle  \langle A, D \rangle \langle B, C \rangle 
% + 4 \langle A, B \rangle  \langle A, C \rangle \langle B, D \rangle \label{diag2} \\
\overline{\left\langle A,w\right\rangle ^{3}\left\langle B,w\right\rangle \left\langle C,w\right\rangle \left\langle D,w\right\rangle }^{c}=6\langle A,B\rangle\langle A,C\rangle\langle A,D\rangle
\label{diag3} 
\ee
from which we now need to "split" the cubic vertex in three distinct points, each
with a line exiting. It gives
\be \label{D} 
D_{i; j, k ,\ell} = \sum_{i' \neq i} \sum_{i'' \neq i} 
\frac{1}{(\zeta_i-\zeta_{i'}) (\zeta_i-\zeta_{i''})} 
\left( \langle A_i A_{i'} , A_j^2 \rangle \langle A_i A_{i''} , A_k^2 \rangle \langle A_{i'} A_{i''} , A_\ell^2 \rangle + \text{5 permutations of} \, j,k,\ell
\right) 
\ee 

The contributions from the second term in \eqref{order3} can easily be obtained
since, if we ignore the denominators, it amounts to identifying $i''=i$.
Hence it reads
\be 
\left.\overline{a_{i}a_{j}a_{k}a_{\ell}}^{c}\right|_{3}=-\frac{64}{\beta^{3}}\left(\tilde{D}_{i;j,k,\ell}+\tilde{D}_{j;i,k,\ell}+\tilde{D}_{k;i,j,\ell}+\tilde{D}_{\ell;i,j,k}\right)
\ee 
where 
\bea \label{Dtilde} 
\tilde D_{i; j, k ,\ell} &:=& 2 \sum_{i' \neq i} 
\frac{1}{(\zeta_i-\zeta_{i'})^2 } 
\bigg( \langle A_i A_{i'} , A_j^2 \rangle \langle A_i A_{i'} , A_k^2 \rangle 
\langle A_{i}^2 , A_\ell^2 \rangle + 
\langle A_i A_{i'} , A_k^2 \rangle \langle A_i A_{i'} , A_\ell^2 \rangle 
\langle A_{i}^2 , A_j^2 \rangle
\nn\\
&+& \langle A_i A_{i'} , A_j^2 \rangle \langle A_i A_{i'} , A_\ell^2 \rangle 
\langle A_{i}^2 , A_k^2 \rangle
\bigg) \, .
\eea 
In total one finds the fourth cumulant
\bea \label{4cumtensor} 
\overline{a_i a_j a_k a_\ell}^c &=& 
\frac{256}{\beta^3} (C_{i, j;k ,\ell} + C_{i, k;j ,\ell} + C_{i ,\ell;j , k}
+ C_{j , k;i , \ell} + C_{j ,\ell ;i , k} + C_{k, \ell ;i ,j} ) 
\nn\\
&+& \frac{64}{\beta^3} (\hat D_{i; j, k, \ell} + \hat D_{j; i , k , \ell} + \hat D_{k;i,j,\ell} + 
\hat D_{\ell;i,j,k} ) \, ,
\eea 
where $\hat D=D-\tilde D$, $C$ is given in \eqref{C}, $D$ in \eqref{D} and 
$\tilde D$ in \eqref{Dtilde}. 
\\

{\bf One point result}. 
There is some rearrangement and one obtains 
\be
\label{4cumapp}
\overline{a_i^4 }^c = \frac{1536}{\beta^3} 
\bigg( 3 \sum_{i' \neq i } \sum_{i'' \neq i } \frac{ \langle A_i^2 , A_{i'} A_{i''} \rangle 
}{(\zeta_i - \zeta_{i'}) (\zeta_i - \zeta_{i''}) } 
\langle A_i^3 , A_{i'} \rangle \langle A_i^3 , A_{i''} \rangle 
- \sum_{i' \neq i} 
\frac{1}{(\zeta_i-\zeta_{i'})^2 } 
\langle A_i^3 , A_{i'}  \rangle^2 \langle A_i^4 , 1  \rangle  
 \bigg) 
\ee 
where we recall that $A_i(x):= \frac{{\rm Ai}(x+\zeta_i)}{{\rm Ai}'(\zeta_i)}$.
We found that for the ground state, $i=1$, a numerical computation of these sums is possible but computationally heavy, due to the rather large number of summands that must be taken into account to obtain good accuracy.
However, an alternative formula, which we give in Appendix \ref{app:Greenfunction}, Eq.~\eqref{4cumapp3}, is numerically easier to compute, and it
yields Eq.~\eqref{FourthCumKurtosis} of the main text.
%\sout{For the purpose of the computation, we summed over $2 \le i', i'' \le 80$ (this was sufficient to %yield the accuracy reported in the main text).}

\subsection{Finite $\beta$ correction to the variance}

As reported in the main text, the variance of $a_i$ is, in the leading order, $O(1/\beta)$, with the coefficient given in Eqs.~\eqref{VaraSol} and \eqref{tildes2sol} of the main text.
The next order correction to the variance is $O(1/\beta^2)$. 
It is the sum of terms of the form $\overline{w^2 w^2}^c$ and
of the form $\overline{w^3 w}^c$. Using a similar derivation to that of the calculation of the fourth cumulant, we obtain
\bea 
\label{varianceSubleadingij}
 \left.\overline{a_{i}a_{j}}^{c}\right|_{\beta^{-2}} &=& 
\frac{32}{\beta^2} \sum_{i' \neq i} \sum_{j' \neq j}
\frac{\langle A_i A_j , A_{i'} A_{j'}\rangle^2}{(\zeta_i-\zeta_{i'})(\zeta_j-\zeta_{j'}) } \nn\\
&+& 
\frac{16}{\beta^{2}}\bigg(\sum_{i'\neq i}\sum_{i''\neq i}\frac{1}{(\zeta_{i}-\zeta_{i'})(\zeta_{i}-\zeta_{i''})}\left(2\langle A_{i}A_{i'}^{2}A_{i''}\rangle\langle A_{i}A_{i''}A_{j}^{2}\rangle+\langle A_{i}^{2}A_{i'}A_{i''}\rangle\langle A_{j}^{2}A_{i'}A_{i''}\rangle\right) \nn\\
&-& \sum_{i'\neq i}\frac{1}{(\zeta_{i}-\zeta_{i'})^{2}}\left(\langle A_{i}^{2}A_{i'}^{2}\rangle\langle A_{i}^{2}A_{j}^{2}\rangle+2\langle A_{i}^{3}A_{i'}\rangle\langle A_{i}A_{j}^{2}A_{i'}\rangle\right)+i\leftrightarrow j\bigg)\,.
\eea 
The first line are terms of the form $\overline{w^2 w^2}^c$,
for which we use $\overline{ \langle A , w \rangle^2 \langle B ,w \rangle^2  }^c 
= 2 \langle A, B \rangle^2$. The second and third line
are terms of the form $\overline{w^3 w}^c$. To 
evaluate them we note that there 
are no disconnected terms (since the averages of $w^3$ and of $w$ both vanish)
and so the cumulant equals the moment, and we can apply \eqref{wick4} 
(equivalently, there are three diagrams each with a tadpole).
For the correction to the one-point variance we find (by taking $i=j$ in \eqref{varianceSubleadingij})
\bea 
\label{varianceSubleading}
 \left.\overline{a_{i}^{2}}^{c}\right|_{\beta^{-2}} &=& \frac{64}{\beta^{2}}\bigg[\sum_{i'\neq i}\sum_{i''\neq i}\frac{1}{(\zeta_{i}-\zeta_{i'})(\zeta_{i}-\zeta_{i''})}\left(\left\langle A_{i}^{2}A_{i'}A_{i''}\right\rangle ^{2}+\left\langle A_{i}A_{i'}^{2}A_{i''}\right\rangle \left\langle A_{i}^{3}A_{i''}\right\rangle \right) \nn\\
&-& \sum_{i'\neq i}\frac{1}{(\zeta_{i}-\zeta_{i'})^{2}}\left(\frac{1}{2}\left\langle A_{i}^{2}A_{i'}^{2}\right\rangle \left\langle A_{i}^{4}\right\rangle +\left\langle A_{i}^{3}A_{i'}\right\rangle ^{2}\right)\bigg]\,.
\eea 
It is technically difficult to accurately compute this sum numerically since it converges rather slowly: Indeed we found that, for $i=1$, truncating the sum at $i'=100$ and $i''=100$, we were able not able to obtain accuracy better than $\sim 1\%$. However, in Appendix \ref{app:Greenfunction} we give an alternative form of this result, Eq.~\eqref{varianceSubleading2}, together with its numerical value, which is technically easier to compute.

% The first type of terms
% are simple to evaluate and give a first contribution
% \bea 
% \overline{a_i a_j}^c|_{\beta^{-2},1} = 
% \frac{16}{\beta^2} \sum_{i' \neq i} \frac{1}{(\zeta_i-\zeta_{i'})(\zeta_j-\zeta_{j'}) }
% \overline{\langle A_i A_{i'}, w \rangle^2 \langle A_j A_{j'}, w \rangle^2 }^c
% = \frac{32}{\beta^2} \sum_{i' \neq i} \sum_{j' \neq j}
% \frac{\langle A_i A_j , A_{i'} A_{j'}\rangle^2}{(\zeta_i-\zeta_{i'})(\zeta_j-\zeta_{j'}) } 
% \eea 
% where we used that 
% $\overline{ \langle A , w \rangle^2 \langle B ,w \rangle^2  }^c 
% = 2 \langle A, B \rangle^2$.
% \bea 
% && \overline{a_i a_j}^c|_{\beta^{-2},2} =
% \frac{16}{\beta^2} \bigg( 
% \sum_{i' \neq i} \sum_{i'' \neq i} \frac{1}{(\zeta_i-\zeta_{i'}) (\zeta_i-\zeta_{i''})}
% ( 2 \langle A_i A_{i'}^2 A_{i''} \rangle \langle A_i A_{i''} A_j^2 \rangle 
% + \langle A_i^2 A_{i'} A_{i''} \rangle \langle A_j^2 A_{i'} A_{i''} \rangle ) \\
% && - \sum_{i' \neq i} \frac{1}{(\zeta_i-\zeta_{i'})^2} 
% (\langle A_i^2 A_{i'}^2 \rangle \langle A_i^2 A_{j}^2 \rangle 
% + 2  \langle A_i^3 A_{i'} \rangle \langle A_i A_j^2 A_{i'}  \rangle ) \bigg) 
% \eea 

\section{Systematic perturbation of the cubic equation near the typical value}
\label{app:perturbation}

In this Appendix, we obtain the leading-order behavior of the rate function $s(E_i)$ and the CGF $\mu(\lambda)$ up to fourth order in $E_i$ and $\lambda$ respectively, by a systematic perturbative treatment of the cubic equation \eqref{NonLinearPsiEq}.
In compact notations the cubic equation \eqref{NonLinearPsiEq} reads 
\be \label{cubicapp1}
(H_0 - E_i) \psi_i = - 4 \lambda \psi_i^3 
\ee 
where $H_0= - \partial_x^2 + x $. 
In this Appendix (as in Appendix \ref{app:thirdcum}) we use the compact notations $\langle f,g \rangle = \int_0^{+\infty} dx f(x) g(x)$,
and $\langle f, 1 \rangle = \langle f \rangle$.
We recall that $\lambda=s'(E_i)$.
Let us write $E_i = - \zeta_i + \delta E_i$ and
\be \label{decomposition} 
\psi_i= \alpha_i \psi_i^{(0)} + \delta \psi_i \quad , \quad \alpha_i= \sqrt{1- \langle \delta \psi_i^2 \rangle}
\ee 
where $\delta \psi_i$ is in the space orthogonal to $\psi^{(0)}_i$
and the last equation is obtained from normalization.
We rewrite \eqref{cubicapp1} as 
\be \label{cubicapp2}
(H_0 + \zeta_i) \delta \psi_i = \delta E_i \psi_i - 4 s'(E_i) \psi_i^3 \, .
\ee
Since $\delta \psi_i$
is in the space orthogonal to $\psi^{(0)}_i$,
this is inverted as 
\be \label{cubicapp3}
 \delta \psi_i = \delta E_i G_i \delta \psi_i - 4 s'(E_i) G_i \psi_i^3   \, ,
\ee
where we introduced the Green's function on the orthogonal subspace
% {\red N: I corrected $\psi_j^{(0)}(x) \to \psi_i^{(0)}(x)$ in the following equation}
% {\red P: By doing so you introduced an error. Have you then changed any of my equations
% below ??? }
\be
G_i = (H_0 + \zeta_i)^{-1,\perp} \quad , \quad G_i\left(x,x'\right)= \sum_{j \neq i} \frac{\psi_j^{(0)}(x) 
\psi_j^{(0)}\left(x'\right) }{\zeta_i-\zeta_j} \, .
\ee 
 Although this Green's function can be explicitly computed \cite{EPS14}, we will
ignore this fact for now, and defer that discussion to Appendix \ref{app:Greenfunction}.
% {\magenta This Green's function $G_i(x,y)$ can in fact be calculated explicitly, by solving the assoicated homogeneous equation on the intervals $x \in (0,y)$ and $x \in (y,\infty)$, and then using the boundary conditions at $x=0$ and $x\to\infty$ and the matching conditions at $x=y$. This procedure was indeed performed in Ref.~\cite{EPS14}, and the result is given by%
% \footnote{\magenta Note that the Green's function as defined in \cite{EPS14} differs from the function $G_i(x,y)$ defined here by an overall minus sign.}
% \bea
% \label{GiSol}
% G_{i}\left(x,y\right) &=& \frac{\text{Ai}\left(x+\zeta_{i}\right)\text{Ai}'\left(y+\zeta_{i}\right)+\text{Ai}'\left(x+\zeta_{i}\right)\text{Ai}\left(y+\zeta_{i}\right)}{\text{Ai}'\left(\zeta_{i}\right)^{2}} - \frac{\pi\text{Bi}'\left(\zeta_{i}\right)\text{Ai}\left(x+\zeta_{i}\right)\text{Ai}\left(y+\zeta_{i}\right)}{\text{Ai}'\left(\zeta_{i}\right)} \nn\\[1mm]
% &+&\pi\times\begin{cases}
% \text{Bi}\left(x+\zeta_{i}\right)\text{Ai}\left(y+\zeta_{i}\right), & x\le y,\\[1mm]
% \text{Ai}\left(x+\zeta_{i}\right)\text{Bi}\left(y+\zeta_{i}\right), & x>y,
% \end{cases}
% \eea
% {\red N: I verified numerically that these two expressions for the Green's function agree (this is now after I restored (C5) to its correct form).}}
Note that the r.h.s. of \eqref{cubicapp2} 
must be orthogonal to $\psi_i^{(0)}$
which implies 
\be \label{ortho}
\delta E_i \alpha_i = 4 s'(E_i) \langle \psi_i^3 , \psi_i^{(0)} \rangle \, .
\ee 
The two equations \eqref{cubicapp3} and \eqref{ortho}, upon expansion in powers of $\delta E_i$, allow to determine order by order the derivatives of the function $s(E_i)$ at $E_i=-\zeta_i$, and from there, the cumulants. 

We now illustrate the procedure to the fourth cumulant order. 
For notational simplicity we now denote 
$\delta E_i=\epsilon$, $\delta \psi_i = \sum_{n \geq 1} \epsilon^n \psi_{i,n}$
and change notation to $\psi_i^{(0)}=\psi_{i,0}$. We denote $s_n = s^{(n)}(-\zeta_i)$
the $n$-the derivative of $s(E_i)$ at $E_i=-\zeta_i$, so that we 
insert $s(E_i)= \sum_{n \geq 2} \frac{s_n}{n!} \epsilon^n$ (this notation for $s_n$ coincides with the one from \eqref{sofEPowerSeries}). We recall that the cumulants
are obtained in terms of the $s_n$ by the involution
\be 
\kappa_2= \frac{1}{s_2} \quad , \quad \kappa_3 = - \frac{s_3}{s_2^3} 
\quad , \quad \kappa_4 = \frac{3 s_3^2 - s_2 s_4}{s_2^5} 
\ee 
with the same relation $s_n \leftrightarrow \kappa_n$.

We need Eq. \eqref{cubicapp3} to order $O(\epsilon^2)$. 
Hence, inserting the decomposition \eqref{decomposition}, we need
the expansion of $\psi_i^3$ to order $O(\epsilon)$, which reads
\be
\psi_i^3  = \psi_{i,0}^3 + 3  \psi_{i,0}^2 \psi_{i,1} \epsilon  \, .
\ee
This leads to 
\bea 
\label{psii1sol}
&& \psi_{i,1}= - 4 s_2 G_i \psi_{i,0}^3 \, , \\[1mm]
\label{psii2sol}
&& \psi_{i,2}= G_i \psi_{i,1} - 12 s_2 G_i (\psi_{i,0}^2 \psi_{i,1}) - 2 s_3 G_i \psi_{i,0}^3 \, .
\eea 
We need however \eqref{ortho} to order $O(\epsilon^3)$ so we can determine the $s_n$ up to $s_4$.
Hence, inserting the decomposition \eqref{decomposition}, we now need
the expansion of $\psi_i^3$ to order $O(\epsilon^2)$, which reads
\be 
\psi_i^3 
% = \alpha^3 \psi_{i,0}^3 + 3 \alpha^2 \psi_{i,0}^2 \psi_{i,1} \epsilon + 
% 3 \alpha \psi_{i,0} \psi_{i,1}^2 \epsilon^2 
% + 3 \alpha^2 \psi_{i,0}^2 \psi_{i,2} \epsilon^2 + O(\epsilon^3)  \\
%&& 
=\left(1-\frac{3}{2}\left\langle \psi_{i,1}^{2}\right\rangle \epsilon^{2}\right)\psi_{i,0}^{3}+3\psi_{i,0}^{2}\psi_{i,1}\epsilon+3\psi_{i,0}\psi_{i,1}^{2}\epsilon^{2}+3\psi_{i,0}^{2}\psi_{i,2}\epsilon^{2}+O\left(\epsilon^{3}\right) \, .
\ee
The equation \eqref{ortho}, expanded to order $O(\epsilon^3)$ 
reads 
\bea  
&&   \epsilon\left(1-\frac{1}{2}\left\langle \psi_{i,1}^{2}\right\rangle \epsilon^{2}\right)
\nn\\
&& =4\epsilon\left(s_{2}+\frac{s_{3}}{2}\epsilon+\frac{s_{4}}{6}\epsilon^{2}\right)\left(\left\langle \psi_{i,0}^{4}\right\rangle +3\left\langle \psi_{i,0}^{3},\psi_{i,1}\right\rangle \epsilon+3\left\langle \psi_{i,0}^{2},\psi_{i,1}^{2}\right\rangle \epsilon^{2}+3\left\langle \psi_{i,0}^{3},\psi_{i,2}\right\rangle \epsilon^{2}-\frac{3}{2}\left\langle \psi_{i,1}^{2}\right\rangle \left\langle \psi_{i,0}^{4}\right\rangle \epsilon^{2}\right)+O\left(\epsilon^{4}\right)\, .\nn\\
\eea
Matching order by order in $\epsilon$, we obtain
\bea
\label{s2eq}
&& s_{2}=1/\left(4\left\langle \psi_{i,0}^{4}\right\rangle \right)\,, \\[1mm]
\label{s3eq}
&& 2s_{3}\left\langle \psi_{i,0}^{4}\right\rangle +12s_{2}\left\langle \psi_{i,0}^{3},\psi_{i,1}\right\rangle =0\,, \\[1mm]
\label{s4eq}
&& \frac{1}{2}\left\langle \psi_{i,1}^{2}\right\rangle +\frac{2}{3}s_{4}\left\langle \psi_{i,0}^{4}\right\rangle +6s_{3}\left\langle \psi_{i,0}^{3},\psi_{i,1}\right\rangle +12s_{2}\left(\left\langle \psi_{i,0}^{2},\psi_{i,1}^{2}\right\rangle +\left\langle \psi_{i,0}^{3},\psi_{i,2}\right\rangle -\frac{1}{2}\left\langle \psi_{i,1}^{2}\right\rangle \left\langle \psi_{i,0}^{4}\right\rangle \right)=0 \, .
\eea 

The first equation \eqref{s2eq} recovers the reduced variance given in the text,
$C_2=\kappa_2=\frac{1}{s_2}= 4  \int_0^{+\infty} dx \psi_i^{(0)}(x)^4$.  
The second equation \eqref{s3eq}, using the result for $s_2$ and \eqref{psii1sol}, determines $s_3$ as 
\be
\label{s3sol}
s_3 %= 24 \frac{s_2^2}{\langle \psi_{i,0}^4 \rangle}  \langle  \psi_{i,0}^3 , G_i \psi_{i,0}^3  \rangle
=\frac{3}{2}\frac{\left\langle \psi_{i,0}^{3},G_{i}\psi_{i,0}^{3}\right\rangle }{\left\langle \psi_{i,0}^{4}\right\rangle ^{3}}=\frac{3}{2\left\langle \psi_{i,0}^{4}\right\rangle ^{3}}\sum_{i'\neq i}\frac{\left\langle \psi_{i,0}^{3},\psi_{i',0}\right\rangle ^{2}}{\zeta_{i}-\zeta_{i'}}
\ee 
where we have used that 
\be 
\langle A , G_i B \rangle = \langle B , G_i A \rangle =
\sum_{i' \neq i} \frac{\langle A, \psi_{i',0} \rangle \langle B , \psi_{i',0} \rangle }{\zeta_i-\zeta_{i'}} \, .
\ee 
The result is in agreement with 
Eq.~\eqref{sofECubicOrder} in the text (together with \eqref{tildes2sol} and \eqref{tildes3sol}).

Using Eqs.~\eqref{psii1sol} and \eqref{psii2sol}, the last equation \eqref{s4eq} gives
\bea
&& \antiquad \frac{2}{3}s_{4}\left\langle \psi_{i,0}^{4}\right\rangle -16s_{2}^{2}\left\langle G_{i}\psi_{i,0}^{3},G_{i}\psi_{i,0}^{3}\right\rangle -24s_{2}s_{3}\left\langle \psi_{i,0}^{3},G_{i}\psi_{i,0}^{3}\right\rangle +192s_{2}^{3}\left\langle \psi_{i,0}^{2},\left(G_{i}\psi_{i,0}^{3}\right)^{2}\right\rangle 
\nn\\[1mm]
&& \qquad\qquad-48s_{2}^{2}\left\langle \psi_{i,0}^{3},G_{i}G_{i}\psi_{i,0}^{3}\right\rangle +576s_{2}^{3}\left\langle \psi_{i,0}^{3},G_{i}\left(\psi_{i,0}^{2}G_{i}\psi_{i,0}^{3}\right)\right\rangle -24s_{2}s_{3}\left\langle \psi_{i,0}^{3},G_{i}\psi_{i,0}^{3}\right\rangle =0 \, .
\eea 
Some terms combine using $\langle A , G_i B \rangle = \langle B , G_i A \rangle$ and we get 
\be
 \frac{2}{3}s_{4}\left\langle \psi_{i,0}^{4}\right\rangle =48s_{2}s_{3}\left\langle \psi_{i,0}^{3},G_{i}\psi_{i,0}^{3}\right\rangle -768s_{2}^{3}\left\langle \psi_{i,0}^{2},\left(G_{i}\psi_{i,0}^{3}\right)^{2}\right\rangle +64s_{2}^{2}\left\langle G_{i}\psi_{i,0}^{3},G_{i}\psi_{i,0}^{3}\right\rangle 
%- 576 s_2^3  \langle  \psi_{i,0}^3, G_i (\psi_{i,0}^2 G_i \psi_{i,0}^3 ) \rangle 
\ee 
which, using \eqref{s2eq} and \eqref{s3sol}, leads to (the first term simplifies using the expression for $s_3$)
\be
 s_4  =  \frac{3 s_3^2}{s_2} 
%288 s_2^2 s_3 \langle  \psi_{i,0}^3 , G_i \psi_{i,0}^3    \rangle 
-4608s_{2}^{4}\left\langle \psi_{i,0}^{2},\left(G_{i}\psi_{i,0}^{3}\right)^{2}\right\rangle +384s_{2}^{3}\left\langle G_{i}\psi_{i,0}^{3},G_{i}\psi_{i,0}^{3}\right\rangle \, .
%- 576 s_2^3  \langle  \psi_{i,0}^3, G_i (\psi_{i,0}^2 G_i \psi_{i,0}^3 ) \rangle 
\ee 
Using the relations between the $\kappa_n$'s and $s_n$'s, we obtain 
\be
\kappa_{4}=4608\left\langle \psi_{i,0}^{2},\left(G_{i}\psi_{i,0}^{3}\right)^{2}\right\rangle -\frac{384}{s_{2}}\left\langle G_{i}\psi_{i,0}^{3},G_{i}\psi_{i,0}^{3}\right\rangle \, .
\ee 
One has  (denoting $\zeta_{ii'}=\zeta_i-\zeta_{i'}$) 
% {\red P: we could comment the lines below to save a bit of space, just mention what is used}
\bea 
\left\langle \psi_{i,0}^{3},G_{i}\psi_{i,0}^{3}\right\rangle &=&\sum_{i'\neq i}\frac{1}{\zeta_{ii'}}\left\langle \psi_{i,0}^{3}\psi_{i'}\right\rangle ^{2}\\
\left\langle \psi_{i,0}^{2},\left(G_{i}\psi_{i,0}^{3}\right)^{2}\right\rangle &=&\sum_{i',i''\neq i}\frac{1}{\zeta_{ii'}\zeta_{ii''}}\left\langle \psi_{i'}\psi_{i,0}^{3}\right\rangle \left\langle \psi_{i''}\psi_{i,0}^{3}\right\rangle \left\langle \psi_{i'}\psi_{i''}\psi_{i,0}^{2}\right\rangle \\
\left\langle G_{i}\psi_{i,0}^{3},G_{i}\psi_{i,0}^{3}\right\rangle &=&\sum_{i',i''\neq i}\frac{1}{\zeta_{ii'}\zeta_{ii''}}\left\langle \psi_{i'}\psi_{i,0}^{3}\right\rangle \left\langle \psi_{i''}\psi_{i,0}^{3}\right\rangle \left\langle \psi_{i'}\psi_{i''}\right\rangle =\sum_{i'\neq i}\frac{1}{\zeta_{ii'}^{2}}\left\langle \psi_{i'}\psi_{i,0}^{3}\right\rangle ^{2}
\eea 
where in the last line we used the orthonormality relation
$\langle \psi_{i'} \psi_{i''} \rangle=\delta_{i',i''}$.
Hence we find 
\be
\kappa_{4}=1536\left(3\sum_{i',i''\neq i}\frac{1}{\zeta_{ii'}\zeta_{ii''}}\left\langle \psi_{i'}\psi_{i,0}^{3}\right\rangle \left\langle \psi_{i''}\psi_{i,0}^{3}\right\rangle \left\langle \psi_{i'}\psi_{i''}\psi_{i,0}^{2}\right\rangle -\left\langle \psi_{i,0}^{4}\right\rangle \sum_{i'\neq i}\frac{1}{\zeta_{ii'}^{2}}\left\langle \psi_{i'}\psi_{i,0}^{3}\right\rangle ^{2}\right) \, ,
\ee
which is identical to the result \eqref{4cumapp} obtained by a
different perturbation method.

\section{Alternative expressions using the Green's function}
\label{app:Greenfunction}

 In Ref.~\cite{EPS14} it was noted that the Green's function $G_i(x,y)$, whose
definition we recall (in both notations used above)
\be \label{GF1} 
G_i\left(x,x'\right)= \sum_{j \neq i} \frac{\psi_j^{(0)}(x) 
\psi_j^{(0)}\left(x'\right) }{\zeta_i-\zeta_j} = \sum_{j \neq i} \frac{A_j(x) 
A_j\left(x'\right) }{\zeta_i-\zeta_j} \quad , \quad A_i(x)= \frac{{\rm Ai}(x+\zeta_i)}{{\rm Ai}'(\zeta_i)}
\, 
\ee 
can in fact be calculated explicitly, by solving the associated homogeneous equation on the intervals $x \in (0,y)$ and $x \in (y,\infty)$, and then using the boundary conditions at $x=0$ and $x\to\infty$ and the matching conditions at $x=y$. The result obtained in Ref.~\cite{EPS14} reads%
\footnote{Note that the Green's function as defined in \cite{EPS14} differs from the function $G_i(x,y)$ defined here by an overall minus sign.}
\bea
\label{GiSol2}
G_{i}\left(x,y\right) &=& \frac{\text{Ai}\left(x+\zeta_{i}\right)\text{Ai}'\left(y+\zeta_{i}\right)+\text{Ai}'\left(x+\zeta_{i}\right)\text{Ai}\left(y+\zeta_{i}\right)}{\text{Ai}'\left(\zeta_{i}\right)^{2}} - \frac{\pi\text{Bi}'\left(\zeta_{i}\right)\text{Ai}\left(x+\zeta_{i}\right)\text{Ai}\left(y+\zeta_{i}\right)}{\text{Ai}'\left(\zeta_{i}\right)} \nn\\[1mm]
&+&\pi\times\begin{cases}
\text{Bi}\left(x+\zeta_{i}\right)\text{Ai}\left(y+\zeta_{i}\right), & x\le y,\\[1mm]
\text{Ai}\left(x+\zeta_{i}\right)\text{Bi}\left(y+\zeta_{i}\right), & x>y.
\end{cases}
\eea
% {\red N: I verified numerically that these two expressions for the Green's function agree (this is now after I restored (C5) to its correct form).}
Since it contains no summation or integral it leads to numerically more efficient 
alternative formula with a smaller 
number of sums.

We will now translate the main results of Appendix \ref{app:thirdcum} substituting summations of the form of \eqref{GF1} by \eqref{GiSol2}. The correction to the mean \eqref{meanCorrection} may be conveniently rewritten as
\be 
\label{meanCorrection2}
\overline{a_{i}} %= \zeta_{i}+\frac{4}{\beta}\sum_{i'\neq i}\frac{\left\langle A_{i}^{2},A_{i'}^{2}\right\rangle }{\zeta_{i}-\zeta_{i'}} 
= \zeta_{i}+\frac{4}{\beta} \int_0^{+\infty} dx \, G_i(x,x) A_i(x)^2 
\,,
\ee
coinciding with the result given in \cite{EPS14}.
The three-point third cumulant \eqref{3point3rdCumulant} reads
\bea 
\overline{ a_i a_j a_k }^c &=& C_{i;j,k} + C_{j;i,k} + C_{k;i,j} \, , \\[2mm]
C_{i;j,k} &=& %=\frac{32}{\beta^{2}}\sum_{i'\neq i}\frac{\left\langle A_{i}A_{i'},A_{j}^{2}\right\rangle \left\langle A_{i}A_{i'},A_{k}^{2}\right\rangle }{\zeta_{i}-\zeta_{i'}}
= \frac{32}{\beta^{2}} \int_0^{+\infty} dx \int_0^{+\infty} dy \, G_i(x,y) A_i(x) A_j(x)^2 A_i(y) A_k^2(y)  
\,.
\eea 
and the equal point third cumulant \eqref{1point3rdCumulant} reads 
%{\red N: I changed $A_j(y)^3 \to A_i(y)^3$ and checked this formula numerically for $i=1$}
\be \label{1point3rdCumulant2}
\overline{ a_i^3 }^c %= \frac{96}{\beta^2}  \sum_{i' \neq i} \frac{\left\langle A_i^3 , A_{i'} \right\rangle^2 }{\zeta_i-\zeta_{i'}} 
= \frac{96}{\beta^2}  
\int_0^{+\infty} dx \int_0^{+\infty} dy \, G_i(x,y) A_i(x)^3 A_i(y)^3 \, .
\ee 
The third cumulant of the gap \eqref{ThirdCumulantGap} reads
\bea 
\label{ThirdCumulantGap2}
&& \overline{\left(a_{i}-a_{j}\right)^{3}}^{c}
=\frac{32}{\beta^{2}}\left(f_{i,j}-f_{j,i}\right) \, , \\
&& f_{i,j} 
% =\sum_{i'\neq i}\frac{3\left\langle A_{i}^{3},A_{i'}\right\rangle ^{2}-2\left\langle A_{i}^{3},A_{i'}\right\rangle \left\langle A_{i}A_{j}^{2},A_{i'}\right\rangle +\left\langle A_{i}A_{j}^{2},A_{i'}\right\rangle ^{2}}{\zeta_{i}-\zeta_{i'}} 
=\int_{0}^{+\infty}dx\int_{0}^{+\infty}dy\,G_{i}(x,y)A_{i}(x)A_{i}(y)\left[3A_{i}(x)^{2}A_{i}(y)^{2}-2A_{i}(x)^{2}A_{j}(y)^{2}+A_{j}(x)^{2}A_{j}(y)^{2}\right] \, .
\eea  

For the fourth cumulant we give only the one point result \eqref{4cumapp} (the result for
$\overline{a_i a_j a_k a_\ell}^c$ can similarly be translated from 
\eqref{4cumtensor} but the result is bulky) 
% \be \label{4cumapp2}
% \overline{a_i^4 }^c 
% % = \frac{1536}{\beta^3} 
% % \bigg( 3 \sum_{i' \neq i } \sum_{i'' \neq i } \frac{ \langle A_i^2 , A_{i'} A_{i''} \rangle 
% % }{(\zeta_i - \zeta_{i'}) (\zeta_i - \zeta_{i''}) } 
% % \langle A_i^3 , A_{i'} \rangle \langle A_i^3 , A_{i''} \rangle 
% % - \sum_{i' \neq i} 
% % \frac{1}{(\zeta_i-\zeta_{i'})^2 } 
% % \langle A_i^3 , A_{i'}  \rangle^2 \langle A_i^4 , 1  \rangle  
% %  \bigg) 
% = \frac{1536}{\beta^3} 
%  \bigg( 3 
%  \int_0^{+\infty} dx \int_0^{+\infty} dy \int_0^{+\infty} dz \, 
%  G_i(x,y) G_i(x,z) A_i(x)^2 A_i(y)^3 A_i(z)^3 - \sum_{i' \neq i} 
%  \frac{1}{(\zeta_i-\zeta_{i'})^2 } 
%  \langle A_i^3 , A_{i'}  \rangle^2 \langle A_i^4 , 1  \rangle  
%   \bigg) 
% \ee 
\be \label{4cumapp2}
\overline{a_{i}^{4}}^{c}=\frac{1536}\beta^{3}\int_{0}^{+\infty}dx\int_{0}^{+\infty}dy\int_{0}^{+\infty}dz\,G_{i}(x,y)G_{i}(x,z)A_{i}\left(y\right)^{3}A_{i}\left(z\right)^{3}\left[3A_{i}\left(x\right)^{2}-\left\langle A_{i}^{4},1\right\rangle \right] \, ,
\ee 
where, to obtain the second term in the square brackets, we used that
\be
\label{Gorthogonality}
\int_{0}^{+\infty}dxG_{i}\left(x,y\right)G_{i}\left(x,z\right)=\sum_{j\neq i}\frac{A_{j}\left(y\right)A_{j}\left(z\right)}{\left(\zeta_{i}-\zeta_{j}\right)^{2}} \, ,
\ee
which follows from the orthogonality $\int_{0}^{\infty}dx A_{j}\left(x\right)A_{k}\left(x\right) = \delta_{jk}$ of the eigenfunctions.
 The formula \eqref{4cumapp2} can be further simplified by noticing that the integrations over $y$ and $z$ may be rewritten as the square of a single integral:
\be
\label{4cumapp3}
\overline{a_{i}^{4}}^{c}=\frac{1536}{\beta}^{3}\int_{0}^{+\infty}dx\left[3A_{i}\left(x\right)^{2}-\left\langle A_{i}^{4},1\right\rangle \right]\left[\int_{0}^{+\infty}dw\,G_{i}(x,w)A_{i}\left(w\right)^{3}\right]^{2}\,.
\ee

Finally, the correction \eqref{varianceSubleading} to the one-point variance reads
% \bea 
% \label{varianceSubleading2}
%  \left.\overline{a_{i}^{2}}^{c}\right|_{\beta^{-2}} &=& 
% %  \frac{64}{\beta^2} \bigg( 
% % \sum_{i' \neq i} \sum_{i'' \neq i} \frac{1}{(\zeta_i-\zeta_{i'}) (\zeta_i-\zeta_{i''})} 
% % ( \langle A_i^2 A_{i'} A_{i''} \rangle^2 + 
% % \langle A_i A_{i'}^2 A_{i''} \rangle \langle A_i^3 A_{i''} \rangle ) \nn\\
% % &-& \sum_{i'\neq i}\frac{1}{(\zeta_{i}-\zeta_{i'})^{2}}\left(\frac{1}{2}\left\langle A_{i}^{2}A_{i'}^{2}\right\rangle \left\langle A_{i}^{4}\right\rangle +\left\langle A_{i}^{3}A_{i'}\right\rangle ^{2}\right)\bigg)\,.
% \frac{64}{\beta^2} \bigg[  
%  \int_0^{+\infty} dx \int_0^{+\infty} dy \left( G_i(x,y)^2 A_i(x)^2 A_i(y)^2 + 
%  G_i(x,x) G_i(x,y) A_i(x) A_i(y)^3 \right) \nn\\
%  &-& \sum_{i'\neq i}\frac{1}{(\zeta_{i}-\zeta_{i'})^{2}}\left(\frac{1}{2}\left\langle A_{i}^{2}A_{i'}^{2}\right\rangle \left\langle A_{i}^{4}\right\rangle +\left\langle A_{i}^{3}A_{i'}\right\rangle ^{2}\right)\bigg]\,.
% \eea 
\bea
\label{varianceSubleading2}
\left.\overline{a_{i}^{2}}^{c}\right|_{\beta^{-2}} &=& \frac{64}{\beta^{2}}\int_{0}^{+\infty}dx\int_{0}^{+\infty}dy\bigg[G_{i}(x,y)^{2}A_{i}(x)^{2}\left(A_{i}(y)^{2}-\frac{1}{2}\left\langle A_{i}^{4}\right\rangle \right)+G_{i}(x,x)G_{i}(x,y)A_{i}(x)A_{i}(y)^{3} \nn\\
 &-& \int_{0}^{+\infty}dz\,G_{i}(x,y)G_{i}(x,z)A_{i}\left(y\right)^{3}A_{i}\left(z\right)^{3}\bigg] \nn\\[1mm]
 &=&  \frac{64}{\beta^{2}} \Biggl\{\int_{0}^{+\infty}dx\int_{0}^{+\infty}dy\left[G_{i}(x,y)^{2}A_{i}(x)^{2}\left(A_{i}(y)^{2}-\frac{1}{2}\left\langle A_{i}^{4}\right\rangle \right)+G_{i}(x,x)G_{i}(x,y)A_{i}(x)A_{i}(y)^{3}\right]  \nn\\
 &-&  \left[\int_{0}^{+\infty}dx\int_{0}^{+\infty}dw\,G_{i}(x,w)A_{i}\left(w\right)^{3}\right]^{2} \Biggr\}  =-\frac{0.12482\dots}{\beta^{2}}\,,%= -\frac{0.12482\dots}{\beta^2} \, ,
\eea
where again we used \eqref{Gorthogonality}.
The form of the results as given in this Appendix are more convenient, at least from the point of view of numerical computation, since they do not contain infinite sums of integrals as in Appendix \ref{app:thirdcum}.
However, it is worth noting that the form of the results given in Appendix \ref{app:thirdcum} is nevertheless useful since they may straightforwardly be extended to an arbitrary trapping potential $V_0(x)$ (by replacing the $A_i(x)$'s by its associated energy eigenfunctions). The form of the results as given in the current appendix is very useful for the particular case where $V_0(x)$ is given by \eqref{V0def}, but for a general $V_0(x)$ it would be challenging to calculate the Green's function.

\section{Some identities involving sums of integrals of Airy functions}
\label{app:identities}

Let us examine the equations  \eqref{sEphiIdentity1} and \eqref{sEphiIdentity2} %\eqref{2id} 
near the typical value,  $E_i \simeq E_i^{(0)}$. Using
$\phi_{i}(x)^{2}=4\left|\lambda\right|\psi_{i}(x)^{2}=4\left|s'\left(E_{i}\right)\right|\psi_{i}(x)^{2}$
they read
\bea %\label{2id}
&& \frac{2}{3} E_i  - \frac{s(E_i)}{s'(E_i)} =  \int_0^{+\infty} dx x \psi_i(x)^2 \, , \\[1mm]
&& E_i  - 3 \frac{s(E_i)}{s'(E_i)} = 3  \int_0^{+\infty} dx  \psi_i'(x)^2 \, .
\eea 
Let us recall that $\psi_{i,0}(x)= {\rm Ai}(x+\zeta_i)/{\rm Ai}'(\zeta_i)$.
To lowest order they give identities obeyed by Airy functions
\bea 
&& \int_0^{+\infty} dx \, x \left( \frac{{\rm Ai}(x+\zeta_i)}{{\rm Ai}'(\zeta_i)} \right)^2 = - \frac{2}{3} \zeta_i \, ,  \\[1mm]
&& \int_0^{+\infty} dx \, \left( \frac{{\rm Ai}'(x+\zeta_i)}{{\rm Ai}'(\zeta_i)} \right)^2 =
- \frac{1}{3} \zeta_i \, .
\eea 
Expanding to higher orders using 
$\frac{s(E_i)}{s'(E_i)}= \frac{\epsilon}{2} - \frac{s_3}{12 s_2} \epsilon^2 + O(\epsilon^3)$
gives different formula for the $s_n$ from those obtained so far. For instance one obtains 
% \be 
% \frac{s(E_i)}{s'(E_i)}= \frac{\epsilon}{2} - \frac{s_3}{12 s_2} \epsilon^2 
% + \frac{s_3^2- s_2 s_4}{24 s_2^2} \epsilon^3 + O(\epsilon^4) 
% \ee 
% The next order gives 
\be 
\frac{1}{6}=2\left\langle x\psi_{i,0}\psi_{i,1}\right\rangle =-8s_{2}\left\langle x\psi_{i,0},G_{i}\psi_{i,0}^{3}\right\rangle =-8s_{2}\sum_{i'\neq i}\frac{\left\langle x\psi_{i,0},\psi_{i',0}\right\rangle \left\langle \psi_{i,0}^{3},\psi_{i',0}\right\rangle }{\zeta_{i}-\zeta_{i'}}
\ee 
which gives another determination of $s_2$. Comparing with our previous result
$s_{2}=1/\left(4\left\langle \psi_{i,0}^{4}\right\rangle \right)$,
it gives a non-trivial sum rule
\be 
\sum_{i'\neq i}\frac{\left\langle x\psi_{i,0},\psi_{i',0}\right\rangle \left\langle \psi_{i,0}^{3},\psi_{i',0}\right\rangle }{\zeta_{i}-\zeta_{i'}}=-\frac{1}{12}\left\langle \psi_{i,0}^{4}\right\rangle \,,
\ee 
and more sum rules can be obtained by considering higher orders.

\section{Legendre transform properties}
\label{appendix:Legendre}

%{\red P: I added some missing tilde please recheck}

Using the Legendre transform relations $s'(E) = ds/dE = \lambda$ and $\mu'(\lambda) = d\mu/d\lambda = E$ between the rate function and the CGF, we have that $s'$ and $\mu'$ are inverse functions
\be 
s'( \mu'(\lambda)) = \lambda \quad , \quad \mu'(s'(E))= E \, .
\ee 
Centering the function around $E_{\rm typ}=\mu'(0)$, we define
$\tilde \mu(\lambda)=\mu(\lambda)-\mu'(0) \lambda$
and $s(E)= \tilde s(\epsilon = E-E_{\rm typ})$. One still has
\be 
\label{sPrimeMuPrime}
\tilde s'( \tilde \mu'(\lambda)) = \lambda \quad , \quad \tilde \mu'(\tilde s'(\epsilon))= \epsilon \, .
\ee 
We now expand these functions as power series $\tilde \mu(\lambda)=\sum_{n \geq 2} \frac{\kappa_n}{n!} \lambda^n$
and $\tilde s(\epsilon) =\sum_{n \geq 2} \frac{s_n}{n!} \epsilon^n$, where the sums start at $n=2$ due to the centering. 
By plugging these series expansions into \eqref{sPrimeMuPrime}, we can express coefficients of one series using those of the other. For instance, keeping terms up to fourth order, we obtain
\be
\epsilon = \tilde \mu'(\tilde s'(\epsilon)) = \kappa_{2}s_{2}\epsilon+\frac{1}{2}\left(\kappa_{3}s_{2}^{2}+\kappa_{2}s_{3}\right)\epsilon^{2}+\frac{1}{6}\left(\kappa_{4}s_{2}^{3}+3\kappa_{3}s_{3}s_{2}\right)\epsilon^{3}+O\left(\epsilon^{4}\right) \, ,
\ee
and solving for the $\kappa_n$'s, we find
\be 
\kappa_2= \frac{1}{s_2} \quad , \quad \kappa_3 = - \frac{s_3}{s_2^3} 
\quad , \quad \kappa_4 = \frac{3 s_3^2 - s_2 s_4}{s_2^5} \, .
\ee 
The relation between the $\kappa_n$'s and $s_n$'s is an involution, we similarly have
\be 
s_2= \frac{1}{\kappa_2} \quad , \quad s_3 = - \frac{\kappa_3}{\kappa_2^3} 
\quad , \quad s_4 = \frac{3 \kappa_3^2 - \kappa_2 \kappa_4}{\kappa_2^5} \, ,
\ee 
which is Eq.~\eqref{sOfKappa}  of the main text.

\section{Saddle-point equations for gap distributions}
\label{app:gap}

Following similar steps to the derivation of the saddle-point equation for a single energy level $E_i$ (see Sec.~\ref{sec:SaddlePointGeneral}), we minimize the action functional \eqref{sVdef} and add a Lagrange multiplier to enforce the constraint on a given value of $E_{ij} = E_i - E_j$, i.e., we minimize the modified action
\be
\mathcal{S}_{\lambda}\left[V\right]=\mathcal{S}\left[V\right]-\lambda\left(\mathcal{E}_{i}\left[V\right]-\mathcal{E}_{j}\left[V\right]-E_{ij}\right)\,.
\ee
The variation of the modified action is given by
\be
\delta\mathcal{S}_{\lambda}=\frac{1}{4}\int_{-\infty}^{+\infty}\left[V(x)-V_{0}(x)-4\lambda\psi_{i}(x)^{2}+4\lambda\psi_{j}(x)^{2}\right]\delta V\left(x\right)dx+O\left(\delta V^{2}\right)\,,
\ee
and requiring it to vanish yields Eq.~\eqref{V1psiipsijlambda} of the main text, which must be solved together with the Schr\"{o}dinger equations \eqref{psiieq} and \eqref{psijeq} (with values of $E_i$ and $E_j$ that are a priori unknown, but are determined by $E_{ij}$ and $\lambda$).
Finally, after solving the saddle-point equations, the rate function $s(E_{ij})$ is obtained by evaluating the action integral \eqref{sVdef} on the optimal realization of the disorder $V_1(x)$.

These saddle-point equations are numerically solvable using the  algorithm described around Eq.~\eqref{Vnplus1}. The input of the algorithm is $\lambda$. At each iteration, given a candidate for the realization of the disorder $V_{1,n}(x)$ from the previous iteration, one numerically computes the wave functions corresponding to the $i$th and $j$th energy levels. One then computes $V_{1,n+1}(x)$ from Eq.~\eqref{Vnplus1} [or Eq.~\eqref{Vnplus1Convex} with some $0 < \alpha < 1$ to improve the stability] with $\lambda_i=-\lambda_j=\lambda$. This is the algorithm we used to plot the numerical data in Fig.~\ref{figSofaGap}.

Finally, we note that the saddle-point equations for the gap distributions may also be derived by minimizing the rate function $s(E_i,E_j)$ of the joint distribution constrained on the difference $E_i - E_j$, with the same result, but we do not present this alternative derivation here.

\section{Behavior of $v(t)$ near $t=0$}
\label{app:vt0}

We rewrite the differential equation \eqref{NonLinearvEq} as %with the + sign gives
\begin {equation}
\frac{d}{dt}\left( \dot v^{2}- \sigma v^{4}+Ev^{2}-tv^{2}\right)=-v^{2} \, ,
\end {equation}
where we recall that $\sigma=\pm 1$ depending whether we are studying the
$E$ larger or smaller than the typical value.
Integrating over the time interval $[t,+\infty[$  and using the fact that $v(t)$ vanishes for  large $t$ we get
\begin {equation}
\dot v^{2}-\sigma v^{4}+Ev^{2}-tv^{2}=\int_{t}^{+\infty}v^{2}(\tau)d\tau \, .
\end {equation}
Plugging in $t=0$ and using the Dirichlet boundary condition, %at $t=0$ 
one obtains
\begin {equation}
\label{dvdt0}
\dot v(0)^{2}=\int_{0}^{+\infty}v^{2}(\tau)d\tau \, .
\end {equation}
Eq.~\eqref{NonLinearvEq} gives
\begin {equation}
\frac{\ddot v}{v} -(t-E)= 2 \sigma v^2 \, .
\end {equation}
Differentiating with respect to $E$  and mutipling by $v^2$ we get
\begin {equation}
v\frac{d^3v}{dt^2dE}- \frac{d^2 v}{dt^2}\frac{dv}{dE} +v^2=4 \sigma v^3\frac{dv}{dE} \, .
\end {equation}
The first two terms can be written as a total derivative with respect to $t$. Integrating over $[0,\infty]$ gives
\begin {equation}
-v(0)\frac{d^2v}{dtdE}(0)+ \frac{dv}{dt}(0)\frac{dv}{dE}(0) +\int_0^{+\infty} v^2 d\tau=4  \sigma \int_0^{+\infty}v^3\frac{dv}{dE} d\tau \, .
\end {equation}
The right hand side can be expressed in terms of  the derivative with respect to $E$ of the classical action $S(E)=\frac{1}{2}\int_0^{+\infty} v^4(\tau) d\tau$.
Then, by using the small time expansion of $ v(t,E) $
\begin{equation}
v(t,E)=a(E)t+b(E)t^3+...
\end{equation}
together with Eq.~\eqref{dvdt0}, one obtains our final result
\begin{equation}
a^2(E)=2 \sigma S'(E) =  2 |S'(E)|\, .
\end{equation}
where in the last line we have used $\sigma=\sigma_E={\rm sgn} (S'(E))$
which we know from
the first part of the paper.

\end{widetext}

\end{document}